\documentclass[showpacs,preprintnumbers,amsmath,amssymb]{revtex4}

\usepackage{graphicx}
\usepackage{dcolumn}
\usepackage{bm}

\newcommand{\uhs}{{\rm uhs}}
\def\Sumprime{\mathop{{\sum}'}}
\newtheorem{theorem}{Theorem}

\begin{document}

\title{Statistics of Fourier Modes in Non-Gaussian Fields}

\author{Takahiko Matsubara}
 \email{taka@nagoya-u.jp}
\affiliation{%
Department of Physics, Nagoya University,
Chikusa, Nagoya, 464-8602, Japan, and\\
Institute for Advanced Research, Nagoya University,
Chikusa, Nagoya, 464-8602, Japan
}%

\date{\today}

\begin{abstract}
    Fourier methods are fundamental tools to analyze random fields.
    Statistical structures of homogeneous Gaussian random fields are
    completely characterized by the power spectrum. In non-Gaussian
    random fields, polyspectra, higher-order counterparts of the power
    spectrum, are usually considered to characterize statistical
    information contained in the fields. However, it is not trivial
    how the Fourier modes are distributed in general non-Gaussian
    fields. In this paper, distribution functions of Fourier modes are
    directly considered and their explicit relations to the
    polyspectra are given. Under the condition that any of the
    polyspectra does not diverge, the distribution function is
    expanded by dimensionless combinations of polyspectra and a total
    volume in which the Fourier transforms are performed. The
    expression up to second order is generally given, and higher-order
    results are also derived in some cases. A concept of $N$-point
    distribution function of Fourier modes are introduced and
    explicitly calculated. Among them, the one-point distribution
    function is completely given in a closed form up to arbitrary
    order. As an application, statistics of Fourier phases are explored
    in detail. A number of aspects regarding statistical properties of
    phases are found. It is clarified, for the first time, how phase
    correlations arise in general non-Gaussian fields. Some of our
    analytic results are tested against numerical realizations of
    non-Gaussian fields, showing good agreements.
\end{abstract}

\pacs{
02.30.Nw,
05.40.-a,
98.65.Dx
}
\maketitle

\section{\label{sec:intro}
Introduction
}

The Fourier analysis has long been one of the most important methods
in various fields of study. Among them, randomly varying data are
efficiently analyzed by Fourier transform methods, or spectral
methods. In random processes as functions of time, the Fourier
transform plays one of the central roles in unveiling the statistical
nature of the process \cite{rice1944, rice1945, ya86, th92}. A random
process is represented by a random function defined over 1-dimensional
space. Generalizations to multi-dimensional spaces are
straightforward, and random functions defined over some space are
called random fields \cite{long1957, adler1981}. A random field is
called {\it homogeneous} when all the statistical properties are
translationally invariant and independent on a position in space. In
a one-dimensional case, such as a random process, homogeneous random
fields are also called {\em stationary}. Fourier analyses of random
processes or random fields are especially useful when they are
stationary or homogeneous.

There is a class of random fields which are called Gaussian random
fields. When a field arises from the superposition of a large number
of independent random effects, the resulting field is a Gaussian
random field under very weak conditions, by virtue of the so-called
{\em central limit theorem}. There are many situations encountered in
physical and engineering problems that random fields are accurately or
approximately Gaussian. Statistical properties of a homogeneous
Gaussian random field is completely specified by a correlation
function, or equivalently, its Fourier transform, a power spectrum.
Therefore, Gaussian random fields are analytically easier to treat
than other random fields, and many aspects of Gaussian random fields
have been investigated before \cite{adler1981}.

In cosmological physics, Gaussian random fields are very important,
since the initial density fluctuations are thought to be, at least
approximately, a Gaussian random field. In inflationary models, the
curvature perturbations generated by quantum fluctuations in the very
early universe would yield a Gaussian random density field. Statistics
of Gaussian random fields in cosmological contexts have been explored
in detail \cite{dor1970, bbks1986}.

However, non-Gaussianity naturally appears in reality, especially when
nonlinear dynamics are involved. While linear evolution of an initial
density field in cosmology keeps the Gaussianity, nonlinear evolution
raises non-Gaussianity in the density field \cite{pe1980}. Moreover,
whether or not the initial density field is a Gaussian random field is
an important question to investigate the beginning of the universe,
and understanding non-Gaussian random fields has a great importance. 

The non-Gaussian features in Fourier analyses are usually
characterized by polyspectra, i.e., higher-order counterparts of the
power spectra. For a given set of random variables, a set of all
moments of the variables carries complete information of the
statistical distribution. However, it is still not trivial to
understand the statistical properties of the Fourier modes, even when
a hierarchy of all the polyspectra are known. For example, statistical
behaviors of Fourier phases in non-Gaussian random fields has long
been studied empirically, lacking any well-established statistical
framework to understand information contained in the Fourier phases
\cite{ss1991, sms1991, jb1998, cc2000, wc2003, wcm2003}.

The statistical structures of Fourier phases are recently unveiled by
the present author, properly calculating the joint distribution
function of phases \cite{mat2003}. The analytical results are tested
against numerical simulations of cosmological structure formation
\cite{hik2004}, and applied to an analysis of observational data
\cite{hik2005}. The analytical calculations involve an expansion by a
certain parameter, and only lowest-order terms were discussed in the
previous work.

In this paper, the previous analytical calculations are extended to
give a framework for studying statistics of Fourier modes in general.
The statistical structures of not only Fourier phases but also Fourier
moduli are simultaneously considered. We give a general joint
distribution function of Fourier modes in general non-Gaussian fields.
The joint distribution function of Fourier modes has complete
information of statistical properties. Since a set of all the
polyspectra also has the complete information, the joint distribution
function can, in principle, be expressible by a set of all the
polyspectra. Such relation has not been known, and it is surely
difficult to obtain the complete relation. However, the distribution
function is found to inevitably depend on a total volume in which
Fourier transforms are applied. Some dimensionless combinations of a
polyspectrum and a total volume, each of which is a dimensional
quantity, can actually be arbitrary small for a large volume limit.
Therefore, it is natural to expand the distribution function by some
powers of a volume. The expression is found to be expanded by an
inverse of square root of the volume as a small parameter. The way to
calculate the distribution function up to arbitrary order is
established in this paper. The general expression up to second order
is actually given. More general properties involving higher-order
effects are also obtained and discussed in some cases. Remarkably, an
analytic expression of one-point distribution function of a Fourier
mode is obtained to arbitrary order in a closed form.

This paper is organized as follows. In Sec.~\ref{sec:Gaussian}, our
notations of Fourier transforms are defined, and the distribution
function of Fourier modes in Gaussian random fields are reviewed. In
Sec.~\ref{sec:GenDist}, methods to derive distribution function in
general non-Gaussian fields are detailed, and an explicit expression
up to second order is obtained. In Sec.~\ref{sec:NptDist}, $N$-point
distribution functions of Fourier modes are introduced, and calculated
in several cases. An analytic expression of the one-point distribution
function is completely given in a closed form. General expression of
the two-, three-, four-point distribution functions are given up to
second order. In Sec.~\ref{sec:PhaseCorr}, statistical structures of
Fourier phases are revealed. $N$-point distribution functions of
phases are derived, and a number of theorems regarding general
properties of phase distributions are proven. In
Sec.~\ref{sec:numerical}, some of the analytic results are tested
against numerical realizations of non-Gaussian fields. The complete form
of the joint distribution function up to second order is given in
Appendix~\ref{app:a}.

\section{\label{sec:Gaussian}
Gaussian Random Fields and Fourier Modes}

\subsection{\label{subsec:power_spectrum}
Fourier Transform and Power spectrum
}

For a given random field $f(\bm{x})$ in a $d$-dimensional, Euclidean
configuration space, $\bm{x} \in \mathbb{R}^d$, the Fourier
coefficient $\tilde{f}(\bm{k})$ is defined by
\begin{equation}
  \tilde{f}(\bm{k}) = \int d^d\!x f(\bm{x}) e^{-i\bm{k}\cdot\bm{x}}.
\label{eq:1-1}
\end{equation}
Without loss of generality, we assume the field has a zero mean:
$\langle f(\bm{x}) \rangle = 0$. The power spectrum, $P(\bm{k})$, is
defined by the relation
\begin{equation}
  \left\langle \tilde{f}^*(\bm{k}) \tilde{f}(\bm{k}') \right\rangle =
  (2\pi)^d \delta^d(\bm{k} - \bm{k}') P(\bm{k}),
\label{eq:1-2}
\end{equation}
where $\delta^d$ is the $d$-dimensional Dirac's delta function. The
average $\langle \cdots \rangle$ represents an ensemble average. In an
isotropic field, the power spectrum is an function of only an absolute
length of the wavevector $\bm{k}$. However, we do not assume the
isotropy of space in principle, so that the configuration space can be
anisotropic in general. The appearance of the Dirac's delta function
in equation (\ref{eq:1-2}) is a consequence of statistical homogeneity
of space. In this paper, we only assume the homogeneity of
configuration space. Since the right hand side of equation
(\ref{eq:1-2}) is non-zero only when $\bm{k}=\bm{k}'$, the power
spectrum is always a real function.

In this paper, we consider the field $f(\bm{x})$ to be real:
$f(\bm{x}) \in \mathbb{R}$. In this case, the Fourier coefficient of
Eq.~(\ref{eq:1-1}) satisfies
\begin{equation}
  \tilde{f}^*(\bm{k}) = \tilde{f}(-\bm{k}).
\label{eq:1-3}
\end{equation}
Therefore, the power spectrum of Eq.~(\ref{eq:1-2}) is equivalently
given by
\begin{equation}
  \left\langle \tilde{f}(\bm{k}) \tilde{f}(\bm{k}') \right\rangle =
  (2\pi)^d \delta^d(\bm{k} + \bm{k}') P(\bm{k}).
\label{eq:1-4}
\end{equation}


\subsection{\label{subsec:gaussian_pdf}
Probability Distribution
Function of Fourier Modes in Random Gaussian Fields }

In random Gaussian fields, statistical properties of field
distribution are completely specified by the two-point correlation
function,
\begin{equation}
  \xi(\bm{x}-\bm{x}') = 
  \left\langle f(\bm{x}) f(\bm{x}') \right\rangle.
\label{eq:1-5}
\end{equation}
In a homogeneous field, the two-point correlation function is a
function of only a displacement of the two points. The power spectrum
is a Fourier transform of the two-point correlation function
(Wiener-Khinchin theorem):
\begin{equation}
  P(\bm{k}) = \int d^d\!x e^{-i\bm{k}\cdot\bm{x}} \xi(\bm{x}).
\label{eq:1-6}
\end{equation}
Therefore, statistical properties of random Gaussian fields are
completely contained in the power spectrum, since the power spectrum
and the two-point correlation function are mathematically equivalent.
Higher-order moments in Gaussian fields are fully reduced to
combinations of two-point correlation functions, and higher-order
cumulants are zero:
\begin{equation}
  \left\langle 
    f(\bm{x}_1) f(\bm{x}_2) \cdots f(\bm{x}_N)
  \right\rangle_{\rm c} = 0,
  \qquad (N \geq 3).
\label{eq:1-7}
\end{equation}
The probability of taking particular values of a random
Gaussian field $f(\bm{x})$ is given by a functional,
\begin{equation}
  \mathcal{P}[f(\bm{x})] =
  A \exp\left[
    - \frac12 \int d^d\!x d^d\!x'
    f(\bm{x}) \xi^{-1}(\bm{x} - \bm{x}') f(\bm{x}')
  \right],
\label{eq:1-8}
\end{equation}
where $A$ is a normalization constant, and the function
$\xi^{-1}(\bm{x}-\bm{x}')$ is the ``inverse'' of the two-point
correlation function, which is implicitly defined by
\begin{equation}
  \int d^d\!x'' \xi(\bm{x} - \bm{x}'') \xi^{-1}(\bm{x}'' - \bm{x}') 
  = \delta^d (\bm{x} - \bm{x}').
\label{eq:1-9}
\end{equation}
Using the Fourier representation, this ``inverse'' correlation
function is explicitly represented by 
\begin{equation}
  \xi^{-1}(\bm{x}) = 
  \int \frac{d^d\! k}{(2\pi)^d}
  \frac{e^{i\bm{k}\cdot\bm{x}}}{P(k)}.
\label{eq:1-10}
\end{equation}
The functional of Eq.~(\ref{eq:1-8}) is a generalization of the
multi-variate Gaussian distribution function to the continuum case.
This functional contains all the statistical information of the field,
and it is obvious from Eq.~(\ref{eq:1-8}) that the statistical properties
of a random Gaussian field are fully described by the two-point
correlation function, as claimed above.

The distribution function of the Fourier coefficients is
straightforwardly obtained from the Eq.~(\ref{eq:1-8}) by changing the
variables. Since the Fourier transform is a linear
transform, the Jacobian of the transform is a constant.
The integral in Eq.~(\ref{eq:1-8}) is given by
\begin{equation}
  \int d^d\!x d^d\!x'
    f(\bm{x}) \xi^{-1}(\bm{x} - \bm{x}') f(\bm{x}') =
  2 \int_{\rm uhs} \frac{d^d\!k}{(2\pi)^d}
  \frac{|\tilde{f}(\bm{k})|^2}{P(\bm{k})},
\label{eq:1-11}
\end{equation}
where ``uhs'' indicates the integration over the independent modes
taking the reality condition of Eq.~(\ref{eq:1-3}) into account.
Usually, one can take the ``upper half sphere'', $k_z \geq 0$ of the
$\bm{k}$-space.

At this point, introducing a regularization box of volume $V=L_1 L_2
\cdots L_d$ with periodic boundary condition is useful to discretize
the Fourier space. The size of fundamental cells in Fourier space is
given by $\Delta k_i = 2\pi/L_i$ for the $i$-th direction ($i = 1, 2,
\ldots, d$), and the right hand side of Eq.~(\ref{eq:1-11}) becomes a
sum over discrete set of $\bm{k}$ in this representation. Defining the
volume-normalized Fourier coefficient,
\begin{equation}
  f_{\bm{k}} \equiv \frac{\tilde{f}(\bm{k})}{\sqrt{V}},
\label{eq:1-12}
\end{equation}
the joint probability distribution function of the Fourier
coefficients are given by
\begin{equation}
  \mathcal{P}[f_{\bm{k}}] =
  \tilde{A} \prod_{\bm{k}}^{\rm uhs}
  \exp\left[
  - \frac{|f_{\bm{k}}|^2}{P(\bm{k})}
  \right],
\label{eq:1-13}
\end{equation}
where $\tilde{A}$ is a normalization constant, which differs from $A$
by a Jacobian of the Fourier transform. The square bracket indicates
the function of all Fourier modes in uhs, and not of only a particular
mode. It is obvious in Eq.~(\ref{eq:1-13}) that each Fourier mode in
uhs is independent to each other in random Gaussian fields. Denoting
$f_{\bm{k}} = a_{\bm{k}} + i b_{\bm{k}}$, where $a_{\bm{k}} \in
\mathbb{R}$ and $b_{\bm{k}} \in \mathbb{R}$ are real and imaginary
parts respectively, their distribution functions are given by
\begin{subequations}
\begin{equation}
  P(a_{\bm{k}}) = \frac{1}{\sqrt{\pi P(\bm{k})}}
  \exp\left[ - \frac{{a_{\bm{k}}}^2}{P(\bm{k})} \right],
\label{eq:1-14a}
\end{equation}
\begin{equation}
  P(b_{\bm{k}}) = \frac{1}{\sqrt{\pi P(\bm{k})}}
  \exp\left[ - \frac{{b_{\bm{k}}}^2}{P(\bm{k})} \right].
\label{eq:1-14b}
\end{equation}
\end{subequations}
The real and imaginary parts of Fourier modes are independently
distributed with rms $P(\bm{k})/2$ in Gaussian random fields. In terms
of the polar representation, $f_{\bm{k}} =
|f_{\bm{k}}|e^{i\theta_{\bm{k}}}$, the distribution function of each
mode is given by
\begin{equation}
  P(|f_{\bm{k}}|,\theta_{\bm{k}}) d|f_{\bm{k}}| d\theta_{\bm{k}} =
  \exp\left[
  - \frac{|f_{\bm{k}}|^2}{P(\bm{k})}
  \right]
  \frac{2 |f_{\bm{k}}| d|f_{\bm{k}}|}{P(\bm{k})}
  \frac{d\theta_{\bm{k}}}{2\pi}
\label{eq:1-15}
\end{equation}
This distribution function does not depend on the Fourier phase
$\theta_{\bm{k}}$, so that the Gaussian random fields always have
random phases. The Fourier modulus $|f_{\bm{k}}|$ obeys the Rayleigh
distributions in Gaussian random fields.

Thus, the statistics of Fourier modes in Gaussian random fields are
particularly simple: Distributions of every real and imaginary parts
of Fourier modes are completely independent to each other, and
their distributions are Gaussian. In other words, the
distribution of Fourier modulus is given by a Rayleigh distribution and
the Fourier phase is completely random.

In what follows, we will investigate how this simplicity changes in
non-Gaussian fields. We will see both independence of Fourier modes
and Gaussianity of Fourier coefficients (and therefore randomness of
phases) are broken in general.


\section{\label{sec:GenDist}
General Distribution Function of Fourier Modes
}

\subsection{
A formal expansion of non-Gaussian distribution functions
}

One of the main purpose of this paper is to elucidate the general form
of the joint probability distribution function of the Fourier
coefficients $P[f_{\bm{k}}]$. For this purpose, we use the celebrated
cumulant expansion theorem \cite{ma85}. In general, the cumulant
expansion theorem states that the following identity holds for a
random variable $X$:
\begin{equation}
  \left\langle e^{-i X} \right\rangle =
  \exp\left(
    \sum_{N=1}^\infty \frac{(-i)^N}{N!}
    \left\langle X^N \right\rangle_{\rm c}
  \right),
\label{eq:2-1}
\end{equation}
where $\langle \cdots \rangle$ indicates taking a usual average value,
and $\langle \cdots \rangle_{\rm c}$ indicates taking cumulants. We
put
\begin{equation}
  X = \sum_{\bm{k}}^{\rm uhs}
  K_{\bm{k}} a_{\bm{k}} +  L_{\bm{k}} b_{\bm{k}},
\label{eq:2-2}
\end{equation}
in the above equation, where $K_{\bm{k}}$ and $L_{\bm{k}}$ are just
usual numbers, and variables $a_{\bm{k}}$ and $b_{\bm{k}}$ are the
real and imaginary parts of the Fourier coefficients, $f_{\bm{k}} =
a_{\bm{k}} + i b_{\bm{k}}$ ($a_{\bm{k}} \in \mathbb{R}$, $b_{\bm{k}}
\in \mathbb{R}$). The summation is taken over any subset of
independent Fourier modes. Since a mode with a wavevector $\bm{k}$ in
a lower half sphere is always given by a mode with a wavevector
$-\bm{k}$ in real fields, the wavevectors in the summation are taken
only from uhs.

With this substitution, the left hand side
of Eq.~(\ref{eq:2-1}) has the form of the Fourier transform of the
probability distribution function $P[a_{\bm{k}}, b_{\bm{k}}]$.
Therefore, performing inverse Fourier transforms, we obtain
\begin{equation}
  \mathcal{P}[a_{\bm{k}}, b_{\bm{k}}] =
  \int \prod_{\bm{k}}^{\rm uhs}
  \frac{dK_{\bm{k}}}{2\pi} \frac{dL_{\bm{k}}}{2\pi} \cdot
  e^{i X}
  \exp\left(
    \sum_{N=1}^\infty \frac{(-i)^N}{N!}
    \left\langle X^N \right\rangle_{\rm c}
  \right),
\label{eq:2-3}
\end{equation}

Since the original field $f(\bm{x})$ is assumed to have a zero mean,
the first moments and cumulants vanishes: $\langle a_{\bm{k}} \rangle
= \langle a_{\bm{k}} \rangle_{\rm c} = \langle b_{\bm{k}} \rangle =
\langle b_{\bm{k}} \rangle_{\rm c} = 0$. Thus the term of $N=1$ in the
last exponent of Eq.~(\ref{eq:2-3}) vanishes. Substituting
Eq.~(\ref{eq:2-2}) into Eq.~(\ref{eq:2-3}), using the binomial
theorem, and expanding the sum over wavenumbers, we obtain
\begin{multline}
  \mathcal{P}[a_{\bm{k}},b_{\bm{k}}]
  = 
  \int \prod_{\bm{k}}^{\rm uhs}
  \frac{dK_{\bm{k}}}{2\pi} \frac{dL_{\bm{k}}}{2\pi}
\\
  \times
  \exp\left[
    \sum_{m + n \geq 2}
    \frac{(-i)^{m+n}}{m! n!}
    \sum_{\bm{k}_1}^{\rm uhs} \cdots \sum_{\bm{k}_m}^{\rm uhs}
    \sum_{\bm{l}_1}^{\rm uhs} \cdots \sum_{\bm{l}_n}^{\rm uhs}
    \left\langle
      a_{\bm{k}_1} \cdots a_{\bm{k}_m}
      b_{\bm{l}_1} \cdots b_{\bm{l}_n}
    \right\rangle_{\rm c}
    K_{\bm{k}_1} \cdots K_{\bm{k}_m}
    L_{\bm{l}_1} \cdots L_{\bm{l}_n}
  \right]
\\
  \times
  \exp\left[
    i \sum_{\bm{k}}^{\rm uhs}
    \left(
      K_{\bm{k}} a_{\bm{k}} +  L_{\bm{k}} b_{\bm{k}}
    \right)
  \right].
\label{eq:2-4}
\end{multline}
where the sum over all non-negative integers $m$ and $n$ which
satisfies $m+n\geq 2$ is taken. We single out the term of $n+m=2$ in
the first exponent, adopt substitutions $K_{\bm{k}} \rightarrow
-i\partial/\partial a_{\bm{k}}$,  $L_{\bm{k}} \rightarrow
-i\partial/\partial b_{\bm{k}}$ in the rest terms, and perform the
Fourier integrations. We obtain
\begin{multline}
  \mathcal{P}[a_{\bm{k}},b_{\bm{k}}]
  = 
  \exp\left[
    \sum_{m + n \geq 3}
    \frac{(-1)^{m+n}}{m! n!}
    \sum_{\bm{k}_1}^{\rm uhs} \cdots \sum_{\bm{k}_m}^{\rm uhs}
    \sum_{\bm{l}_1}^{\rm uhs} \cdots \sum_{\bm{l}_n}^{\rm uhs}
    \left\langle
      a_{\bm{k}_1} \cdots a_{\bm{k}_m}
      b_{\bm{l}_1} \cdots b_{\bm{l}_n}
    \right\rangle_{\rm c}
  \right.
\\
  \left. \times
    \frac{\partial}{\partial a_{\bm{k}_1}} \cdots
    \frac{\partial}{\partial a_{\bm{k}_m}}
    \frac{\partial}{\partial b_{\bm{k}_1}} \cdots
    \frac{\partial}{\partial b_{\bm{k}_n}}
  \right]
  \mathcal{P}_{\rm G}[a_{\bm{k}},b_{\bm{k}}],
\label{eq:2-5}
\end{multline}
where $P_{\rm G}$ is a Gaussian distribution function, given by
products of Eq.~(\ref{eq:1-14a}) and (\ref{eq:1-14b}):
\begin{equation}
  \mathcal{P}_{\rm G}[a_{\bm{k}},b_{\bm{k}}] =
  \prod_{\bm{k}}^{\rm uhs}
  \frac{1}{\pi P(\bm{k})}
  \exp\left[-\frac{{a_{\bm{k}}}^2 + {b_{\bm{k}}}^2}{P(\bm{k})}\right].
\label{eq:2-6}
\end{equation}
The Eqs.~(\ref{eq:2-5}) and (\ref{eq:2-6}) are the fundamental
equations of expressing the non-Gaussian distribution function of
Fourier coefficients in terms of higher-order cumulants. 

At this point, it is more convenient to use complex variables
$f_{\bm{k}}$ as an independent variables instead of $a_{\bm{k}}$ and
$b_{\bm{k}}$. One can consider simultaneous linear transformations of
independent variables:
\begin{equation}
  \left\{
  \begin{array}{l}
    f_{\bm{k}} = a_{\bm{k}} + i b_{\bm{k}}, \\
    f_{-\bm{k}} = a_{\bm{k}} - i b_{\bm{k}},
  \end{array}
  \right.\ \ 
  \left(\bm{k} \in {\rm uhs} \right).
\label{eq:2-7}
\end{equation}
With this definitions, $f_{\bm{k}}$ is defined in all $\bm{k}$ space,
while $a_{\bm{k}}$ and $b_{\bm{k}}$ are defined in uhs space, so that
the degrees of freedom is the same for both sets of variables.
Carefully changing independent variables in the expression of
Eq.~(\ref{eq:2-5}), we obtain
\begin{equation}
  \mathcal{P}[f_{\bm{k}}]
  = 
  \exp\left[
    \sum_{N=3}^\infty
    \frac{(-1)^N}{N!}
    \sum_{\bm{k}_1}^{\rm both} \cdots \sum_{\bm{k}_N}^{\rm both}
    \left\langle
      f_{\bm{k}_1} \cdots f_{\bm{k}_N}
    \right\rangle_{\rm c}
    \frac{\partial}{\partial f_{\bm{k}_1}} \cdots
    \frac{\partial}{\partial f_{\bm{k}_N}}
  \right]
  \mathcal{P}_{\rm G}[f_{\bm{k}}],
\label{eq:2-8}
\end{equation}
where the function $P[f_{\bm{k}}]$ and $P_{\rm G}[f_{\bm{k}}]$ are the
formal probability distribution functions, defined by
\begin{eqnarray}
&&
  \mathcal{P}[f_{\bm{k}}] \prod_{\bm{k}}^{\rm both} df_{\bm{k}} =
  \mathcal{P}[a_{\bm{k}},b_{\bm{k}}] \prod_{\bm{k}}^{\rm uhs} da_{\bm{k}}
  db_{\bm{k}}, 
\label{eq:2-9a}\\
&&
  P_{\rm G}[f_{\bm{k}}] \prod_{\bm{k}}^{\rm both} df_{\bm{k}} =
  P_{\rm G}[a_{\bm{k}},b_{\bm{k}}]
  \prod_{\bm{k}}^{\rm uhs} da_{\bm{k}} db_{\bm{k}}.
\label{eq:2-9b}
\end{eqnarray}
The summation in Eq.~(\ref{eq:2-8}) is taken over any subset of
independent Fourier modes, provided that any two modes with $\pm
\bm{k}$ are always included simultaneously (``both'' stands for ``uhs
$+$ lhs''). By calculating the Jacobian of the transformation of
Eq.~(\ref{eq:2-7}), we obtain
\begin{eqnarray}
&&
  \mathcal{P}[f_{\bm{k}}] =
  2^{-N_{\rm uhs}}
  \mathcal{P}[a_{\bm{k}},b_{\bm{k}}]
\label{eq:2-10a}\\
&&
  \mathcal{P}_{\rm G}[f_{\bm{k}}] =
  \prod_{\bm{k}}^{\rm uhs}
  \frac{1}{2\pi P(\bm{k})}
  \exp\left[-\frac{f_{-\bm{k}} f_{\bm{k}}}{P(\bm{k})}\right]
\label{eq:2-10b}
\end{eqnarray}

The Eq.~(\ref{eq:2-8}) with Eq.~(\ref{eq:2-10b}) can
also be obtained by formally putting
\begin{equation}
  X = \sum_{\bm{k}}^{\rm both} J_{\bm{k}} f_{\bm{k}}
\label{eq:2-11}
\end{equation}
in the Eq.~(\ref{eq:2-1}) and formally following the similar
calculations above as if $f_{\bm{k}}$'s are real numbers. Therefore,
the Eq.~(\ref{eq:2-8}) is considered as an analytic continuation of a
general relation to the case of complex variables. In any case, the
physical interpretation of the function $P[f_{\bm{k}}]$ is given by
Eq.~(\ref{eq:2-9a}), the right hand side of which is a well-defined
quantity.

The higher-order cumulants in the exponent of Eq.~(\ref{eq:2-8}) are
non-zero only when $\bm{k}_1 + \cdots + \bm{k}_N = 0$ because of the
statistical homogeneity of the random field. In an infinitely large
space, the polyspectra $P^{(N)}(\bm{k}_1,\ldots,\bm{k}_{N-1})$, which
are the higher-order counterparts of the power spectrum, are defined
by cumulants of the Fourier coefficients as
\begin{eqnarray}
&&
  \left\langle
    \tilde{f}(\bm{k}_1) \cdots \tilde{f}(\bm{k}_N)
  \right\rangle_{\rm c}
\nonumber\\
&&\quad = 
  (2\pi)^d \delta^d\left(\bm{k}_1 + \cdots + \bm{k}_N\right)
  P^{(N)}(\bm{k}_1,\ldots,\bm{k}_{N-1}).
\label{eq:2-12}
\end{eqnarray}
In finite-volume cases, 
\begin{equation}
  \left\langle
    f_{\bm{k}_1} \cdots f_{\bm{k}_N}
  \right\rangle_{\rm c} = 
  V^{1-N/2} \delta^{\rm K}_{\bm{k}_1 + \cdots + \bm{k}_N}
  P^{(N)}(\bm{k}_1,\ldots,\bm{k}_{N-1}),
\label{eq:2-13}
\end{equation}
where $\delta^{\rm (K)}_{\bm{k}}$ is a Kronecker's delta defined by
\begin{equation}
  \delta^{\rm (K)}_{\bm{k}} = 
  \left\{
  \begin{array}{ll}
    1, & (\bm{k} = 0),\\
    0, & (\bm{k} \neq 0).
  \end{array}
  \right.
\label{eq:2-14}
\end{equation}
The polyspectra $P^{(N)}(\bm{k}_1,\cdots,\bm{k}_{N-1})$ are completely
symmetric about permutations of arguments, and have symmetries with
respect to their arguments:
\begin{eqnarray}
&&
  P^{(N)}(-\bm{k}_1,\cdots,-\bm{k}_{N-1}) =
  P^{(N)}(\bm{k}_1,\cdots,\bm{k}_{N-1}),
\label{eq:2-15a}\\
&&
  P^{(N)}(\bm{k}_1,\bm{k}_2,\cdots,\bm{k}_{N-1}) =
  P^{(N)}(\bm{k}_2,\cdots,\bm{k}_{N-1},-\bm{k}_1-\bm{k}_2-\cdots-\bm{k}_{N-1}),
\label{eq:2-15b}
\end{eqnarray}
and so on.

Substituting Eq.~(\ref{eq:2-13}) into Eq.~(\ref{eq:2-8}), we obtain
\begin{equation}
  \mathcal{P}[f_{\bm{k}}]
  = 
  \exp\left[
    \sum_{N=3}^\infty
    \frac{(-1)^N}{N!\ V^{N/2-1} }
    \sum_{\bm{k}_1}^{\rm both} \cdots \sum_{\bm{k}_N}^{\rm both}
    \delta^{\rm K}_{\bm{k}_1 + \cdots + \bm{k}_N}
    P^{(N)}(\bm{k}_1,\ldots,\bm{k}_{N-1})
    \frac{\partial}{\partial f_{\bm{k}_1}} \cdots
    \frac{\partial}{\partial f_{\bm{k}_N}}
  \right]
  \mathcal{P}_{\rm G}[f_{\bm{k}}].
\label{eq:2-16}
\end{equation}
This equation is a fundamental equation to derive the statistical
distributions of Fourier coefficients in terms of higher-order
polyspectra. One can formally take a limit of $V \rightarrow \infty$
in this expression, when all the Fourier modes are considered and
included in summations. There are correspondences
\begin{eqnarray}
&&
  V \delta^{\rm K}_{\bm{k}_1 + \cdots \bm{k}_N} \rightarrow
  (2\pi)^d\delta^d(\bm{k}_1 + \cdots \bm{k}_N),
\label{eq:2-17a}\\
&&
  \frac{1}{V} \sum_{\bm k} \rightarrow
  \int \frac{d^d k}{(2\pi)^d},
\label{eq:2-17b}\\
&&
  \sqrt{V}\frac{\partial}{\partial f_{\bm{k}}} \rightarrow
  (2\pi)^d \frac{\delta}{\delta \tilde{f}(\bm{k})},
\label{eq:2-17c}
\end{eqnarray}
where $\delta/\delta \tilde{f}(\bm{k})$ is the functional derivative.
In this limit, Eq.~(\ref{eq:2-16}) reduces to
\begin{multline}
  \mathcal{P}[\tilde{f}(\bm{k})] = 
  \exp\left[
    \sum_{N=3}^\infty
    \frac{(-1)^N}{N!}
    \int d^d k_1 \cdots d^d k_N
    (2\pi)^d \delta^d(\bm{k}_1 + \cdots + \bm{k}_N)
    P^{(N)}(\bm{k}_1,\ldots,\bm{k}_{N-1})
  \right.
\\
  \left. \times
    \frac{\delta}{\delta \tilde{f}(\bm{k}_1)} \cdots
    \frac{\delta}{\delta \tilde{f}(\bm{k}_N)}
  \right]
  \mathcal{P}_{\rm G}[\tilde{f}(\bm{k})],
\label{eq:2-18}
\end{multline}
where $P[\tilde{f}(\bm{k})]$ is a probability distribution functional
of generally non-Gaussian random fields, and $P_{\rm
  G}[\tilde{f}(\bm{k})]$ is that of Gaussian random fields which share
the same power spectrum with $P[\tilde{f}(\bm{k})]$.

In the following, it will be convenient to introduce normalized
variables,
\begin{equation}
  \alpha_{\bm{k}} \equiv \frac{f_{\bm{k}}}{\sqrt{P(\bm{k})}},
\label{eq:2-19}
\end{equation}
which have a simple covariance matrix,
\begin{equation}
  \left\langle \alpha_{\bm{k}} \alpha_{\bm{k}'} \right\rangle =
  \delta^{\rm K}_{\bm{k} + \bm{k}'}.
\label{eq:2-20}
\end{equation}
The probability distribution function of these normalized variables is
given by
\begin{equation}
  \mathcal{P}[\alpha_{\bm{k}}]
  = 
  \exp\left[
    \sum_{N=3}^\infty
    \frac{(-1)^N}{N!\ V^{N/2-1} }
    \sum_{\bm{k}_1}^{\rm both} \cdots \sum_{\bm{k}_N}^{\rm both}
    p^{(N)}(\bm{k}_1,\ldots,\bm{k}_N)
    \frac{\partial}{\partial \alpha_{\bm{k}_1}} \cdots
    \frac{\partial}{\partial \alpha_{\bm{k}_N}}
  \right]
  \mathcal{P}_{\rm G}[\alpha_{\bm{k}}],
\label{eq:2-21}
\end{equation}
where 
\begin{equation}
  p^{(N)}(\bm{k}_1,\ldots,\bm{k}_N) \equiv
  \frac{\delta^{\rm K}_{\bm{k}_1 + \cdots + \bm{k}_N}
     P^{(N)}(\bm{k}_1,\ldots,\bm{k}_{N-1})}
    {\sqrt{P(\bm{k}_1)\cdots P(\bm{k}_{N-1})
      P(\bm{k}_N)}}
\label{eq:2-22}
\end{equation}
are normalized polyspectra, and
\begin{equation}
  \mathcal{P}_{\rm G}[\alpha_{\bm{k}}] =
  \prod_{\bm{k}}^{\rm uhs}
  \frac{1}{2\pi}
  \exp\left(-\alpha_{-\bm{k}} \alpha_{\bm{k}}\right).
\label{eq:2-23}
\end{equation}
The normalized polyspectra of Eq.~(\ref{eq:2-22}) are non-zero only if
$\bm{k}_1 + \cdots \bm{k}_N =0$, and satisfy the following relation
\begin{equation}
  \left\langle
    \alpha_{\bm{k}_1} \cdots \alpha_{\bm{k}_N}
  \right\rangle_{\rm c} = 
  V^{1-N/2} p^{(N)}(\bm{k}_1,\ldots,\bm{k}_N).
\label{eq:2-23-1}
\end{equation}

The polyspectra do not depend explicitly
on the volume $V$. This can be seen by the fact that the polyspectra
are obtained by Fourier transforms of $N$-point correlation functions:
\begin{equation}
  P^{(N)}(\bm{k}_1,\ldots,\bm{k}_{N-1}) =
  \int d^d\!x_1 \cdots d^d\!x_{N-1}
  e^{-i(\bm{k}_1\cdot\bm{x}_1 + \cdots + \bm{k}_{N-1}\cdot\bm{x}_{N-1})}
  \xi^{(N)}(\bm{x}_1,\ldots,\bm{x}_{N-1}),
\label{eq:2-24}
\end{equation}
where
\begin{equation}
  \xi^{(N)}(\bm{x}_1-\bm{x}_N,\ldots,\bm{x}_{N-1}-\bm{x}_N) \equiv
  \left\langle
    f(\bm{x}_1)\cdots f(\bm{x}_{N-1}) f(\bm{x}_N)
  \right\rangle_{\rm c},
\label{eq:2-24-1}
\end{equation}
is an $N$-point correlation function. The case $N=2$ of
Eq.~(\ref{eq:2-24}) is nothing but the old Wiener-Khinchin theorem of
Eq.~(\ref{eq:1-6}). Since $N$-point correlation functions do
not explicitly depend on the volume, the polyspectra also do not
depend on the volume.

Therefore, the expression of Eq.~(\ref{eq:2-21}) can naturally be
expanded by $V^{-1/2}$. After some algebra, we obtain
\begin{multline}
  \mathcal{P}[\alpha_{\bm{k}}]
  = 
  \Biggl[
    1 +
    \sum_{n=1}^\infty V^{-n/2}
    \sum_{m=1}^\infty \frac{1}{m!}
    \sum_{\stackrel{\scriptstyle n_1,\ldots,n_m \geq 1}{n_1 + \cdots + n_m = n}}
    \frac{1}{(n_1+2)! \cdots (n_m+2)!}
\\
    \times
    \sum_{\bm{k}^{(1)}_1,\ldots,\bm{k}^{(1)}_{n_1+2}}^{\rm both} \cdots
    \sum_{\bm{k}^{(m)}_1,\ldots,\bm{k}^{(m)}_{n_m+2}}^{\rm both}
    p^{(n_1+2)}(\bm{k}^{(1)}_1,\ldots,\bm{k}^{(1)}_{n_1 + 2}) \cdots
    p^{(n_m+2)}(\bm{k}^{(m)}_1,\ldots,\bm{k}^{(m)}_{n_m + 2})
\\
    \times
    H_{\bm{k}^{(1)}_1 \cdots \bm{k}^{(1)}_{n_1 + 2} \cdots
      \bm{k}^{(m)}_1 \cdots \bm{k}^{(m)}_{n_m + 2}}
  \Biggr]
  \mathcal{P}_{\rm G}[\alpha_{\bm{k}}],
\label{eq:2-25}
\end{multline}
where
\begin{equation}
  H_{\bm{k}_1 \cdots \bm{k}_n} \equiv
  \frac{(-1)^N}{P_{\rm G}[\alpha_{\bm{k}}]}
  \frac{\partial}{\partial \alpha_{\bm{k}_1}} \cdots
  \frac{\partial}{\partial \alpha_{\bm{k}_n}}
  \mathcal{P}_{\rm G}[\alpha_{\bm{k}}].
\label{eq:2-26}
\end{equation}
The last quantities are generalization of the Hermite polynomials
\cite{appel1926, mat1995}.
They are explicitly given by
\begin{align}
  H_{\bm{k}_1} &=
  \alpha_{-\bm{k}_1}
\label{eq:2-27a}\\
  H_{\bm{k}_1 \bm{k}_2} &= 
  \alpha_{-\bm{k}_1}\alpha_{-\bm{k}_2} - \delta^{\rm K}_{\bm{k}_1+\bm{k}_2},
\label{eq:2-27b}\\
  H_{\bm{k}_1 \bm{k}_2 \bm{k}_3} &=
  \alpha_{-\bm{k}_1}\alpha_{-\bm{k}_2}\alpha_{-\bm{k}_3} -
  \left[
    \delta^{\rm K}_{\bm{k}_1+\bm{k}_2}\alpha_{-\bm{k}_3} +
    {\rm sym.}(3)
  \right],
\label{eq:2-27c}\\
  H_{\bm{k}_1 \bm{k}_2 \bm{k}_3 \bm{k}_4} &=
  \alpha_{-\bm{k}_1}\alpha_{-\bm{k}_2}\alpha_{-\bm{k}_3}\alpha_{-\bm{k}_4} - 
  \left[
    \delta^{\rm K}_{\bm{k}_1+\bm{k}_2}\alpha_{-\bm{k}_3}\alpha_{-\bm{k}_4} +
    {\rm sym.}(6)
  \right]
\nonumber\\
  & \quad + 
  \left[
    \delta^{\rm K}_{\bm{k}_1+\bm{k}_2} \delta^{\rm K}_{\bm{k}_3+\bm{k}_4} +
    {\rm sym.}(3)
  \right],
\label{eq:2-27d}\\
  H_{\bm{k}_1 \bm{k}_2 \bm{k}_3 \bm{k}_4 \bm{k}_5} &=
  \alpha_{-\bm{k}_1}\alpha_{-\bm{k}_2}\alpha_{-\bm{k}_3}
  \alpha_{-\bm{k}_4}\alpha_{-\bm{k}_5} - 
  \left[
    \delta^{\rm K}_{\bm{k}_1+\bm{k}_2}
    \alpha_{-\bm{k}_3}\alpha_{-\bm{k}_4}\alpha_{-\bm{k}_5} +
    {\rm sym.}(10)
  \right]
\nonumber\\
  & \quad + 
  \left[
    \delta^{\rm K}_{\bm{k}_1+\bm{k}_2}
    \delta^{\rm K}_{\bm{k}_3+\bm{k}_4} \alpha_{-\bm{k}_5} +
    {\rm sym.}(15)
  \right],
\label{eq:2-27e}\\
  H_{\bm{k}_1 \bm{k}_2 \bm{k}_3 \bm{k}_4 \bm{k}_5 \bm{k}_6} &=
  \alpha_{-\bm{k}_1}\alpha_{-\bm{k}_2}\alpha_{-\bm{k}_3}
  \alpha_{-\bm{k}_4}\alpha_{-\bm{k}_5}\alpha_{-\bm{k}_6} - 
  \left[
    \delta^{\rm K}_{\bm{k}_1+\bm{k}_2}
    \alpha_{-\bm{k}_3}\alpha_{-\bm{k}_4}
    \alpha_{-\bm{k}_5}\alpha_{-\bm{k}_6} +
    {\rm sym.}(15)
  \right]
\nonumber\\
  &\quad  +
  \left[
    \delta^{\rm K}_{\bm{k}_1+\bm{k}_2}
    \delta^{\rm K}_{\bm{k}_3+\bm{k}_4}
    \alpha_{-\bm{k}_5}\alpha_{-\bm{k}_6} +
    {\rm sym.}(45)
  \right] -
  \left[
    \delta^{\rm K}_{\bm{k}_1+\bm{k}_2}
    \delta^{\rm K}_{\bm{k}_3+\bm{k}_4}
    \delta^{\rm K}_{\bm{k}_5+\bm{k}_6} +
    {\rm sym.}(15)
  \right],
\label{eq:2-27f}
\end{align}
and so forth. In the above equations, ``$+\;{\rm sym.}(n)$'' indicates
that $n-1$ terms are added to symmetrize the preceding term with
respect to $\bm{k}$'s. For example, $ \delta^{\rm
  K}_{\bm{k}_1+\bm{k}_2}\alpha_{-\bm{k}_3} + {\rm sym.}(3) =
\delta^{\rm K}_{\bm{k}_1+\bm{k}_2}\alpha_{-\bm{k}_3} + \delta^{\rm
  K}_{\bm{k}_2+\bm{k}_3}\alpha_{-\bm{k}_1} + \delta^{\rm
  K}_{\bm{k}_3+\bm{k}_1}\alpha_{-\bm{k}_2}$, etc. 
The Eq.~(\ref{eq:2-25}) is formally a series expansion by a
dimensional quantity $V^{-1/2}$. The meaning of which is discussed in
the following subsection.

\subsection{
Explicit expansion up to second order
}

Substituting the explicit representation of $H_{\bm{k}_1 \bm{k}_2
  \cdots}$ into the expression of Eq.~(\ref{eq:2-25}), the
general distribution function of the Fourier coefficients of a
non-Gaussian field is obtained. In the following, we consider a
Fourier amplitude $A_{\bm{k}}$ and a Fourier phase $\theta_{\bm{k}}$
instead of a complex variable $\alpha_{\bm{k}}$:
\begin{equation}
  \left\{
  \begin{array}{l}
    \alpha_{\bm{k}} = A_{\bm{k}} e^{i\theta_{\bm{k}}}, \\
    \alpha_{-\bm{k}} = A_{\bm{k}} e^{-i\theta_{\bm{k}}},
  \end{array}
  \right.\ \ 
  \left(\bm{k} \in {\rm uhs} \right).
\label{eq:2-28}
\end{equation}
where $A_{\bm{k}} \geq 0$ and $0 \leq \theta_{\bm{k}} < 2\pi$. With
these new variables, the distribution function of a Gaussian field is
given by [cf. Eq.~(\ref{eq:1-15})]
\begin{equation}
  \mathcal{P}_{\rm G}[A_{\bm{k}},\theta_{\bm{k}}]
  \prod_{\bm{k}}^{\rm uhs} dA_{\bm{k}} d\theta_{\bm{k}}
 = 
  \prod_{\bm{k}}^{\rm uhs}
  2 A_{\bm{k}} e^{-{A_{\bm{k}}}^2}
  dA_{\bm{k}} \frac{d\theta_{\bm{k}}}{2\pi}.
\label{eq:2-29}
\end{equation}

With these new variables, the Eq.~(\ref{eq:2-25}), up to second order
in $V^{-1/2}$, reduces to
\begin{align}
  \frac{\mathcal{P}}{\mathcal{P}_{\rm G}} =
  1 + &
  V^{-1/2}
  \sum_{\bm{k}_1, \bm{k}_2, \bm{k}_3}^\uhs
  A_{\bm{k}_1} A_{\bm{k}_2} A_{\bm{k}_3}
  \cos\left(\theta_{\bm{k}_1} + \theta_{\bm{k}_2} - \theta_{\bm{k}_3}\right)
  p^{(3)}\left(\bm{k}_1,\bm{k}_2,-\bm{k}_3\right)
\nonumber\\
 +
  V^{-1}
  \Biggl\{ &
  \sum_{\bm{k}_1,\bm{k}_2,\bm{k}_3,\bm{k}_4}^\uhs
  \frac13 {A_{\bm{k}_1}}{A_{\bm{k}_2}}{A_{\bm{k}_3}}{A_{\bm{k}_4}}
  \cos\left(\theta_{\bm{k}_1} + \theta_{\bm{k}_2} + \theta_{\bm{k}_3}
    - \theta_{\bm{k}_4}\right)
  p^{(4)}\left(\bm{k}_1,\bm{k}_2,\bm{k}_3,-\bm{k}_4\right)
\nonumber\\
& +
  \sum_{\bm{k}_1,\bm{k}_2,\bm{k}_3,\bm{k}_4}^\uhs
  \frac14
  {A_{\bm{k}_1}}{A_{\bm{k}_2}}{A_{\bm{k}_3}}{A_{\bm{k}_4}}
  \cos\left(\theta_{\bm{k}_1} + \theta_{\bm{k}_2} - \theta_{\bm{k}_3}
    - \theta_{\bm{k}_4}\right)
  p^{(4)}\left(\bm{k}_1,\bm{k}_2,-\bm{k}_3,-\bm{k}_4\right)
\nonumber\\
&  -
  \sum_{\bm{k}_1,\bm{k}_2}^\uhs
  \frac12 
  \left(
    {A_{\bm{k}_1}}^2 + {A_{\bm{k}_2}}^2 - 1
  \right)
  p^{(4)}\left(\bm{k}_1,\bm{k}_2,-\bm{k}_1,-\bm{k}_2\right)
\nonumber\\
& +
  \sum_{\bm{k}_1,\bm{k}_2,\bm{k}_3,\bm{k}_4,\bm{k}_5,\bm{k}_6}^\uhs
  \frac12 {A_{\bm{k}_1}}{A_{\bm{k}_2}}{A_{\bm{k}_3}}{A_{\bm{k}_4}}
  {A_{\bm{k}_5}}{A_{\bm{k}_6}}
  \cos\left(\theta_{\bm{k}_1} + \theta_{\bm{k}_2} - \theta_{\bm{k}_3}\right)
  \cos\left(\theta_{\bm{k}_4} + \theta_{\bm{k}_5} - \theta_{\bm{k}_6}\right)
\nonumber\\
& \qquad\qquad\qquad\quad   \times
  p^{(3)}\left(\bm{k}_1,\bm{k}_2,-\bm{k}_3\right)
  p^{(3)}\left(\bm{k}_4,\bm{k}_5,-\bm{k}_6\right)
\nonumber\\
&  -
  \sum_{\bm{k}_1,\bm{k}_2,\bm{k}_3,\bm{k}_4}^\uhs
  {A_{\bm{k}_1}}{A_{\bm{k}_2}}{A_{\bm{k}_3}}{A_{\bm{k}_4}}
  \cos\left(\theta_{\bm{k}_1} + \theta_{\bm{k}_2} + \theta_{\bm{k}_3}
    - \theta_{\bm{k}_4}\right)
\nonumber\\
& \qquad\qquad\qquad\quad   \times
  \sum_{\bm{k}_5}^\uhs
  p^{(3)}\left(\bm{k}_1,\bm{k}_2,-\bm{k}_5\right)
  p^{(3)}\left(\bm{k}_3,-\bm{k}_4,\bm{k}_5\right)
\nonumber\\
& -
  \sum_{\bm{k}_1,\bm{k}_2,\bm{k}_3,\bm{k}_4}^\uhs
  \frac14
  {A_{\bm{k}_1}}{A_{\bm{k}_2}}{A_{\bm{k}_3}}{A_{\bm{k}_4}}
  \cos\left(\theta_{\bm{k}_1} + \theta_{\bm{k}_2} - \theta_{\bm{k}_3}
    - \theta_{\bm{k}_4}\right)
\nonumber\\
& \qquad\qquad\qquad\quad   \times
  \sum_{\bm{k}_5}^\uhs
  \left[
    p^{(3)}\left(\bm{k}_1,\bm{k}_2,-\bm{k}_5\right)
    p^{(3)}\left(\bm{k}_3,\bm{k}_4,-\bm{k}_5\right)
  \right.
\nonumber\\
& 
  \left.
  \qquad\qquad\qquad\qquad\qquad\qquad\qquad +
    4 p^{(3)}\left(\bm{k}_1,-\bm{k}_3,\bm{k}_5\right)
    p^{(3)}\left(\bm{k}_2,-\bm{k}_4,-\bm{k}_5\right)
  \right]
\nonumber\\
& +
  \sum_{\bm{k}_1,\bm{k}_2,\bm{k}_3}^\uhs
  \frac12 
  \left(
    {A_{\bm{k}_1}}^2 + {A_{\bm{k}_2}}^2 + {A_{\bm{k}_3}}^2 - 1
  \right)
  \left[
  p^{(3)}\left(\bm{k}_1,\bm{k}_2,-\bm{k}_3\right)\right]^2
\Biggr\}
+ \mathcal{O}(V^{-3/2}),
\label{eq:2-30}
\end{align}
where the wavevectors which appear as indices of $A_{\bm{k}}$ and
$\theta_{\bm{k}}$ are only summed over the uhs.

The above expression, however, is still inconvenient for further
investigations, because the modes in each summation with different
$\bm{k}$-labels can be the same. We should expand each summation in
Eq.~(\ref{eq:2-30}) so that the different labels refer different
modes. For example, in the second term the modes $\bm{k}_1$ and
$\bm{k}_2$ can be the same and also can be different. In the same
term, the modes $\bm{k}_1$ and $\bm{k}_3$ should always different,
because $\bm{k}_2 = \bm{k}_3$ requires $\bm{k}_2 = \bm{0}$, which
contradicts $\bm{k}_2 \in \uhs$. In the following we are going to
integrate over some modes, in which case it is convenient when
different labels refer to different modes. We carefully expand each
summation in Eq.~(\ref{eq:2-30}) to summations in which the modes are
mutually different. For example, the summation in the second term in
Eq.~(\ref{eq:2-30}) should be separated into a summation with
$\bm{k}_1 = \bm{k}_2$ and $\bm{k}_1 \neq \bm{k}_2$, resulting in
\begin{multline}
  \sum_{\bm{k}_1, \bm{k}_2, \bm{k}_3}^\uhs
  A_{\bm{k}_1} A_{\bm{k}_2} A_{\bm{k}_3}
  \cos\left(\theta_{\bm{k}_1} + \theta_{\bm{k}_2} - \theta_{\bm{k}_3}\right)
  p^{(3)}\left(\bm{k}_1,\bm{k}_2,-\bm{k}_3\right)
\\
  =
  \Sumprime_{\bm{k}_1,\bm{k}_2}^\uhs
  {A_{\bm{k}_1}}^2 A_{\bm{k}_2}
  \cos\left(2\theta_{\bm{k}_1} - \theta_{\bm{k}_2}\right)
  p^{(3)}\left(\bm{k}_1,\bm{k}_1,-\bm{k}_2\right)
\\
  +
  \Sumprime_{\bm{k}_1,\bm{k}_2,\bm{k}_3}^\uhs
  A_{\bm{k}_1} A_{\bm{k}_2} A_{\bm{k}_3}
  \cos\left(\theta_{\bm{k}_1} + \theta_{\bm{k}_2} - \theta_{\bm{k}_3}\right)
  p^{(3)}\left(\bm{k}_1,\bm{k}_2,-\bm{k}_3\right),
\label{eq:2-31}
\end{multline}
where the summation $\Sumprime_{\bm{k}_1,\cdots}^\uhs$ indicates that
all the modes are different to each other. On the left hand side, the
wavevector $\bm{k}_3$ is identical to $\bm{k}_1 + \bm{k}_2$ because of
the Kronecker's delta in the normalized bispectrum. Therefore, the
mode $\bm{k}_3$ is automatically different from $\bm{k}_1$ and
$\bm{k}_2$ because $\bm{k}_1, \bm{k}_2 \ne \bm{0}$. The zero mode
$\bm{k}=\bm{0}$ does not appear because it is a purely homogeneous
mode which is excluded by imposing $\langle f(\bm{x})\rangle = 0$. It
is only when $\bm{k}_1 = \bm{k}_2$ that two of three modes on the left
hand side are identical. The case $\bm{k}_1 = \bm{k}_2$ corresponds to
the first term on the right hand side, and other case corresponds to
the second term.

Likewise, one can carefully expand other summations into summations of
mutually different modes. After tedious, but straightforward
manipulations, the result has a form,
\begin{equation}
  \frac{\mathcal{P}}{\mathcal{P}_{\rm G}} =
    1 + 
    \frac{1}{\sqrt{V}} \sum_{i=1}^2 \mathcal{Q}^{(1)}_i +
    \frac{1}{V} \sum_{i=1}^{22} \mathcal{Q}^{(2)}_i
    + \mathcal{O}(V^{-3/2}),
\label{eq:2-32}
\end{equation}
where $\mathcal{Q}^{(1)}_i$ and $\mathcal{Q}^{(2)}_i$ are the summations
given in Appendix~\ref{app:a}.

\subsection{Meaning of  the expansion by $V^{-1/2}$}

We obtained the distribution function of the Fourier coefficients in a
formal series by powers of $V^{-1/2}$. The volume $V$ is a dimensional
quantity, and the expansion seems meaningless if we naively choose the
unit of length by $V=1$. However, powers of $V^{-1/2}$ is always
accompanied by the normalized polyspectra $p^{(n)}$, which are also
dimensional quantities. The dimensionless quantities in our expansion
is actually $V^{1-n/2} p^{(n)}$, which can be seen in
Eq.~(\ref{eq:2-21}). Therefore, when these quantities are small for
large $n$ and higher-order terms are negligible, one can approximately
obtain the distribution function by truncating the series, which is
practically useful.

In fact, the polyspectra are not dependent on the volume $V$, and so
are the normalized polyspectra, when the wavevectors in the
polyspectra are fixed. Thus the quantity $V^{1-n/2}
p^{(n)}(\bm{k}_1,\ldots,\bm{k}_n)$ can be arbitrarily small by taking
large $V$ for a fixed set of wavevectors $\bm{k}_1,\ldots,\bm{k}_n$.
Our expansion is actually an expansion by these quantities, assuming
all the normalized polyspectra do not diverge.

We should note that the number of possible wavevectors are large when
the volume is large. This means that the number of terms which have a
same order of $V^{-1/2}$ can be infinitely large if $V \rightarrow
\infty$. However, one does not have to consider all the possible
modes in Eq.~(\ref{eq:2-25}). This equation is valid even in a case
only a set of particular modes are considered, in which case the
summation is taken over only wavevectors of that set of modes and the
number of terms of a same order is finite. After the next section we
consider distribution functions in which only particular set of modes
are considered.

Absolute values of the quantities $V^{1-n/2} p^{(n)}$ are usually very
small when the scales of wavevectors in the arguments of polyspectra
are sufficiently smaller than the box size. We will see this property
by some examples of non-Gaussian fields in section~\ref{sec:numerical}
below. Therefore, our expansion is actually efficient and turns
out to be very useful for most of the cases.



\section{\label{sec:NptDist}
$N$-point Distribution Functions of Fourier Modes
}

\subsection{Zero-point Distributions}

As a consistency check, the probability distribution function
$P[A_{\bm{k}}, \theta_{\bm{k}}]$ in the form of Eq.~(\ref{eq:2-32}) is
shown to have a correct normalization,
\begin{equation}
  \int \prod_{\bm{k}}^{\rm uhs} dA_{\bm{k}} d\theta_{\bm{k}}
  \cdot
  \mathcal{P} = 1,
\label{eq:2-33}
\end{equation}
even when the expansion is truncated. The integration on the right
hand side can be explicitly performed using the expansion of ${\cal
  P}$ given in Appendix~\ref{app:a} with the Gaussian distribution
$\mathcal{P}_{\rm G}$ of Eq.~(\ref{eq:2-29}). The following integrals are
useful:
\begin{align}
&
  I(n) \equiv \int_0^\infty A^n \cdot 2A e^{-A^2} dA =
  \Gamma\left(1 + \frac{n}{2}\right),
\label{eq:2-34a}\\
&
  \int_0^{2\pi} \frac{d\theta}{2\pi} \cos(n\theta + \alpha) = 0,
\label{eq:2-34b}\\
&
  \int_0^{2\pi} \frac{d\theta}{2\pi}
    \cos(n\theta + \alpha) \cos(m\theta + \beta)
  =
  \left\{
  \begin{array}{ll}
    \frac12 \cos(\alpha - \beta), & (n = m), \\
    0, & (n \ne m),
  \end{array}
  \right.
\label{eq:2-34c}
\end{align}
where $n$ and $m$ are non-negative integers. We use
Eq.~(\ref{eq:2-34a}) with $I(0) = 1$, $I(1) = \sqrt{\pi}/2$, $I(2) =
1$, $I(3) = 3\sqrt{\pi}/4$, $I(4) = 2$. Using these integrals, it is
straightforward to show
\begin{equation}
  \int dA_{\bm{k}} d\theta_{\bm{k}}
  \mathcal{Q}^{(j)}_i \mathcal{P}_{\rm G} = 0, 
\label{eq:2-35}
\end{equation}
for any mode $\bm{k} \in \uhs$.
The Eq.~(\ref{eq:2-33}) is a direct consequence of this property.

\subsection{One-point Distributions}

The distribution function of a particular Fourier mode $f_{\bm{k}}$
can be obtained by integrating all modes but one in the general
distribution function of Eq.~(\ref{eq:2-32}). The one-point
distribution function $P_1(A_{\bm{k}}, \theta_{\bm{k}})$ for a
particular mode $\bm{k}$ is therefore given by
\begin{equation}
  P_1(A_{\bm{k}}, \theta_{\bm{k}}) =
  \int
  \prod_{\stackrel{\scriptstyle \bm{k}'}{\bm{k}' \ne \bm{k}}}^{\rm uhs}
   dA_{\bm{k}'} d\theta_{\bm{k}'} \cdot \mathcal{P},
\label{eq:2-36}
\end{equation}
where all the modes $\bm{k}'$ but a particular mode $\bm{k}$ are
integrated over.

Because of Eq.~(\ref{eq:2-35}), the $\mathcal{Q}^{(j)}_i$ terms
which have more than two different modes does not contribute to the
above Eq.~(\ref{eq:2-36}). The only contribution comes from
$\mathcal{Q}^{(2)}_1$, which has only one mode, and the
Eq.~(\ref{eq:2-36}) reduces to
\begin{equation}
  P_1(A_{\bm{k}}, \theta_{\bm{k}}) = 
  \left[
    1 + \frac{1}{V}
    \left(
      \frac14 {A_{\bm{k}}}^4 - {A_{\bm{k}}}^2 + \frac12
    \right)
     p^{(4)}\left(\bm{k},\bm{k},-\bm{k},-\bm{k}\right)
      + \mathcal{O}(V^{-2})
  \right]
  P_{\rm G}(A_{\bm{k}}, \theta_{\bm{k}}).
\label{eq:2-37}
\end{equation}
where
\begin{equation}
  P_{\rm G}(A_{\bm{k}}, \theta_{\bm{k}}) dA_{\bm{k}} d\theta_{\bm{k}}
  = 
  2 A_{\bm{k}} e^{-{A_{\bm{k}}}^2}
  dA_{\bm{k}} \frac{d\theta_{\bm{k}}}{2\pi},
\label{eq:2-38}
\end{equation}
is a one-point distribution function of a Gaussian random field.

There is a reason why only $p^{(4)}(\bm{k},\bm{k},-\bm{k},-\bm{k})$
contributes to the non-Gaussian correction and the terms with
half-integer power of $V$ do not contribute at all. The one-point
distribution for a particular mode $\bm{k}$ is determined once the
hierarchy of higher-order cumulants is provided, because of the
cumulant expansion theorem. In the present case, we have two
independent variables $\alpha_{\bm{k}}$ and $\alpha_{-\bm{k}}$ for a
single mode $\bm{k} \in {\rm uhs}$. Therefore all the cumulants have
the form, $\langle (\alpha_{\bm{k}})^n (\alpha_{-\bm{k}})^m
\rangle_{\rm c}$. Because of Eq.~(\ref{eq:2-23-1}), which is a
manifestation of the translational invariance, such cumulants are
non-zero only when $n=m$, and they are given by
\begin{equation}
  \langle (\alpha_{\bm{k}})^n (\alpha_{-\bm{k}})^m \rangle_{\rm c}
  = \delta_{nm} V^{1-n} q^{(2n)}(\bm{k}),
\label{eq:2-40}
\end{equation}
where
\begin{equation}
  q^{(2n)}(\bm{k}) \equiv
  p^{(2n)}(
    \underbrace{\bm{k},\ldots,\bm{k}}_{n\,{\rm elements}},
    \underbrace{-\bm{k},\ldots,-\bm{k}}_{n\,{\rm elements}}),
\label{eq:2-41}
\end{equation}
is a collapsed polyspectrum. Because the cumulants of
Eq.~(\ref{eq:2-40}) have integer powers of $V$, the terms with
half-integer power do not appear in the expansion of
Eq.~(\ref{eq:2-37}).

In the case of one-point distribution function, obtaining higher-order
terms in Eq.(\ref{eq:2-37}) is not so difficult. To this end, we
re-start the calculation from Eq.~(\ref{eq:2-21}). When we consider
only a particular mode $\bm{k} \in {\rm uhs}$, the sums over
$\bm{k}_1, \ldots \bm{k}_N$ in the exponent are only taken for
$\bm{k}$ and $-\bm{k}$. Because of the Kronecker's delta, the number
of $\bm{k}$ and that of $-\bm{k}$ should be the same, so that $N$
should be an even number, $N = 2n$. Taking proper combinatorial weights
into account, we obtain a joint distribution function of
$\alpha_{\bm{k}}$ and $\alpha_{-\bm{k}}$,
\begin{equation}
  P(\alpha_{\bm{k}},\alpha_{-\bm{k}})
  = 
  \frac{1}{2\pi}
  \exp\left[
    \sum_{n=2}^\infty
    \frac{1}{(n!)^2 V^{n-1} }
    q^{(2n)}(\bm{k})
    \left(
      \frac{\partial^2}
        {\partial \alpha_{\bm{k}} \partial \alpha_{-\bm{k}}}
    \right)^n
  \right]
  \exp(-\alpha_{-\bm{k}}\alpha_{\bm{k}}).
\label{eq:2-42}
\end{equation}
In terms of variables $A_{\bm{k}}$ and $\theta_{\bm{k}}$ defined in
Eq.~(\ref{eq:2-28}), the differential operator in the exponent is
given by
\begin{equation}
  \frac{\partial^2}
    {\partial \alpha_{\bm{k}} \partial \alpha_{-\bm{k}}} =
  \frac{1}{4}
  \left(
    \frac{\partial^2}{\partial {A_{\bm{k}}^2}} +
    \frac{1}{A_{\bm{k}}}\frac{\partial}{\partial A_{\bm{k}}} +
    \frac{1}{{A_{\bm{k}}^2}}\frac{\partial^2}{\partial{\theta_{\bm{k}}}^2}
  \right).
\label{eq:2-43}
\end{equation}
The last factor in the new variables is given by
$\exp(-{A_{\bm{k}}}^2)$ which does not depend on the phase
$\theta_{\bm{k}}$. The Jacobian of the transformation is
$\partial (\alpha_{\bm{k}}, \alpha_{-\bm{k}}) / \partial
(A_{\bm{k}}, \theta_{\bm{k}}) = 2 A_{\bm{k}}$. Therefore, the general
form of the one-point distribution function in terms of
$(A_{\bm{k}},\theta_{\bm{k}})$ is given by
\begin{equation}
  P_1(A_{\bm{k}},\theta_{\bm{k}}) =
  \frac{A_{\bm{k}}}{\pi}
  \exp\left[
    \sum_{n=2}^\infty \frac{1}{4 (n!)^2 V^{n-1}}
    q^{(2n)}(\bm{k})
    \left(
      \frac{\partial^2}{\partial {A_{\bm{k}}}^2} +
      \frac{1}{A_{\bm{k}}} \frac{\partial}{\partial A_{\bm{k}}}
    \right)^n
  \right]
  \exp(-{A_{\bm{k}}}^2).
\label{eq:2-44}
\end{equation}
This equation provides a general expression of the one-point
distribution function of a Fourier mode. There is an important theorem
which is directly derived from Eq.~(\ref{eq:2-44}):
\begin{theorem}
    For a random field in a spatially homogeneous space, a one-point
    distribution function of Fourier phase $P(\theta_{\bm{k}})$ is
    always homogeneous:
  \begin{equation}
    P(\theta_{\bm{k}}) = \frac{1}{2\pi}.
  \label{eq:2-45}
  \end{equation}
\label{th:2-1}
\end{theorem}
The proof of this theorem is trivial since the right hand side of
Eq.~(\ref{eq:2-44}) does not depend on the phase $\theta_{\bm{k}}$.
The spatial homogeneity is crucial in this theorem. If the statistics
of the random field is spatially inhomogeneous, the number of operator
$\partial/\partial\alpha_{\bm{k}}$ and that of
$\partial/\partial\alpha_{-\bm{k}}$ does not necessarily agree because
the lack of Kronecker's deltas in the exponent of
Eq.~(\ref{eq:2-21}). In which case, the exponent of Eq.~(\ref{eq:2-44})
explicitly depends on the phase $\theta_{\bm{k}}$, and the theorem does
not hold.

The value of a phase $\theta_{\bm{k}}$ determines positions of peaks
and troughs of the Fourier mode $\bm{k}$, since the particular Fourier
mode has the spatial dependence,
$f_{\bm{k}}e^{i\bm{k}\cdot\bm{x}} = |f_{\bm{k}}|
e^{i\bm{k}\cdot\bm{x} + i\theta_{\bm{k}}}$, so that a spatial
translation $\bm{x} \rightarrow \bm{x}_0$ is equivalent to a phase
shift $\theta_{\bm{k}}\rightarrow \theta_{\bm{k}} +
\bm{k}\cdot\bm{x}_0$. Therefore, spatial homogeneity implies that
there should not be preferred values in phases. This is an intuitive
meaning of Theorem~\ref{th:2-1}. However, when the other modes are
simultaneously considered, the phases are not independently
distributed and there are phase correlations among Fourier modes, as
shown in the following sections.

Another important consequence of Eq.~(\ref{eq:2-44}) is that the
non-Gaussian corrections always vanish in the limit of $V\rightarrow
0$ as long as all $q^{(2n)}(\bm{k})$ are finite. Therefore, the
following theorem holds:
\begin{theorem}
    For a random field in a spatially homogeneous space, a one-point
    distribution function of a Fourier mode approaches to be Gaussian
    when the spacial volume $V$ is sufficiently large:
  \begin{equation}
     P(|f_{\bm{k}}|,\theta_{\bm{k}})
     d|f_{\bm{k}}| d\theta_{\bm{k}} \rightarrow 
    \exp\left[
    - \frac{|f_{\bm{k}}|^2}{P(\bm{k})}
    \right]
    \frac{2 |f_{\bm{k}}| d|f_{\bm{k}}|}{P(\bm{k})}
    \frac{d\theta_{\bm{k}}}{2\pi},
  \qquad
    \text{when}\quad V \rightarrow \infty,
  \label{eq:2-45-1}
  \end{equation}
  provided that all the polyspectra of type
  $P^{(2n)}(\bm{k},\dots,\bm{k},-\bm{k},\dots,-\bm{k})$
  are finite for any positive integer $n$.
\label{th:2-2}
\end{theorem}

This theorem is related to the central limit theorem. To illustrate
the relation, we note the Fourier coefficients in the whole space are
considered as superimposition of the Fourier coefficients of finite
sub-volumes. A pair of sub-volumes which is sufficiently separated in
space is almost independent to each other as long as the spatial
correlations are not so strong on large scales. Expected separations
of arbitrary pairs of such sub-volumes diverges in the limit $V
\rightarrow \infty$. The central limit theorem tells that
superimposition of infinitely many independent random variables is
normally distributed and the distribution function is Gaussian,
irrespective to the distribution functions of the original random
variables. The Fourier coefficients of the whole space is considered
as superimposition of Fourier coefficients of many sub-volumes. Since
these sub-volumes are almost independent, the distribution function of
a Fourier coefficient of the whole space is expected to be Gaussian,
even when the distribution in the sub-volumes are non-Gaussian.
According to the Theorem~\ref{th:2-2}, the one-point distribution
function of a Fourier mode in a sufficiently large volume does not
distinguish non-Gaussianity of the distribution.

Expanding Eq.~(\ref{eq:2-44}) to arbitrary
order in $V^{-1}$ is straightforward. To second order, for example, we
obtain
\begin{align}
  \frac{P_1(A_{\bm{k}},\theta_{\bm{k}})}
     {P_{\rm G}(A_{\bm{k}},\theta_{\bm{k}})} = &
    1 +
    \frac{1}{V} \left(\frac{1}{4}{A_{\bm{k}}}^4 - {A_{\bm{k}}}^2 +
        \frac{1}{2}\right) q^{(4)}(\bm{k})
\nonumber\\
  & + \frac{1}{V^2}
    \left\{
    \left(\frac{1}{36}{A_{\bm{k}}}^6 - \frac{1}{4} {A_{\bm{k}}}^4 + 
        \frac{1}{2} {A_{\bm{k}}}^2 - \frac{1}{6}
    \right) q^{(6)}(\bm{k})
    \right.
\nonumber\\
 & \qquad 
    \left.
  + \left(\frac{1}{32}{A_{\bm{k}}}^8 - \frac{1}{2} {A_{\bm{k}}}^6 +
        \frac{9}{4}{A_{\bm{k}}}^4 - 3 {A_{\bm{k}}}^2 + \frac{3}{4}
    \right) \left[q^{(4)}(\bm{k})\right]^2
    \right\}
  + \mathcal{O}(V^{-3}).
\label{eq:2-46}
\end{align}
The first-order term agrees with Eq.~(\ref{eq:2-37}) as it should be.
It is straightforward to obtain higher-order terms by expanding
Eq.~(\ref{eq:2-44}).

The distribution function of the original variables, $f_{\bm{k}} =
|f_{\bm{k}}|\exp(i\theta_{\bm{k}})$ is simply obtained from
Eq.~(\ref{eq:2-46}), since the Jacobians of $P_1$ and $P_{\rm G}$ are
identical. In fact, $P_1(|f_{\bm{k}}|,\theta_{\bm{k}})/P_{\rm
  G1}(|f_{\bm{k}}|,\theta_{\bm{k}})$ is just given by the right hand
side of Eq.~(\ref{eq:2-46}) with a substitution $A_{\bm{k}} =
|f_{\bm{k}}|/\sqrt{P(k)}$.

\subsection{Two-point Distributions}

The two-point distribution function for a couple of particular modes
$\bm{k}_1$ and $\bm{k}_2$ is given by
\begin{equation}
  P_2(A_{\bm{k}_1}, \theta_{\bm{k}_1}; A_{\bm{k}_2}, \theta_{\bm{k}_2}) =
  \int
  \prod^{\rm uhs}_{\stackrel
    {\scriptstyle \bm{k}}{\bm{k} \ne \bm{k}_1, \bm{k}_2}}
   dA_{\bm{k}} d\theta_{\bm{k}} \cdot \mathcal{P},
\label{eq:2-47}
\end{equation}
where all the modes $\bm{k}$ but two particular modes $\bm{k}_1$ and
$\bm{k}_2$ are integrated over. Instead of considering
the full two-point distribution function of Eq.~(\ref{eq:2-47}), it is
convenient to consider a reduced distribution function $R_2$
defined by
\begin{equation}
  P_2(A_{\bm{k}_1},\theta_{\bm{k}_1};A_{\bm{k}_2},\theta_{\bm{k}_2})
  = 
  P_1(A_{\bm{k}_1},\theta_{\bm{k}_1})
  P_1(A_{\bm{k}_2},\theta_{\bm{k}_2})
  \left[
    1 + 
    R_2(A_{\bm{k}_1},\theta_{\bm{k}_1};A_{\bm{k}_2},\theta_{\bm{k}_2})
  \right].
\label{eq:2-48}
\end{equation}
If there is no correlation between the two modes $\bm{k}_1$ and
$\bm{k}_2$, the reduced distribution function vanishes. The reduced
distribution function represents the additional contribution that is
not represented by one-point distribution functions. In the following,
the reduced distribution function
$R_2(A_{\bm{k}_1},\theta_{\bm{k}_1};A_{\bm{k}_2},\theta_{\bm{k}_2})$
is abbreviated as $R_2(\bm{k}_1,\bm{k}_2)$. The reduced distribution
function does not change when the independent variables are changed
since the Jacobian is the same for both sides of Eq.~(\ref{eq:2-48}).
Therefore, the reduced distribution function for arbitrary set of
independent variables, such as $(\mathrm{Re}f_{\bm{k}}$,
${\mathrm{Im}}f_{\bm{k}})$, $(|f_{\bm{k}}|,\theta_{\bm{k}})$, is
simply obtained by represent the function in terms of the new
variables.

Most of the terms in Eq.~(\ref{eq:2-32}) vanish in the
Eq.~(\ref{eq:2-47}), because of the integrations of
Eqs.~(\ref{eq:2-34a})--(\ref{eq:2-34c}), or Eq.~(\ref{eq:2-35}). The
$\mathcal{Q}^{(j)}_i$ terms with more than three different modes
vanish in the integration of Eq.~(\ref{eq:2-47}). The survived terms
are only $\mathcal{Q}^{(1)}_1$, $\mathcal{Q}^{(2)}_1$,
$\mathcal{Q}^{(2)}_{2}$, $\mathcal{Q}^{(2)}_{3}$, and
$\mathcal{Q}^{(2)}_{8}$. The terms $\mathcal{Q}^{(1)}_1$,
$\mathcal{Q}^{(2)}_8$ survive only when $\bm{k}_2 = 2\bm{k}_1$ or
$\bm{k}_1 = 2\bm{k}_2$, and the term $\mathcal{Q}^{(2)}_{3}$ survives
only when $\bm{k}_2 = 3\bm{k}_1$ or $\bm{k}_1 = 3\bm{k}_2$. The
remaining terms $\mathcal{Q}^{(2)}_{1}$, $\mathcal{Q}^{(2)}_{2}$
always survive. However, the term $\mathcal{Q}^{(2)}_{1}$ does not
contribute to the reduced function $R_2$, because the contribution is
absorbed in the one-point distribution functions. In the term
$\mathcal{Q}^{(2)}_{2}$, there is a symmetry $\bm{k}_1 \leftrightarrow
\bm{k}_2$ in the summation, so that the symmetrization factor 2 should
be taken into account in the integration of Eq.~(\ref{eq:2-47}). As a
result, the reduced function $R_2$ for two-point distributions is
given by
\begin{align}
  R_2(\bm{k}_1, \bm{k}_2) =&
  \frac{1}{\sqrt{V}}
  {A_{\bm{k}_1}}^2 A_{\bm{k}_2}
  \cos(2\theta_{\bm{k}_1} - \theta_{\bm{k}_2})
  p^{(3)}\left(\bm{k}_1,\bm{k}_1,-\bm{k}_2\right)
  + {\rm sym.}\left(\bm{k}_1 \leftrightarrow \bm{k}_2\right)
\nonumber\\
  &+ \frac{1}{2V}
  \left[
    {A_{\bm{k}_1}}^4 {A_{\bm{k}_2}}^2
    \cos^2(2\theta_{\bm{k}_1} - \theta_{\bm{k}_2}) -
      2 {A_{\bm{k}_1}}^2 {A_{\bm{k}_2}}^2 -
      \frac12 {A_{\bm{k}_1}}^4 +
      2 {A_{\bm{k}_1}}^2 + {A_{\bm{k}_2}}^2 - 1
  \right]
\nonumber\\
  & \hspace{15pc} \times
  \left[
    p^{(3)}\left(\bm{k}_1,\bm{k}_1,-\bm{k}_2\right)
  \right]^2
    + {\rm sym.}\left(\bm{k}_1 \leftrightarrow \bm{k}_2\right)
\nonumber\\
  &+ \frac{1}{3V}
    {A_{\bm{k}_1}}^3 A_{\bm{k}_2}
    \cos(3\theta_{\bm{k}_1} - \theta_{\bm{k}_2})
    p^{(4)}\left(\bm{k}_1,\bm{k}_1,\bm{k}_1,-\bm{k}_2\right)
    + {\rm sym.}\left(\bm{k}_1 \leftrightarrow \bm{k}_2\right)
\nonumber\\
  &+ \frac{1}{V}
    \left(
      {A_{\bm{k}_1}}^2 {A_{\bm{k}_2}}^2 -
      {A_{\bm{k}_1}}^2 - {A_{\bm{k}_2}}^2 + 1
    \right)
    p^{(4)}\left(\bm{k}_1,\bm{k}_2,-\bm{k}_1,-\bm{k}_2\right)
\nonumber\\
  &+ \mathcal{O}(V^{-3/2}),
\label{eq:2-50}
\end{align}
where $\bm{k}_1, \bm{k}_2 \in \uhs$. The symbol ${\rm
  sym.}\left(\bm{k}_1 \leftrightarrow \bm{k}_2\right)$ means an
additional term which are needed to symmetrize each preceding term.
The first and second terms of the lhs contribute only when $\bm{k}_2 =
2\bm{k}_1$ or $\bm{k}_1 = 2\bm{k}_2$, and the third term contribute
only when $\bm{k}_1 = 3\bm{k}_2$ or $\bm{k}_2 = 3\bm{k}_1$. When the third
term contributes, the first and second terms do not contribute, and vice
versa. The forth term always contributes.

The two-point distribution depends on phases only when the
vector $\bm{k}_2$ is proportional to $\bm{k}_1$. This property is
understood by translational invariance of the statistical
distribution. As considered in the previous subsection, phases appear
in the distribution functions only in the form which is invariant under
the phase shift $\theta_{\bm{k}}\rightarrow \theta_{\bm{k}} +
\bm{k}\cdot\bm{x}_0$. If $\bm{k}_2$ is not proportional to $\bm{k}_1$,
there is not any way of making an invariant combination out of the two
phases $\theta_{\bm{k}_1}$ and $\theta_{\bm{k}_2}$. If $\bm{k}_2 = c
\bm{k}_1$, a phase combination $c \theta_{\bm{k}_1} -
\theta_{\bm{k}_2}$ is invariant.

\subsection{Three-point distributions}

Next we consider the three-point distributions. The reduced
three-point distribution function $R_3$ is defined by
\begin{multline}
  P_3(\bm{k}_1,\bm{k}_2,\bm{k}_3) =
  P_1(\bm{k}_1) P_1(\bm{k}_2) P_1(\bm{k}_3)
\\
  \times\left[1
  + R_2(\bm{k}_1,\bm{k}_2) + R_2(\bm{k}_2,\bm{k}_3)
  + R_2(\bm{k}_1,\bm{k}_3) 
  + R_3(\bm{k}_1,\bm{k}_2,\bm{k}_3) \right],
\label{eq:2-51}
\end{multline}
where we adopt notational abbreviations such as $P_1(\bm{k}_1) =
P_1(A_{\rm{k}_1},\theta_{\bm{k}_1})$, etc. Three modes, $\bm{k}_1$,
$\bm{k}_2$, $\bm{k}_3$ are all different from each other. The reduced
distribution function is a component that is not represented by
lower-point distribution functions.

Contributions to the reduced three-point distribution function from
Eq.~(\ref{eq:2-32}) come from $\mathcal{Q}^{(1)}_{2}$,
$\mathcal{Q}^{(2)}_{4}$, $\mathcal{Q}^{(2)}_{5}$,
$\mathcal{Q}^{(2)}_{9}$, $\mathcal{Q}^{(2)}_{10}$,
$\mathcal{Q}^{(2)}_{11}$. There are conditions among wavevectors under
which these terms contribute to the distribution function. For
example, the terms $\mathcal{Q}^{(1)}_{2}$, $\mathcal{Q}^{(2)}_{9}$
survive only when there is a relation $\bm{k}_3 = \bm{k}_1 + \bm{k}_2$
or its permutation among the three modes, and so forth. The resulting
reduced function $R_3$ is totally symmetric with respect to its
arguments. It is convenient to define an asymmetric function $R^{\rm
  (a)}_3$, with which the reduced function is obtained by 
\begin{multline}
  R_3(\bm{k}_1,\bm{k}_2,\bm{k}_3) =
  R^{\rm (a)}_3(\bm{k}_1,\bm{k}_2,\bm{k}_3) +
  R^{\rm (a)}_3(\bm{k}_2,\bm{k}_3,\bm{k}_1) +
  R^{\rm (a)}_3(\bm{k}_3,\bm{k}_1,\bm{k}_2)
\\ +
  R^{\rm (a)}_3(\bm{k}_2,\bm{k}_1,\bm{k}_3) +
  R^{\rm (a)}_3(\bm{k}_1,\bm{k}_3,\bm{k}_2) +
  R^{\rm (a)}_3(\bm{k}_3,\bm{k}_2,\bm{k}_1),
\label{eq:2-52}
\end{multline}
i.e., asymmetric functions with all the possible permutations are
summed up to obtain the reduced function. The asymmetric
functions can be extracted from expressions of $\mathcal{Q}^{(j)}_i$
given in the Appendix~\ref{app:a}, resulting in
\begin{align}
&
  R^{\rm (a)}_3(\bm{k}_1,\bm{k}_2,\bm{k}_3) =
  \frac{1}{\sqrt{V}}
  A_{\bm{k}_1} A_{\bm{k}_2} A_{\bm{k}_3}
  \cos\left(
    \theta_{\bm{k}_1} + \theta_{\bm{k}_2} - \theta_{\bm{k}_3}\right)
  p^{(3)}\left(\bm{k}_1,\bm{k}_2,-\bm{k}_3\right)
\nonumber\\
  & \quad
  + \frac{1}{V}
  {A_{\bm{k}_1}}^2 {A_{\bm{k}_2}} {A_{\bm{k}_3}}
  \cos\left(2\theta_{\bm{k}_1} + \theta_{\bm{k}_2} - \theta_{\bm{k}_3}\right)
  p^{(4)}\left(\bm{k}_1,\bm{k}_1,\bm{k}_2,-\bm{k}_3\right)
\nonumber\\
  & \quad
  + \frac{1}{2V}
  {A_{\bm{k}_1}}^2 {A_{\bm{k}_2}} {A_{\bm{k}_3}}
  \cos\left(2\theta_{\bm{k}_1} - \theta_{\bm{k}_2} -
      \theta_{\bm{k}_3}\right)
  p^{(4)}\left(\bm{k}_1,\bm{k}_1,-\bm{k}_2,-\bm{k}_3\right)
\nonumber\\
  & \quad
  + \frac{1}{V}
  \biggl[
    {A_{\bm{k}_1}}^2 {A_{\bm{k}_2}}^2 {A_{\bm{k}_3}}^2
    \cos^2
    \left(\theta_{\bm{k}_1}+\theta_{\bm{k}_2}-\theta_{\bm{k}_3}\right)
    - \frac12
    \left( {A_{\bm{k}_1}}^2 {A_{\bm{k}_2}}^2
      + {A_{\bm{k}_1}}^2 {A_{\bm{k}_3}}^2
      + {A_{\bm{k}_2}}^2 {A_{\bm{k}_3}}^2
    \right.
\nonumber\\
  & \hspace{18pc}
    \left.
      - {A_{\bm{k}_1}}^2 - {A_{\bm{k}_2}}^2
      - {A_{\bm{k}_3}}^2 + 1
    \right)
  \biggr]
  \left[p^{(3)}\left(\bm{k}_1,\bm{k}_2,-\bm{k}_3\right)\right]^2
\nonumber\\
  & \quad
  + \frac{1}{V}
  \left[
    {A_{\bm{k}_1}}^2 {A_{\bm{k}_2}}^3 {A_{\bm{k}_3}}
    \cos\left(2\theta_{\bm{k}_1} - \theta_{\bm{k}_2}\right)
    \cos\left(2\theta_{\bm{k}_2} - \theta_{\bm{k}_3}\right) -
    {A_{\bm{k}_1}}^2 {A_{\bm{k}_2}} {A_{\bm{k}_3}}
    \cos\left(2\theta_{\bm{k}_1}+\theta_{\bm{k}_2} - \theta_{\bm{k}_3}\right)
  \right]
\nonumber \\
& \hspace{23pc} \times
    p^{(3)}\left(\bm{k}_1,\bm{k}_1,-\bm{k}_2\right)
    p^{(3)}\left(\bm{k}_2,\bm{k}_2,-\bm{k}_3\right)
\nonumber\\
  & \quad
  + \frac{1}{V}
  \biggl[
    2 {A_{\bm{k}_1}}^3 {A_{\bm{k}_2}}^2 A_{\bm{k}_3}
    \cos\left(2\theta_{\bm{k}_1} - \theta_{\bm{k}_2}\right)
    \cos\left(\theta_{\bm{k}_1} + \theta_{\bm{k}_2} - \theta_{\bm{k}_3}\right) -
    {A_{\bm{k}_1}}^3 A_{\bm{k}_3}
    \cos\left(3\theta_{\bm{k}_1} - \theta_{\bm{k}_3}\right)
\nonumber \\
& \hspace{8pc}
   -\; 2 {A_{\bm{k}_1}} A_{\bm{k}_2}^2 {A_{\bm{k}_3}}
    \cos\left(\theta_{\bm{k}_1} - 2 \theta_{\bm{k}_2} + \theta_{\bm{k}_3}\right)
  \biggr]
  p^{(3)}\left(\bm{k}_1,\bm{k}_1,-\bm{k}_2\right)
  p^{(3)}\left(\bm{k}_1,\bm{k}_2,-\bm{k}_3\right) 
\nonumber\\
  & \quad
  + \mathcal{O}(V^{-3/2}),
\label{eq:2-53}
\end{align}
where $\bm{k}_1, \bm{k}_2, \bm{k}_3 \in \uhs$. It is only when
$\bm{k}_1 + \bm{k}_2 = \bm{k}_3$, or its permutations is satisfied
that the first and fourth term on the rhs contribute. Similarly,
$2\bm{k}_1 + \bm{k}_2 = \bm{k}_3$ for the second term to contribute,
$2\bm{k}_1 = \bm{k}_2 + \bm{k}_3$ for the third term, $2\bm{k}_1 =
\bm{k}_2$ and $4\bm{k}_1 = \bm{k}_3$ for the fifth term, $3\bm{k}_1 =
\bm{k}_2$ and $2\bm{k}_1 = \bm{k}_3$ for the last term to contribute.
The three wavevectors $\bm{k}_1, \bm{k}_2, \bm{k}_3$ should be in a
2-dimensional plane because there is at least one constraint among
three wavevectors. In the last two terms, three wavevectors should be
in a 1-dimensional line, i.e., three wavevectors should be parallel to
each other.

Specifically, if there is only a relation
$\bm{k}_1 + \bm{k}_2 = \bm{k}_3$, and if there is not any other
relation among these wavenumbers, the reduced three-point function is
simply given by
\begin{align}
&
  R_3(\bm{k}_1,\bm{k}_2,\bm{k}_3) =
  \frac{2}{\sqrt{V}}
  A_{\bm{k}_1} A_{\bm{k}_2} A_{\bm{k}_3}
  \cos\left(
    \theta_{\bm{k}_1} + \theta_{\bm{k}_2} - \theta_{\bm{k}_3}\right)
  p^{(3)}\left(\bm{k}_1,\bm{k}_2,-\bm{k}_3\right)
\nonumber\\
  & \qquad + \frac{2}{V}
  \biggl[
    {A_{\bm{k}_1}}^2 {A_{\bm{k}_2}}^2 {A_{\bm{k}_3}}^2
    \cos^2
    \left(
      \theta_{\bm{k}_1} + \theta_{\bm{k}_2} - \theta_{\bm{k}_3}
    \right)
   - \frac12
    \left( {A_{\bm{k}_1}}^2 {A_{\bm{k}_2}}^2
      + {A_{\bm{k}_1}}^2 {A_{\bm{k}_3}}^2
      + {A_{\bm{k}_2}}^2 {A_{\bm{k}_3}}^2
    \right.
\nonumber\\
  & \hspace{18pc}
    \left.
      - {A_{\bm{k}_1}}^2 - {A_{\bm{k}_2}}^2
      - {A_{\bm{k}_3}}^2 + 1
    \right)
  \biggr]
  \left[p^{(3)}\left(\bm{k}_1,\bm{k}_2,-\bm{k}_3\right)\right]^2
\nonumber\\
  & \qquad + \mathcal{O}(V^{-3/2}),
  \qquad
  \mbox{($\bm{k}_1 + \bm{k}_2 = \bm{k}_3$, no other
    independent relation)}.
\label{eq:2-54}
\end{align}
The factor 2 relative to the asymmetric expression of
Eq.~(\ref{eq:2-53}) corresponds to a symmetry of $\bm{k}_1
\leftrightarrow \bm{k}_2$. If there is another relation among the
three wavevectors, the expression of Eq.~(\ref{eq:2-54}) is
insufficient and corresponding terms are added.

\subsection{Four-point distributions}

The four-point distributions can similarly be obtained. The reduced
four-point distribution function $R_4$ is defined by
\begin{multline}
  P_4(\bm{k}_1,\bm{k}_2,\bm{k}_3,\bm{k}_4) =
  P_1(\bm{k}_1) P_1(\bm{k}_2) P_1(\bm{k}_3) P_1(\bm{k}_4)
  \left[
      1
    + R_2(\bm{k}_1,\bm{k}_2) + \mbox{sym.(6)}
  \right.
\\
  \left.
    + R_2(\bm{k}_1,\bm{k}_2) R_2(\bm{k}_3,\bm{k}_4)
    + \mbox{sym.(3)}
    + R_3(\bm{k}_1,\bm{k}_2,\bm{k}_3)
    + \mbox{sym.(4)}
  + R_4(\bm{k}_1,\bm{k}_2,\bm{k}_3,\bm{k}_4)
 \right],
\label{eq:2-55}
\end{multline}
where $ + \mbox{sym.}(n)$ indicates additional $n-1$ terms which
are necessary to symmetrize the preceding term with respect to
$\bm{k}_1, \ldots, \bm{k}_4$. The same abbreviations for $P_4$ as in
the case of $P_3$, etc., are assumed.

Contributions to the reduced four-point distribution function from
Eq.~(\ref{eq:2-32}) come from $\mathcal{Q}^{(2)}_{6}$,
$\mathcal{Q}^{(2)}_{7}$,
$\mathcal{Q}^{(2)}_{13}$, $\mathcal{Q}^{(2)}_{14}$,
$\mathcal{Q}^{(2)}_{15}$, $\mathcal{Q}^{(2)}_{16}$,
$\mathcal{Q}^{(2)}_{17}$. The term $\mathcal{Q}^{(2)}_{12}$ is
absorbed in the non-reduced part of $R_2 \cdot R_2$ in
Eq.~(\ref{eq:2-55}). As in the previous subsection, we define an
asymmetric function $R^{\rm (a)}_4$, with which
\begin{equation}
  R_4(\bm{k}_1,\bm{k}_2,\bm{k}_3,\bm{k}_3)  =
  R^{\rm (a)}_4(\bm{k}_1,\bm{k}_2,\bm{k}_3,\bm{k}_4) +
  \mbox{sym.(24)},
\label{eq:2-56}
\end{equation}
where the last term indicates additional $23$ ($=4! -1$) terms to
symmetrize the previous term with respect to
$\bm{k}_1,\ldots,\bm{k}_4$. The asymmetric function is given by
\begin{align}
&
  R^{\rm (a)}_4(\bm{k}_1,\bm{k}_2,\bm{k}_3,\bm{k}_4) =
  \frac{1}{3V}
  {A_{\bm{k}_1}} {A_{\bm{k}_2}} {A_{\bm{k}_3}} {A_{\bm{k}_4}}
  \cos\left(\theta_{\bm{k}_1} + \theta_{\bm{k}_2} + \theta_{\bm{k}_3}
    - \theta_{\bm{k}_4}\right)
  p^{(4)}\left(\bm{k}_1,\bm{k}_2,\bm{k}_3,-\bm{k}_4\right)
\nonumber\\
  & \quad + \frac{1}{4V}
  {A_{\bm{k}_1}} {A_{\bm{k}_2}} {A_{\bm{k}_3}} {A_{\bm{k}_4}}
  \cos\left(\theta_{\bm{k}_1} + \theta_{\bm{k}_2} - \theta_{\bm{k}_3}
    - \theta_{\bm{k}_4}\right)
  p^{(4)}\left(\bm{k}_1,\bm{k}_2,-\bm{k}_3,-\bm{k}_4\right)
\nonumber\\
  & \quad + \frac{1}{V}
  \biggl[
    2{A_{\bm{k}_1}}^3 {A_{\bm{k}_2}} {A_{\bm{k}_3}} {A_{\bm{k}_4}}
    \cos\left(2\theta_{\bm{k}_1}-\theta_{\bm{k}_3}\right)
    \cos\left(\theta_{\bm{k}_1}-\theta_{\bm{k}_2}+\theta_{\bm{k}_4}\right)
\nonumber\\
& \qquad\qquad
  -\;
    2{A_{\bm{k}_1}} {A_{\bm{k}_2}} {A_{\bm{k}_3}} {A_{\bm{k}_4}}
    \cos\left(\theta_{\bm{k}_1} + \theta_{\bm{k}_2}
      - \theta_{\bm{k}_3} - \theta_{\bm{k}_4}\right)
  \biggr]
  p^{(3)}\left(\bm{k}_1,\bm{k}_1,-\bm{k}_3\right)
  p^{(3)}\left(\bm{k}_1,-\bm{k}_2,\bm{k}_4\right)
\nonumber\\
  & \quad + \frac{1}{V}
  \Bigl[
    {A_{\bm{k}_1}}^3 {A_{\bm{k}_2}} {A_{\bm{k}_3}}
    {A_{\bm{k}_4}}
    \cos\left(2\theta_{\bm{k}_1}-\theta_{\bm{k}_4}\right)
    \cos\left(\theta_{\bm{k}_1}-\theta_{\bm{k}_2}-\theta_{\bm{k}_3}\right)
&
\nonumber \\
&
\qquad\qquad
    - {A_{\bm{k}_1}} {A_{\bm{k}_2}} {A_{\bm{k}_3}} {A_{\bm{k}_4}}
    \cos\left(
      \theta_{\bm{k}_1}+\theta_{\bm{k}_2}+\theta_{\bm{k}_3} -
      \theta_{\bm{k}_4}
    \right)
  \Bigr]
  p^{(3)}\left(\bm{k}_1,\bm{k}_1,-\bm{k}_4\right)
  p^{(3)}\left(-\bm{k}_1,\bm{k}_2,\bm{k}_3\right)
\nonumber\\
  & \quad + \frac{1}{V}
  \Bigl[
    2{A_{\bm{k}_1}}^2 {A_{\bm{k}_2}} {A_{\bm{k}_3}} {A_{\bm{k}_4}}^2
    \cos\left(2\theta_{\bm{k}_1} - \theta_{\bm{k}_4}\right)
    \cos\left(\theta_{\bm{k}_2} - \theta_{\bm{k}_3}
      + \theta_{\bm{k}_4}\right)
\nonumber \\
& \qquad\qquad
    - {A_{\bm{k}_1}}^2 {A_{\bm{k}_2}} {A_{\bm{k}_3}}
    \cos\left(2\theta_{\bm{k}_1} + \theta_{\bm{k}_2} -
    \theta_{\bm{k}_3}\right)
  \Bigr]
  p^{(3)}\left(\bm{k}_1,\bm{k}_1,-\bm{k}_4\right)
  p^{(3)}\left(\bm{k}_2,-\bm{k}_3,\bm{k}_4\right)
\nonumber\\
  & \quad + \frac{1}{V}
  \biggl[
    {A_{\bm{k}_1}}^2 {A_{\bm{k}_2}} {A_{\bm{k}_3}} {A_{\bm{k}_4}}^2
    \cos\left(2\theta_{\bm{k}_1}-\theta_{\bm{k}_4}\right)
    \cos\left(\theta_{\bm{k}_2}+\theta_{\bm{k}_3}-\theta_{\bm{k}_4}\right)
&
\nonumber \\
&
\qquad\qquad
    - \frac12 {A_{\bm{k}_1}}^2 {A_{\bm{k}_2}} {A_{\bm{k}_3}}
    \cos\left(2\theta_{\bm{k}_1}-\theta_{\bm{k}_2}-\theta_{\bm{k}_3}\right)
  \biggr]
  p^{(3)}\left(\bm{k}_1,\bm{k}_1,-\bm{k}_4\right)
  p^{(3)}\left(\bm{k}_2,\bm{k}_3,-\bm{k}_4\right)
\nonumber\\
  & \quad + \frac{1}{V}
  \biggl[
    4{A_{\bm{k}_1}}^2 {A_{\bm{k}_2}} {A_{\bm{k}_3}} {A_{\bm{k}_4}}^2
    \cos\left(\theta_{\bm{k}_1}+\theta_{\bm{k}_2}-\theta_{\bm{k}_4}\right)
    \cos\left(\theta_{\bm{k}_1}-\theta_{\bm{k}_3}+\theta_{\bm{k}_4}\right)
\nonumber \\
&
\qquad\qquad
  -\; 2{A_{\bm{k}_1}}^2 {A_{\bm{k}_2}} {A_{\bm{k}_3}}
    \cos\left(2\theta_{\bm{k}_1}+\theta_{\bm{k}_2}-\theta_{\bm{k}_3}\right)
  - 2{A_{\bm{k}_2}} {A_{\bm{k}_3}} {A_{\bm{k}_4}}^2
    \cos\left(\theta_{\bm{k}_2}+\theta_{\bm{k}_3}-2\theta_{\bm{k}_4}\right)
  \biggr]
\nonumber \\
&
\hspace{22pc}
  \times\;
  p^{(3)}\left(\bm{k}_1,\bm{k}_2,-\bm{k}_4\right)
  p^{(3)}\left(\bm{k}_1,-\bm{k}_3,\bm{k}_4\right)
\nonumber\\
  & \quad + \mathcal{O}(V^{-3/2}),
\label{eq:2-57}
\end{align}
where $\bm{k}_1, \bm{k}_2, \bm{k}_3, \bm{k}_4 \in \uhs$.

Specifically, if there is only a relation
$\bm{k}_1 + \bm{k}_2 + \bm{k}_3 = \bm{k}_4$, and if there is not any other
relation among these wavevectors, the reduced four-point function is
simply given by
\begin{align}
  R_4(\bm{k}_1,\bm{k}_2,\bm{k}_3,\bm{k}_4) & =
  \frac{2}{V}
  {A_{\bm{k}_1}} {A_{\bm{k}_2}} {A_{\bm{k}_3}} {A_{\bm{k}_4}}
  \cos\left(\theta_{\bm{k}_1} + \theta_{\bm{k}_2} + \theta_{\bm{k}_3}
    - \theta_{\bm{k}_4}\right)
  p^{(4)}\left(\bm{k}_1,\bm{k}_2,\bm{k}_3,-\bm{k}_4\right)
\nonumber\\
  &
  + \mathcal{O}(V^{-3/2}),
  \quad \mbox{($\bm{k}_1 + \bm{k}_2 + \bm{k}_3 = \bm{k}_4$, no
    other independent relation)}.
\label{eq:2-58}
\end{align}
The symmetry factor 6 of permutating $(\bm{k}_1,\bm{k}_2,\bm{k}_3)$ is
taken into account. When there is an additional relation, such as
$2\bm{k}_1 = \bm{k}_4$ etc., a corresponding term, such as the fourth
term in Eq.~(\ref{eq:2-57}), should be added to the above equation.
Similarly, if there is only a relation $\bm{k}_1 + \bm{k}_2 = \bm{k}_3
+ \bm{k}_4$, and if there is not any other relation among these
wavenumbers, the reduced four-point function is given by
\begin{align}
  R_4(\bm{k}_1,\bm{k}_2,\bm{k}_3,\bm{k}_4)
  &=
  \frac{2}{V}
  {A_{\bm{k}_1}} {A_{\bm{k}_2}} {A_{\bm{k}_3}} {A_{\bm{k}_4}}
  \cos\left(\theta_{\bm{k}_1} + \theta_{\bm{k}_2} - \theta_{\bm{k}_3}
    - \theta_{\bm{k}_4}\right)
  p^{(4)}\left(\bm{k}_1,\bm{k}_2,-\bm{k}_3,-\bm{k}_4\right)
\nonumber\\
  & + \mathcal{O}(V^{-3/2}),
  \quad \mbox{($\bm{k}_1 + \bm{k}_2 = \bm{k}_3 + \bm{k}_4$, no
    other independent relation)}.
\label{eq:2-59}
\end{align}

\subsection{Higher-point distributions}

It is also straightforward to obtain expressions of five- and
six-point distribution functions from
Eqs.~(\ref{eq:12j})--(\ref{eq:12o}). Up to the order $V^{-1}$, five-
and six-point reduced distributions are present only when the
wavevectors satisfy at least two conditions. For example, the five
modes $\bm{k}_1,\ldots,\bm{k}_5$ with a condition $\bm{k}_1 + \bm{k}_2
+ \bm{k}_3 + \bm{k}_4 = \bm{k}_5$ do not have a reduced five-point
distribution function up to this order. This is consistent with the
fact that $N$-point correlations of Fourier modes are of order
$V^{1-N/2}$ as given by Eq.~(\ref{eq:2-13}).

\section{\label{sec:PhaseCorr}
Phase Correlations
}

One of the most prominent feature in non-Gaussian fields is that the
Fourier phases are not completely random. Therefore, the
non-Gaussianity is sometimes called as ``phase correlations''.
However, explicit forms of the phase correlations in non-Gaussian
fields was unknown. Unlike in the random Gaussian fields, Fourier
moduli and Fourier phases are not independently distributed, as
seen from the distribution function we derived above. Thus,
correlations only among phases does not have enough information on
non-Gaussianity. Moduli and phases of different modes are
mutually correlated. 

When one is interested only in phase distributions, the distribution
function of phases is obtained by integrating over Fourier moduli.
Such integrations are simply performed by using Eq.~(\ref{eq:2-34a}).
In the expressions for $\mathcal{Q}^{(i)}_j$ in Appendix~\ref{app:a},
the integrations result in replacements
\begin{equation}
  A_{\bm{k}} \rightarrow \frac{\sqrt{\pi}}{2}, \quad
  {A_{\bm{k}}}^2 \rightarrow 1, \quad
  {A_{\bm{k}}}^3 \rightarrow \frac{3\sqrt{\pi}}{4}, \quad
  {A_{\bm{k}}}^4 \rightarrow 2,
\label{eq:3-1}
\end{equation}
since all the different labels of wavevectors corresponds to actually
different modes. The same replacements are applied in the expressions
of the $N$-point distribution functions. The one-point phase
distribution function is constant because of Theorem~\ref{th:2-1}:
\begin{equation}
  2\pi P_1\left(\theta_{\bm{k}}\right) =  1.
\label{eq:3-2}
\end{equation}
This can be explicitly confirmed by Eq.~(\ref{eq:2-37}) and also by
Eq.~(\ref{eq:2-46}).

For higher-point distribution functions of phases, reduced functions
are defined along the lines of defining $R_N$. From the $N$-point
distribution function of phases
$P_N(\theta_{\bm{k}_1},\ldots,\theta_{\bm{k}_N})$, the reduced
distribution functions $C_N$ are iteratively defined by
\begin{align}
  (2\pi)^2 P_2(\theta_{\bm{k}_1},\theta_{\bm{k}_2}) = &
  1 + C_2(\theta_{\bm{k}_1},\theta_{\bm{k}_2})
\label{eq:3-2-1a}\\
  (2\pi)^3 P_3(\theta_{\bm{k}_1},\theta_{\bm{k}_2},\theta_{\bm{k}_3})
  = &
  1 + C_2(\theta_{\bm{k}_1},\theta_{\bm{k}_2})
  + C_2(\theta_{\bm{k}_2},\theta_{\bm{k}_3})
  + C_2(\theta_{\bm{k}_3},\theta_{\bm{k}_1})
  + C_3(\theta_{\bm{k}_1},\theta_{\bm{k}_2},\theta_{\bm{k}_3})
\label{eq:3-2-1b}\\
  (2\pi)^4 P_4(\theta_{\bm{k}_1},\ldots,\theta_{\bm{k}_4}) = &
  1 + C_2(\theta_{\bm{k}_1},\theta_{\bm{k}_2}) + {\rm sym.}(6)
  + C_2(\theta_{\bm{k}_1},\theta_{\bm{k}_2})
    C_2(\theta_{\bm{k}_3},\theta_{\bm{k}_4}) + {\rm sym.}(3)
\nonumber\\
  &
  + C_3(\theta_{\bm{k}_1},\theta_{\bm{k}_2},\theta_{\bm{k}_3})
  + {\rm sym.}(4)
  + C_4(\theta_{\bm{k}_1},\theta_{\bm{k}_2},\theta_{\bm{k}_3},\theta_{\bm{k}_4})
\label{eq:3-2-1c}
\end{align}
Such reduced functions are just given by integrals of the function $R_N$:
\begin{equation}
  C_N(\theta_{\bm{k}_1},\ldots,\theta_{\bm{k}_N}) =
  (2\pi)^N \int dA_{\bm{k}_1} \cdots dA_{\bm{k}_N}
  P_1(\bm{k}_1) \cdots P_1(\bm{k}_N)
  R_N(\bm{k}_1,\ldots,\bm{k}_N).
\label{eq:3-2-2}
\end{equation}

\subsection{Two-point distributions}

The two-point distribution functions of phases are obtained from
Eqs.~(\ref{eq:2-37}), (\ref{eq:2-50}) and (\ref{eq:3-2-2}). Up to the
order $V^{-1}$, the one-point function $P_1(\bm{k})$ in
Eq.~(\ref{eq:3-2-2}) can be replaced by a Gaussian function $P_{\rm
  G}(\bm{k})$. Thus the reduced function $C_2$ is just given by
replacements of Eq.~(\ref{eq:3-1}) in Eq.~(\ref{eq:2-50}). As a
result, the fourth term of Eq.~(\ref{eq:2-50}) vanishes and the phase
correlations between two modes appear only when one of the modes is
parallel to the other and the proportional factor is either 2 or 3.
Specifically, when $2\bm{k}_1 = \bm{k}_2 \equiv \bm{k}$,
\begin{multline}
  C_2\left(\theta_{\bm{k}}, \theta_{2\bm{k}}\right) =
  \frac{\sqrt{\pi}}{2\sqrt{V}}
  \cos(2\theta_{\bm{k}} - \theta_{2\bm{k}})
  p^{(3)}\left(\bm{k},\bm{k},-2\bm{k}\right)
  + \frac{1}{2V}
  \cos\left[2(2\theta_{\bm{k}} - \theta_{2\bm{k}})\right]
  \left[
    p^{(3)}\left(\bm{k},\bm{k},-2\bm{k}\right)
  \right]^2
\\
   + \mathcal{O}(V^{-3/2}),
\label{eq:3-3}
\end{multline}
and when $3\bm{k}_1 = \bm{k}_2 \equiv \bm{k}$,
\begin{equation}
  C_2\left(\theta_{\bm{k}}, \theta_{3\bm{k}}\right) =
    \frac{\pi}{8V}
    \cos(3\theta_{\bm{k}} - \theta_{3\bm{k}})
    p^{(4)}\left(\bm{k},\bm{k},\bm{k},-3\bm{k}\right)
  + \mathcal{O}(V^{-3/2}).
\label{eq:3-4}
\end{equation}
Otherwise, there is not any other two-point correlation between phases
up to this order. In the above two cases, the phase correlations are
present only between two modes in which wavevectors are mutually
parallel and their proportional factor is a simple integer. This
property is not specific for the lower-order expansion. In fact,
there is a following theorem:
\begin{theorem}
    Phase correlations between two different modes with wavevectors
    $\bm{k}_1$ and $\bm{k}_2$, where $\bm{k}_1,\bm{k}_2 \in \uhs$, are
    present only when the two wavevectors are parallel to each other,
    and the proportional factor $c$ is a rational number.
\label{th:3-1}
\end{theorem}
{\it Proof.} 
The two-point distribution function is obtained from
Eq.~(\ref{eq:2-25}) where the wavevectors in the summation take their
values only from four vectors, $\bm{k}_1$, $-\bm{k}_1$, $\bm{k}_2$,
and $-\bm{k}_2$. Let the number of these wavevectors in a term be
$n_1$, $m_1$, $n_2$, $m_2$, respectively. Because of Kronecker's
delta's in the polyspectra, the equation $n_1 \bm{k}_1 - m_1 \bm{k}_1
+ n_2 \bm{k}_2 - m_2 \bm{k}_2 = 0$ should be satisfied in the term.
Therefore, we obtain 
\begin{equation}
  \bm{k}_2 = \frac{n_1 - m_1}{m_2 - n_2} \bm{k}_1
\label{eq:3-5}
\end{equation}
unless $n_1 = m_1$ and $n_2 = m_2$ are simultaneously satisfied. The
theorem is proven if a term in which the conditions $n_1 = m_1$ and
$n_2 = m_2$ are satisfied does not contain phase distributions. In the
term with $n_1 = m_1$ and $n_2 = m_2$, the generalized Hermite
polynomial has a form of $H_{\bm{k}_1 \cdots -\bm{k}_1 \cdots \bm{k}_2
  \cdots -\bm{k}_2}$, where the numbers of $\bm{k}_1$ and $-\bm{k}_1$
are the same, and that of $\bm{k}_2$ and $-\bm{k}_2$ are the same. In
explicit forms of the polynomials, as illustrated in
Eqs.~(\ref{eq:2-27a}) -- (\ref{eq:2-27f}), the terms with Kronecker's
delta's, $\delta^{\rm K}_{\bm{k}_1 + \bm{k}_1}$, $\delta^{\rm
  K}_{-\bm{k}_1 - \bm{k}_1}$, $\delta^{\rm K}_{\bm{k}_2 + \bm{k}_2}$,
$\delta^{\rm K}_{-\bm{k}_2 - \bm{k}_2}$, $\delta^{\rm K}_{\bm{k}_1 +
  \bm{k}_2}$, $\delta^{\rm K}_{-\bm{k}_1 - \bm{k}_2}$, are zero since
$\bm{k}_1, \bm{k}_2 \in \uhs$. Only terms with $\delta^{\rm
  K}_{\bm{k}_1 - \bm{k}_2}$ and $\delta^{\rm K}_{-\bm{k}_1 + \bm{k}_2}$
survive. Therefore, the numbers of $\bm{k}_1$ and $-\bm{k}_1$ are the
same and that of $\bm{k}_2$ and $-\bm{k}_2$ are the same even in each
term of the generalized Hermite polynomials. That means each term
is a function of only combinations of
$\alpha_{\bm{k}_1}\alpha_{-\bm{k}_1} = |\alpha_{\bm{k}_1}|^2$ and
$\alpha_{\bm{k}_2}\alpha_{-\bm{k}_2} = |\alpha_{\bm{k}_2}|^2$, which
do not depend on phases. \hfill $\blacksquare$

According to this theorem, there are phase correlations between modes
$\bm{k}_1$ and $\bm{k}_2$ only when a relation $n \bm{k}_1 = m
\bm{k}_2$ is satisfied for some different integers $n,m$. In terms of
music, this is a relation of ``overtones''. Considering a fundamental
tone $\bm{k}$, the two modes are given by $m$-th and $n$-th overtones,
$\bm{k}_1 = m \bm{k}$, $\bm{k}_2 = n \bm{k}$. In the
Eq.~(\ref{eq:3-3}), for example, these numbers are $m = 1$ and $n =
2$, and they correspond to a fundamental tone and a first overtone (an
octave), respectively. In the Eq.~(\ref{eq:3-4}), the relation is a
perfect twelfth (an octave and fifth). A phase correlation between
$2\bm{k}$ and $3\bm{k}$ corresponds to ``a perfect fifth'', and that
between $3\bm{k}$ and $4\bm{k}$ corresponds to ``a perfect fourth'',
and so on.

In Eqs.~(\ref{eq:3-3}) and (\ref{eq:3-4}), lowest-order deviations
from random distribution of phases are of order
$\mathcal{O}(V^{-1/2})$ for a two-point phase distribution of modes
$\bm{k}$ and $2\bm{k}$, and $\mathcal{O}(V^{-1})$ for modes $\bm{k}$
and $3\bm{k}$. The deviations are of higher order when a proportional
factor of the two wavevectors is not simple. More precisely, we have a
following theorem:
\begin{theorem}
    Deviations from a random distribution of phases for a two-point
    phase distribution of modes $\bm{k}_1$ and $\bm{k}_2$ with a
    relation $n \bm{k}_1 = m \bm{k}_2$, where $n$ and $m$ are
    irreducible positive integers, are of order
    $\mathcal{O}(V^{-(n+m-2)/2})$, or higher.
\label{th:3-2}
\end{theorem} 
{\it Proof.}
In the proof of Theorem~\ref{th:3-1}, the term specified by positive
numbers $n_1$, $m_1$, $n_2$, $m_2$ has a order of
$\mathcal{O}(V^{-(n_1 + m_1 + n_2 + m_2 - 2m')/2})$, where $m'$ is the
number of polyspectra in the term in Eq.~(\ref{eq:2-25}). Firstly, the
lowest-order contribution comes from the terms which has only one
polyspectrum, $m' = 1$. Secondly, from Eq.~(\ref{eq:3-5}) in the
context of the proof of Theorem~\ref{th:3-1}, $n/m = (n_1 - m_1)/(m_2
- n_2)$ should be satisfied to have non-trivial phase distributions.
Under this constraint, the lowest-order term is given by $(n_1, m_1,
n_2, m_2) = (n,0,0,m), (0,n,m,0)$, and the order is
$\mathcal{O}(V^{(2-n-m)/2})$. This term is accompanied by a specific
polyspectrum. If the corresponding polyspectrum vanishes, the lowest
order is higher than that. \hfill $\blacksquare$

In fact, the lowest-order term in this theorem can be explicitly
obtained by inspecting Eq.~(\ref{eq:2-25}). From the proof above, and
counting overlapping factors in Eq.~(\ref{eq:2-25}), we obtain
\begin{multline}
  C_2(\theta_{\bm{k_1}},\theta_{\bm{k_2}}) =
  \frac{V^{(2-n-m)/2}}{n!m!} 
  p^{(n+m)}(
    \underbrace{\bm{k}_1,\ldots,\bm{k}_1}_{n},
    \underbrace{-\bm{k}_2,\ldots,-\bm{k}_2}_{m})
\\
  \times
  \int 2A_{\bm{k}_1} dA_{\bm{k}_1}
     2A_{\bm{k}_2} dA_{\bm{k}_2}
  (
    H_{\scriptsize \underbrace{\bm{k}_1,\ldots,\bm{k}_1}_{n},
     \underbrace{-\bm{k}_2,\ldots,-\bm{k}_2}_{m}} +
    H_{\scriptsize \underbrace{-\bm{k}_1,\ldots,-\bm{k}_1}_{n},
     \underbrace{\bm{k}_2,\ldots,\bm{k}_2}_{m}}
  )
  \exp\left(-{A_{\bm{k}_1}}^2 - {A_{\bm{k}_2}}^2\right)
\\
  + \mathcal{O}(V^{(1-n-m)/2}).
\label{eq:3-6}
\end{multline}
Since $\bm{k}_1$ and $\bm{k}_2$ are in uhs and are mutually
different, all the Kronecker's delta's are zero in expansions of
generalized Hermite polynomials appearing in the above equation.
Therefore, we have
\begin{align}
  H_{\scriptsize \underbrace{\bm{k}_1,\ldots,\bm{k}_1}_{n},
   \underbrace{-\bm{k}_2,\ldots,-\bm{k}_2}_{m}} +
  H_{\scriptsize \underbrace{-\bm{k}_1,\ldots,-\bm{k}_1}_{n},
   \underbrace{\bm{k}_2,\ldots,\bm{k}_2}_{m}}
  &=
  (\alpha_{-\bm{k}_1})^n (\alpha_{\bm{k}_2})^m +
  (\alpha_{\bm{k}_1})^n (\alpha_{-\bm{k}_2})^m
\\
  &=
  2 {A_{\bm{k}_1}}^n {A_{\bm{k}_2}}^m
  \cos\left( n \theta_{\bm{k}_1} - m \theta_{\bm{k}_2} \right),
\label{eq:3-7}
\end{align}
and Eq.~(\ref{eq:3-6}) reduces to
\begin{align}
  C_2(\theta_{\bm{k}_1},\theta_{\bm{k}_2}) &=
  \frac{2V^{(2-n-m)/2}}{n!m!}
  \Gamma\left(1+\frac{n}{2}\right) \Gamma\left(1+\frac{m}{2}\right)
  \cos\left( n \theta_{\bm{k}_1} - m \theta_{\bm{k}_2} \right)
\nonumber\\
& \qquad\qquad \times
  p^{(n+m)}(
    \underbrace{\bm{k}_1,\ldots,\bm{k}_1}_{n},
    \underbrace{-\bm{k}_2,\ldots,-\bm{k}_2}_{m})
  + \mathcal{O}(V^{(1-n-m)/2}),
\nonumber\\
  &\mbox{($n\bm{k}_1 = m\bm{k}_2$ for irreducible integers $m,n$)}.
\label{eq:3-8}
\end{align}
The lowest-order correction terms in Eqs.~(\ref{eq:3-3}) and
(\ref{eq:3-4}) are reproduced from Eq.~(\ref{eq:3-8}) by putting
$m=1,n=2$ and $m=1,n=3$ respectively. Lowest-order corrections of
all the other two-point phase correlations are generally obtained by
Eq.~(\ref{eq:3-8}).

In terms of music again, the phase correlations are stronger for modes
with a simpler overtone relation. For example, a pair of modes with an
octave, $(m,n) = (1,2)$, results in a phase correlation of order
$V^{-1/2}$. That with a perfect twelfth (an octave and fifth), $(m,n)
= (1,3)$ results in a correlation of order $V^{-1}$, Similarly, double
octaves, $(m,n) = (1,4)$, and a perfect fifth, $(m,n) = (2,3)$ result
in correlations of order $V^{-3/2}$. A perfect fourth, $(m,n) =
(3,4)$, corresponds to the order $V^{-5/2}$, and so forth. In other
words, Fourier phases are correlated between modes when the modes have
an harmony. The simpler the harmony sounds, the stronger the phase
correlation is.

\subsection{Three-point distributions}

The three-point distribution functions of phases are obtained from
Eqs.~(\ref{eq:2-37}), (\ref{eq:2-52}), (\ref{eq:2-53}) and
(\ref{eq:3-2-2}). Up to the order $V^{-1}$, the reduced function $C_3$
is just obtained by replacements of Eq.~(\ref{eq:3-1}) in
Eq.~(\ref{eq:2-53}) and symmetrization. The phase correlations among
three modes appear only when the three wavevectors are linearly
related. Specifically, when $\bm{k}_1 + \bm{k}_2 = \bm{k}_3$ and there
is not any other relation among these wavevectors, integrations of
Eq.~(\ref{eq:2-54}) result in
\begin{align}
  C_3(\bm{k}_1,\bm{k}_2,\bm{k}_3) &=
  \frac{\pi^{3/2}}{4\sqrt{V}}
  \cos\left(
    \theta_{\bm{k}_1} + \theta_{\bm{k}_2} - \theta_{\bm{k}_3}\right)
  p^{(3)}\left(\bm{k}_1,\bm{k}_2,-\bm{k}_3\right)
\nonumber\\
  &\quad + \frac{1}{V}
  \cos
  \left[
    2\left(
      \theta_{\bm{k}_1} + \theta_{\bm{k}_2} - \theta_{\bm{k}_3}
    \right)
  \right]
  \left[p^{(3)}\left(\bm{k}_1,\bm{k}_2,-\bm{k}_3\right)\right]^2
  + \mathcal{O}(V^{-3/2}),
\nonumber\\
  &
  \mbox{( $\bm{k}_1 + \bm{k}_2 = \bm{k}_3$, no other
    independent relation)}.
\label{eq:3-10}
\end{align}
Similarly, it is easy to obtain phase correlations when there is
only one linear relation of either $2\bm{k}_1 + \bm{k}_2 = \bm{k}_3$
or $2\bm{k}_1 = \bm{k}_2 + \bm{k}_3$. For the first case,
\begin{align}
  C_3(\bm{k}_1,\bm{k}_2,\bm{k}_3)  &=
  \frac{\pi}{4V}
  \cos\left(
    2\theta_{\bm{k}_1} + \theta_{\bm{k}_2} - \theta_{\bm{k}_3}\right)
  p^{(4)}\left(\bm{k}_1,\bm{k}_1,\bm{k}_2,-\bm{k}_3\right)
  + \mathcal{O}(V^{-3/2}),
\nonumber\\
  &
  \mbox{($2\bm{k}_1 + \bm{k}_2 = \bm{k}_3$, no other
    independent relation)},
\label{eq:3-11}
\end{align}
and for the second case,
\begin{align}
  C_3(\bm{k}_1,\bm{k}_2,\bm{k}_3)  & =
  \frac{\pi}{4V}
  \cos\left(
    2\theta_{\bm{k}_1} - \theta_{\bm{k}_2} - \theta_{\bm{k}_3}\right)
  p^{(4)}\left(\bm{k}_1,\bm{k}_1,-\bm{k}_2,-\bm{k}_3\right)
  + \mathcal{O}(V^{-3/2}),
\nonumber\\
  &
  \mbox{($2\bm{k}_1 = \bm{k}_2 + \bm{k}_3$, no other
    independent relation)}.
\label{eq:3-12}
\end{align}

Another specific example is that the three modes are given by
$\bm{k}$, $2\bm{k}$, $3\bm{k}$. In this case, the first, third,
fourth, and fifth terms in symmetrized Eq.~(\ref{eq:2-53}) are
relevant. The phase correlation in this case is calculated to be
\begin{align}
  C_3(\bm{k},2\bm{k},3\bm{k}) & =
  \frac{\pi^{3/2}}{4\sqrt{V}}
  \cos\left(
    \theta_{\bm{k}} + \theta_{2\bm{k}} - \theta_{3\bm{k}}\right)
  p^{(3)}\left(\bm{k},2\bm{k},-3\bm{k}\right)
\nonumber\\
 & \quad + \frac{1}{V}
  \cos
  \left[
    2\left(
      \theta_{\bm{k}} + \theta_{2\bm{k}} - \theta_{3\bm{k}}
    \right)
  \right]
  \left[p^{(3)}\left(\bm{k},2\bm{k},-3\bm{k}\right)\right]^2
\nonumber\\
  & \quad + \frac{\pi}{8V}
  \cos\left(
    \theta_{\bm{k}} - 2\theta_{2\bm{k}} + \theta_{3\bm{k}}
  \right)
  \left[
    2 p^{(4)}\left(2\bm{k},2\bm{k},-\bm{k},-3\bm{k}\right)
  \right.
\nonumber\\
  & \hspace{15pc}
  \left.
    - p^{(3)}\left(\bm{k},\bm{k},-2\bm{k}\right)
      p^{(3)}\left(\bm{k},2\bm{k},-3\bm{k}\right)
  \right]
\nonumber\\
  & \quad + \mathcal{O}(V^{-3/2}).
\label{eq:3-13}
\end{align}
To obtain the three-point distribution
$P_3(\theta_{\bm{k}},\theta_{2\bm{k}},\theta_{3\bm{k}})$ from the
above expression, it should be noted that the two-point contributions
given by Eq.~(\ref{eq:3-3}) and (\ref{eq:3-4}) have to be taken into
account.

Yet another example is the case that the three modes are given by
$\bm{k}$, $2\bm{k}$, $4\bm{k}$. In this case, 
\begin{align}
  C_3(\bm{k},2\bm{k},4\bm{k}) & =
  \frac{\pi}{4V}
  \cos\left(
    2\theta_{\bm{k}} + \theta_{2\bm{k}} - \theta_{4\bm{k}}
  \right)
  p^{(4)}\left(\bm{k},\bm{k},2\bm{k},-4\bm{k}\right)
\nonumber\\
  &\quad + \frac{\pi}{8V}
  \left[
    3\cos\left( 2\theta_{\bm{k}} - \theta_{2\bm{k}} \right)
     \cos\left( 2\theta_{2\bm{k}} - \theta_{4\bm{k}} \right)
   - 2\cos\left(
      2\theta_{\bm{k}} + \theta_{2\bm{k}} - \theta_{4\bm{k}}
    \right)
  \right]
\nonumber\\
  & \hspace{15pc} \times 
  p^{(3)}\left(\bm{k},\bm{k},-2\bm{k}\right)
  p^{(3)}\left(2\bm{k},2\bm{k},-4\bm{k}\right)
\nonumber\\
  &\quad + \mathcal{O}(V^{-3/2}).
\label{eq:3-14}
\end{align}

The Theorems ~\ref{th:3-1} and \ref{th:3-2} can be generalized to the
case of three-point distributions. Corresponding to the
Theorem~\ref{th:3-1}, we have a following theorem:
\begin{theorem}
    Phase correlations among three different modes with wavevectors
    $\bm{k}_1$, $\bm{k}_2$ and $\bm{k}_3$, where
    $\bm{k}_1,\bm{k}_2,\bm{k}_3 \in \uhs$, can be present only when there
    is a linear relation with integral coefficients among the three
    wavevectors:
    \begin{equation}
      j_1 \bm{k}_1 + j_2 \bm{k}_2 + j_3 \bm{k}_3 = \bm{0},
    \label{eq:3-15}
    \end{equation}
    where $j_i$ ($i=1,2,3$) are integers and at least two
    of $j_1$, $j_2$, $j_3$ are not zero.
\label{th:3-3}
\end{theorem}
{\it Proof.} 
The proof of this theorem is quite similar to that of
Theorem~\ref{th:3-1}. We consider the Eq.~(\ref{eq:2-25}) where there
are six vectors, $\pm\bm{k}_1$, $\pm\bm{k}_2$, $\pm\bm{k}_3$. These
wavevectors should satisfy $n_1 \bm{k}_1 - m_1 \bm{k}_1 + n_2 \bm{k}_2
- m_2 \bm{k}_2 + n_3 \bm{k}_3 - m_3 \bm{k}_3 = \bm{0}$ in each term,
where $n_1$ is a number of $\bm{k}_1$, $m_1$ is a number of
$-\bm{k}_1$, etc., in the corresponding term. Therefore,
Eq.~(\ref{eq:3-15}) should be satisfied with $j_i = n_i - m_i$. If
only one of $j_i$'s is not zero and the other two are zero, the
corresponding wavevector is identically zero, which is excluded by the
assumption. Therefore, the theorem is proven if the terms with $j_1 =
j_2 = j_3 =0$ do not have phase dependence. The rest of the proof is
almost the same as that of Theorem~\ref{th:3-1}. \hfill $\blacksquare$

It is immediately follows from this theorem that the three wavevectors
are laid on a two-dimensional plane even when the spatial dimension is
more than two. In other words, if the three wavevectors are not laid
on a common plane, the phases are not correlated at all.

When there is only one linear relation among the three wavevectors, as
in the cases of Eq.~(\ref{eq:3-10}), (\ref{eq:3-11}), (\ref{eq:3-12}),
the lowest-order term of $C_3$ has a simple form. In fact, this
simplicity is a general property, which can be shown by generalizing
the derivation of Eq.~(\ref{eq:3-8}) to the case of three-point
distributions. We consider the case that the only linear relation
is given by Eq.~(\ref{eq:3-15})
where $j_1$, $j_2$, $j_3$ are non-zero, irreducible integers. We
define the signs of $j_i$ ($i = 1,2,3$) by
\begin{equation}
  s_i = \frac{j_i}{|j_i|}.
\label{eq:3-16}
\end{equation}
Although there is an ambiguity of choosing an overall sign in
Eq.~(\ref{eq:3-15}), the following arguments are not affected by the
choice. Counting overlapping factors in Eq.~(\ref{eq:2-25}), the
lowest-order contributions are given by similar calculation of
Eqs.~(\ref{eq:3-6})--(\ref{eq:3-8}):
\begin{align}
&
  C_3(\theta_{\bm{k}_1},\theta_{\bm{k}_2},\theta_{\bm{k}_3}) =
  \frac{2V^{(2 - |j_1| - |j_2| - |j_3|)/2}}{|j_1|! |j_2|! |j_3|!} 
  \Gamma\left(1+\frac{|j_1|}{2}\right) \Gamma\left(1+\frac{|j_2|}{2}\right)
  \Gamma\left(1+\frac{|j_3|}{2}\right)
\nonumber\\
  & \qquad \times
  \cos\left(
    j_1 \theta_{\bm{k}_1} + j_2 \theta_{\bm{k}_2} +
    j_3 \theta_{\bm{k}_3}
  \right)
  p^{(|j_1|+|j_2|+|j_3|)}(
    \underbrace{s_1 \bm{k}_1,\ldots,s_1 \bm{k}_1}_{|j_1|},
    \underbrace{s_2 \bm{k}_2,\ldots,s_2 \bm{k}_2}_{|j_2|},
    \underbrace{s_3 \bm{k}_3,\ldots,s_3 \bm{k}_3}_{|j_3|}
  )
\nonumber\\
  & \quad + \mathcal{O}(V^{(1 - |j_1| - |j_2| - |j_3|)/2}).
\nonumber\\
  & \quad
  \mbox{($j_1\bm{k}_1 + j_2\bm{k}_2 + j_3\bm{k}_3 = \bm{0}$, $j_1
    j_2 j_3 \ne 0$, 
    no other independent relation)}.
\label{eq:3-17}
\end{align}
The ``unconnected part'' of Eq.~(\ref{eq:3-2-1b}), i.e., contributions
of $C_2$ on the left hand side, are not present in this case, because
it is assumed that there is not any linear relation between two of the
three wavevectors. It is crucial that any of $j_1$, $j_2$ $j_3$ is
assumed to be non-zero in the above equation.

If the spatial dimension is one, the number of independent linear
relation among three wavevectors can not be only one, and the
Eq.~(\ref{eq:3-17}) does not apply at all. For example,
Eq.~(\ref{eq:3-14}) is applicable even in one dimensional space, in
which case the lowest-order term is not given by just one
polyspectrum.

If the spatial dimension is two, any three wavevectors can not be
linearly independent. There is always a linear relation among three
vectors. When any two of the three vectors are not parallel to each
other, there is only one linear relation. The simpler the coefficients
of the linear relation is, the larger the three-point phase
correlation is.

The lowest-order terms in Eqs.~(\ref{eq:3-10})--(\ref{eq:3-12}) are
reproduced by putting $(j_1,j_2,j_3) = (1,1,-1)$, $(2,1,-1)$ and
$(2,-1,-1)$, respectively. There are more than one linear relation in
Eqs.~(\ref{eq:3-13}) and (\ref{eq:3-14}), and the
Eq.~(\ref{eq:3-7}) is not applied to them.

\subsection{Four-point distributions}

The four-point distribution functions of phases are similarly analyzed
as the three-point functions. The phase correlations among four modes
appear only when the four wavevectors are linearly related.
Specifically, when $\bm{k}_1 + \bm{k}_2 + \bm{k}_3 = \bm{k}_4$ and
there is not any other relation among these wavevectors, integrations
of Eq.~(\ref{eq:2-58}) result in
\begin{align}
  C_4(\theta_{\bm{k}_1},\theta_{\bm{k}_2},\theta_{\bm{k}_3},\theta_{\bm{k}_4})
  &=
  \frac{\pi^2}{8V}
  \cos\left(\theta_{\bm{k}_1} + \theta_{\bm{k}_2} + \theta_{\bm{k}_3}
    - \theta_{\bm{k}_4}\right)
  p^{(4)}\left(\bm{k}_1,\bm{k}_2,\bm{k}_3,-\bm{k}_4\right)
  + \mathcal{O}(V^{-3/2}),
\nonumber\\
  & \mbox{($\bm{k}_1 + \bm{k}_2 + \bm{k}_3 = \bm{k}_4$, no
    other independent relation)}.
\label{eq:3-18}
\end{align}
When $\bm{k}_1 + \bm{k}_2 = \bm{k}_3 + \bm{k}_4$ and
there is not any other relation among these wavevectors, integrations
of Eq.~(\ref{eq:2-59}) result in
\begin{align}
  C_4(\theta_{\bm{k}_1},\theta_{\bm{k}_2},\theta_{\bm{k}_3},\theta_{\bm{k}_4})
  &=
  \frac{\pi^2}{8V}
  \cos\left(\theta_{\bm{k}_1} + \theta_{\bm{k}_2} - \theta_{\bm{k}_3}
    - \theta_{\bm{k}_4}\right)
  p^{(4)}\left(\bm{k}_1,\bm{k}_2,-\bm{k}_3,-\bm{k}_4\right)
  + \mathcal{O}(V^{-3/2}),
\nonumber\\
  & \mbox{($\bm{k}_1 + \bm{k}_2 = \bm{k}_3 + \bm{k}_4$, no
    other independent relation)}.
\label{eq:3-19}
\end{align}

When there are more than one linear relation, the phase correlations
have different expressions. The explicit expressions are derived from
integrations of the Eq.~(\ref{eq:2-57}) and symmetrization. The
seventh term in Eq.~(\ref{eq:2-57}) identically vanishes by
integrating over the moduli $A_{\bm{k}_1}, \ldots, A_{\bm{k}_4}$, and
does not contribute to the phase correlations. 

For example, we consider a case that $2\bm{k}_2 - \bm{k}_3 = \bm{0}$
and $\bm{k}_1 - \bm{k}_2 + \bm{k}_4 = \bm{0}$ are simultaneously
satisfied. These equations are equivalent to the expressions $\bm{k}_3
= 2\bm{k}_1$ and $\bm{k}_4 = \bm{k}_2 - \bm{k}_1$. The third term in
Eq.~(\ref{eq:2-57}) contributes to the phase correlations in this
case. Additionally, an equation $\bm{k}_1 + \bm{k}_2 -\bm{k}_3 -
\bm{k}_4 = \bm{0}$ is also satisfied, and the second term contributes
as well. Other terms and their symmetrization are incompatible with
the conditions. Therefore, we have 
\begin{align}
  &C_4(\theta_{\bm{k}_1},\theta_{\bm{k}_2},\theta_{\bm{k}_3},
   \theta_{\bm{k}_4}) =
  \frac{\pi^2}{8V}
  \cos\left(\theta_{\bm{k}_1} + \theta_{\bm{k}_2} - \theta_{\bm{k}_3}
    - \theta_{\bm{k}_4}\right)
  p^{(4)}\left(\bm{k}_1,\bm{k}_2,-\bm{k}_3,-\bm{k}_4\right)
\nonumber\\
& \qquad
  + \frac{\pi^2}{16V}
  \left[
    3\cos\left(2\theta_{\bm{k}_1}-\theta_{\bm{k}_3}\right)
    \cos\left(\theta_{\bm{k}_1}-\theta_{\bm{k}_2}+\theta_{\bm{k}_4}\right)
  - 2 \cos\left(\theta_{\bm{k}_1} + \theta_{\bm{k}_2}
      - \theta_{\bm{k}_3} - \theta_{\bm{k}_4}\right)
  \right]
\nonumber\\
& \hspace{17pc} \times
  p^{(3)}\left(\bm{k}_1,\bm{k}_1,-\bm{k}_3\right)
  p^{(3)}\left(\bm{k}_1,-\bm{k}_2,\bm{k}_4\right)
  + \mathcal{O}(V^{-3/2}),
\nonumber\\
  & \qquad \mbox{($2\bm{k}_1 = \bm{k}_3$ and $\bm{k}_1 + \bm{k}_4 =
    \bm{k}_2$, no other independent relation)}.
\label{eq:3-20}
\end{align}
Similarly, considering other cases when either fourth, fifth, or sixth term in
Eq.~(\ref{eq:2-57}) is effective, we obtain
\begin{align}
  &C_4(\theta_{\bm{k}_1},\theta_{\bm{k}_2},\theta_{\bm{k}_3},
   \theta_{\bm{k}_4}) =
  \frac{\pi^2}{8V}
  \cos\left(\theta_{\bm{k}_1} + \theta_{\bm{k}_2} - \theta_{\bm{k}_3}
    - \theta_{\bm{k}_4}\right)
  p^{(4)}\left(\bm{k}_1,\bm{k}_2,-\bm{k}_3,-\bm{k}_4\right)
\nonumber\\
& \qquad
  + \frac{\pi^2}{16V}
  \left[
    3\cos\left(2\theta_{\bm{k}_1}-\theta_{\bm{k}_4}\right)
    \cos\left(\theta_{\bm{k}_1}-\theta_{\bm{k}_2}-\theta_{\bm{k}_3}\right)
  - 2 \cos\left(\theta_{\bm{k}_1} + \theta_{\bm{k}_2}
      + \theta_{\bm{k}_3} - \theta_{\bm{k}_4}\right)
  \right]
\nonumber\\
& \hspace{17pc} \times
  p^{(3)}\left(\bm{k}_1,\bm{k}_1,-\bm{k}_4\right)
  p^{(3)}\left(-\bm{k}_1,\bm{k}_2,\bm{k}_3\right)
  + \mathcal{O}(V^{-3/2}),
\nonumber\\
&  \qquad \mbox{($2\bm{k}_1 = \bm{k}_4$ and $\bm{k}_1 = \bm{k}_2 +
    \bm{k}_3$, no other independent relation)},
\label{eq:3-21}\\
  &C_4(\theta_{\bm{k}_1},\theta_{\bm{k}_2},\theta_{\bm{k}_3},
   \theta_{\bm{k}_4}) =
  \frac{\pi}{4V}
  \cos\left(2\theta_{\bm{k}_1} - \theta_{\bm{k}_2}
      + \theta_{\bm{k}_3} - 2\theta_{\bm{k}_4}\right)
  p^{(3)}\left(\bm{k}_1,\bm{k}_1,-\bm{k}_4\right)
  p^{(3)}\left(\bm{k}_2,-\bm{k}_3,\bm{k}_4\right)
\nonumber\\
  & \qquad  + \mathcal{O}(V^{-3/2}),
\nonumber\\
  & \qquad \mbox{($2\bm{k}_1 = \bm{k}_4$ and $\bm{k}_2 + \bm{k}_4 =
    \bm{k}_3$, no other independent relation)},
\label{eq:3-22}\\
  &C_4(\theta_{\bm{k}_1},\theta_{\bm{k}_2},\theta_{\bm{k}_3},
   \theta_{\bm{k}_4}) =
  \frac{\pi}{4V}
  \cos\left(2\theta_{\bm{k}_1} + \theta_{\bm{k}_2}
      + \theta_{\bm{k}_3} - 2\theta_{\bm{k}_4}\right)
  p^{(3)}\left(\bm{k}_1,\bm{k}_1,-\bm{k}_4\right)
  p^{(3)}\left(\bm{k}_2,\bm{k}_3,-\bm{k}_4\right)
\nonumber\\
  & \qquad  + \mathcal{O}(V^{-3/2}),
\nonumber\\
  & \qquad \mbox{($2\bm{k}_1 = \bm{k}_4$ and $\bm{k}_2 + \bm{k}_3 =
    \bm{k}_4$, no other independent relation)}.
\label{eq:3-23}
\end{align}

There is a theorem which is similar to Theorem~\ref{th:3-3}. If there
is only one linear relation among the four wavevectors, the
lowest-order term of $C_4$ can also be obtained as a trivial
generalization of Eq.~(\ref{eq:3-17}):
\begin{align}
  &C_4(\theta_{\bm{k}_1},\theta_{\bm{k}_2},\theta_{\bm{k}_3},
    \theta_{\bm{k}_4}) =
  \frac{2V^{(2 - |j_1| - |j_2| - |j_3| - |j_4|)/2}}
    {|j_1|! |j_2|! |j_3|! |j_4|!} 
\nonumber\\
  &\qquad \times
  \Gamma\left(1+\frac{|j_1|}{2}\right) \Gamma\left(1+\frac{|j_2|}{2}\right)
  \Gamma\left(1+\frac{|j_3|}{2}\right) \Gamma\left(1+\frac{|j_4|}{2}\right)
\nonumber\\
  &\qquad \times
  \cos\left(
    j_1 \theta_{\bm{k}_1} + j_2 \theta_{\bm{k}_2} +
    j_3 \theta_{\bm{k}_3} + j_4 \theta_{\bm{k}_4}
  \right)
\nonumber\\
  &\qquad \times
  p^{(|j_1|+|j_2|+|j_3|+|j_4|)}(
    \underbrace{s_1 \bm{k}_1,\ldots,s_1 \bm{k}_1}_{|j_1|},
    \underbrace{s_2 \bm{k}_2,\ldots,s_2 \bm{k}_1}_{|j_2|},
    \underbrace{s_3 \bm{k}_3,\ldots,s_3 \bm{k}_1}_{|j_3|},
    \underbrace{s_4 \bm{k}_4,\ldots,s_4 \bm{k}_4}_{|j_4|}
  )
\nonumber\\
  & \quad + \mathcal{O}(V^{(1 - |j_1| - |j_2| - |j_3| - |j_4|)/2}),
\nonumber\\
  & \mbox{($j_1\bm{k}_1 + j_2\bm{k}_2 + j_3\bm{k}_3 +
    j_4\bm{k}_4 = \bm{0}$, $j_1 j_2 j_3 j_4 \ne 0$,
    no other independent relation)}.
\label{eq:3-24}
\end{align}
Eqs.~(\ref{eq:3-18}) and (\ref{eq:3-19}) are reproduced from the
above formula.

If the spatial dimension is two or smaller, the number of
independent linear relations among four vectors can not be just one,
and the Eq.~(\ref{eq:3-24}) does not apply. For example,
Eqs.~(\ref{eq:3-20})--(\ref{eq:3-23}) is applicable even in two
dimensions, and the lowest-order term is not given by just one
polyspectrum. 

If the spatial dimension is three, any four vectors cannot be linearly
independent. There is always a linear relation among four vectors.
When any three of four vectors do not have a common plane, there is
only one linear relation. The simpler the coefficients of the linear
relation is, the larger the four-point phase correlation is.

\subsection{$N$-point distributions}

Now it is straightforward to generalize the Theorem~\ref{th:3-3} and
Eqs.~(\ref{eq:3-7}) and (\ref{eq:3-24}) to cases of $N$-point
distributions of phases, where $N\geq 5$. As a counterpart of
Theorem~\ref{th:3-3}, we have the following theorem:
\begin{theorem}
    Phase correlations among $N$ different modes with wavevectors
    $\bm{k}_1,\bm{k}_2,\ldots,\bm{k}_N$, where $N \geq 3$ and
    $\bm{k}_1,\bm{k}_2,\ldots,\bm{k}_N \in \uhs$, can be present only
    when there is a linear relation with integral coefficients among
    the $N$ wavevectors:
    \begin{equation}
      j_1 \bm{k}_1 + j_2 \bm{k}_2 + \cdots + j_N \bm{k}_N = \bm{0},
    \label{eq:3-25}
    \end{equation}
    where $j_i$ ($i=1,2,\ldots,N$) are integers and at least two of
    $j_1, j_2, \ldots, j_N$ are not zero.
\label{th:3-4}
\end{theorem}

The
generalization of Eqs.~(\ref{eq:3-17}) and (\ref{eq:3-24}) to $N$-point
distributions is given by
\begin{align}
  &C_N(\theta_{\bm{k}_1},\theta_{\bm{k}_2},\ldots,\theta_{\bm{k}_N}) =
  \frac{2V^{(2 - |j_1| - |j_2| - \cdots - |j_N|)/2}}
    {|j_1|! |j_2|! \cdots |j_N|!} 
\nonumber\\
  &\qquad \times
  \Gamma\left(1+\frac{|j_1|}{2}\right) \Gamma\left(1+\frac{|j_2|}{2}\right)
  \cdots \Gamma\left(1+\frac{|j_N|}{2}\right)
  \cos\left(
    j_1 \theta_{\bm{k}_1} + j_2 \theta_{\bm{k}_2} +
    \cdots + j_N \theta_{\bm{k}_N}
  \right)
\nonumber\\
  &\qquad \times\,
  p^{(|j_1|+|j_2|+\cdots +|j_N|)}(
    \underbrace{s_1 \bm{k}_1,\ldots,s_1 \bm{k}_1}_{|j_1|},
    \underbrace{s_2 \bm{k}_2,\ldots,s_2 \bm{k}_1}_{|j_2|},\ldots,
    \underbrace{s_N \bm{k}_N,\ldots,s_N \bm{k}_N}_{|j_N|}
  )
\nonumber\\
  & \quad + \mathcal{O}(V^{(1 - |j_1| - |j_2| - \cdots - |j_N|)/2}),
\nonumber\\
  & \mbox{($j_1\bm{k}_1 + j_2\bm{k}_2 + \cdots +
    j_N\bm{k}_N = \bm{0}$, $j_1 j_2 \cdots j_N \ne 0$,
    no other independent relation)}.
\label{eq:3-26}
\end{align}
However, if the spatial dimension $d$ is $N-2$ or smaller, the number
of independent linear relations cannot be just one, and the above
equation does not apply for $N \geq d+2$. In three dimensional space,
$N \geq 5$ of Eq.~(\ref{eq:3-26}) does not apply.

\section{\label{sec:numerical}
Numerical Examples}

In this section, analytic results derived in previous sections are
compared with numerical realizations of simple non-Gaussian
fields. One of the purposes of this section is to check the analytic
results, which is derived above by lengthy calculations. Another
purpose is to show how accurately truncated expressions in series of
$V^{-1/2}$ reproduce the numerical distributions of Fourier modes.

We artificially generate realizations of non-Gaussian fields on
regular sites in a three-dimensional cubic box. The number of sites,
which corresponds to the resolution of the fields, are $256^3$.
We generate three types of non-Gaussian fields, i.e., Voronoi
tessellation fields, lognormal fields, and quadratic Gaussian
fields. The first one is purely non-Gaussian, and the last two fields
are generated by simple non-linear transformations of Gaussian fields.

\subsection{Voronoi tessellation fields}

\begin{figure}
\includegraphics[width=18pc]{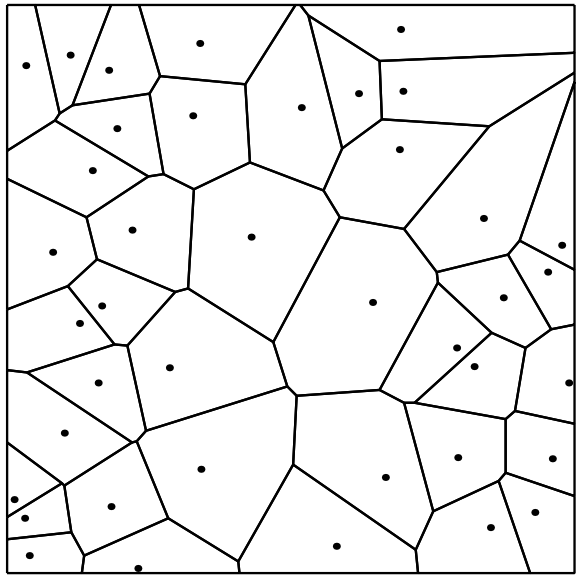}
\caption{\label{fig:voron2d} An example of a Voronoi tessellation in
  2-D space.}
\end{figure}

Our first examples of non-Gaussian fields are Voronoi tessellation
fields. To define a concept of Voronoi tessellation fields, first we
consider a set of points $\mathcal{S}$ which are randomly distributed
in space. A Voronoi tessellation \cite{vo1908} is a set of cells, each
of which delimits a part of space that is closer to a point of
$\mathcal{S}$ than any other point of $\mathcal{S}$. An example of a
Voronoi tessellation in 2-dimensional space is illustrated in
Fig.~\ref{fig:voron2d}.
In 2-dimensional space, a Voronoi tessellation consists of
1-dimensional lines on which the number of nearest points of
$\mathcal{S}$ is more than one. Similarly, in $d$-dimensional space, it
consists of $(d-1)$-dimensional hypersurfaces on which the number of
nearest points of $\mathcal{S}$ is more than one.
In 3-dimensional space, it consists of 2-dimensional surfaces, which
we call {\em Voronoi surfaces} below.

A Voronoi tessellation itself is not a random field. To generate a 3D
random field from a set of Voronoi surfaces, we need to thicken the
surfaces and make them into fuzzy walls in some way. We take a
following procedure in our example: first, we randomly distribute
points of $\mathcal{S}$ in the cubic box. Second, we identify the two
nearest random points of $\mathcal{S}$ from a given point $\bm{x}$ in
the box, and calculate the distances $d_1$ to the nearest random point
and $d_2$ to the second nearest point. These distances are the
functions of $\bm{x}$, and they are defined without ambiguity even if
the number of first and/or second nearest points of $\mathcal{S}$ is
plural. With those distances, we generate a field
$\rho(\bm{x})$ defined by
\begin{equation}
  \rho(\bm{x}) =
  \exp
  \left\{
    - \frac{\left[d_1(\bm{x}) - d_2(\bm{x})\right]^2}{2 \lambda^2}
  \right\},
\label{eq:4-1}
\end{equation}
where a thickness parameter $\lambda$ is an arbitrary length scale,
which corresponds to thickness of the fuzzy walls. In a limit of
$\lambda \rightarrow 0$, the field value is non-zero only on the
Voronoi surfaces. For a finite value of $\lambda$, the Voronoi
surfaces are thickened and we have fuzzy walls with Gaussian-like
profiles. Below, the fields generated by Eq.~(\ref{eq:4-1}) are called
{\em Voronoi tessellation fields}.

There are two parameters for Voronoi tessellation fields defined
above, i.e., the number density $n_{\rm r}$ of the random points
initially set to define Voronoi tessellations, and the thickness
parameter $\lambda$. If the thickness parameter is much larger than
the mean separation of the random points, the generated walls no
longer have planar shapes. We fix the thickness parameter at one-tenth
of the mean separation of points, $\lambda = 0.1\,{n_{\rm r}}^{-1/3}$.
Thus, we have only one parameter, the number density of initial random
points.

\begin{figure}
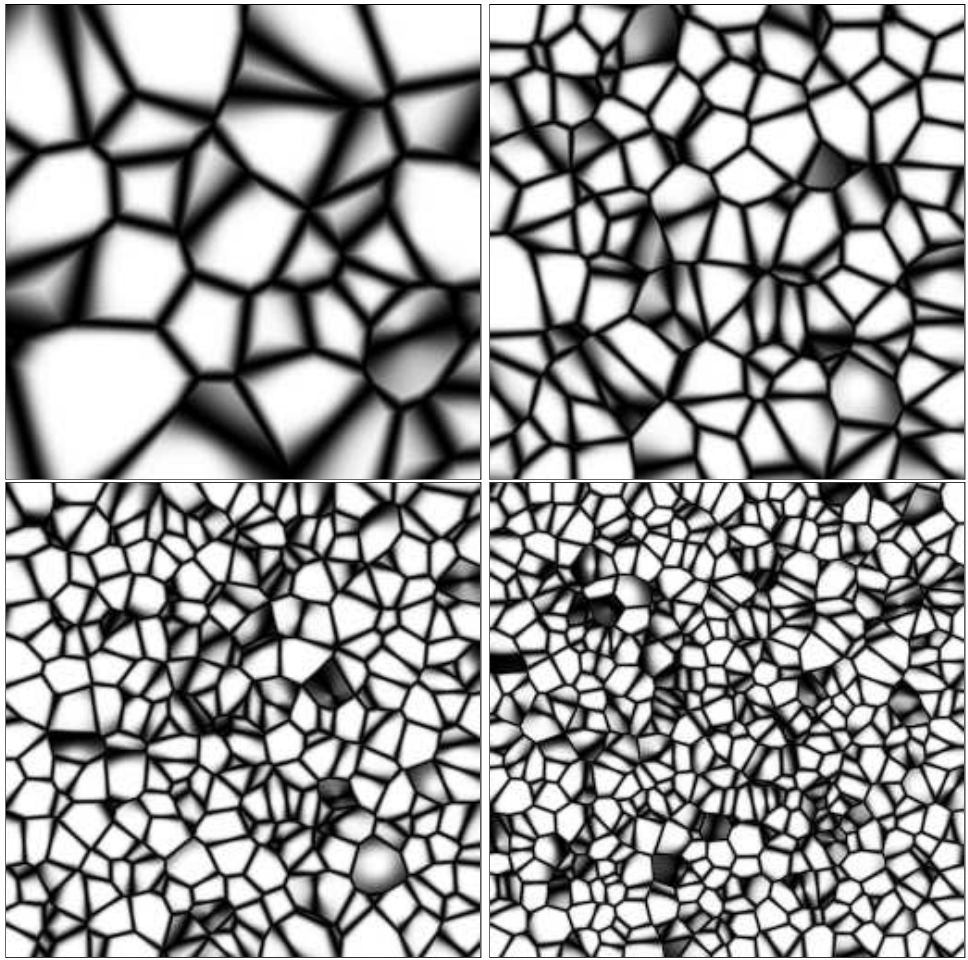

\includegraphics[width=15pc]{fig_2a.ps}
\includegraphics[width=15pc]{fig_2b.ps}\\
\includegraphics[width=15pc]{fig_2c.ps}
\includegraphics[width=15pc]{fig_2d.ps}
\caption{\label{fig:slices_vt} Gray-scale images of 2-dimensional
  slices of 3-dimensional Voronoi tessellation fields. Each panel has
  different number density of the initial random points. Upper left,
  upper right, lower left, and lower right panels respectively
  correspond to $5^3$, $10^3$, $15^3$, $20^3$ points per box.}
\end{figure}

In Fig.~\ref{fig:slices_vt}, gray-scale images of 2-dimensional slices
of generated 3-dimensional fields are plotted.
Four panels have different number densities of the initial random
points. It should be noted that slices of the 3-dimensional Voronoi
tessellation fields are not 2-dimensional Voronoi tessellation fields.
We calculate the spatial mean $\bar{\rho}$ of the generated field and
analyze the shifted field $f = \rho - \bar{\rho}$.

\begin{figure}
\includegraphics[width=25pc]{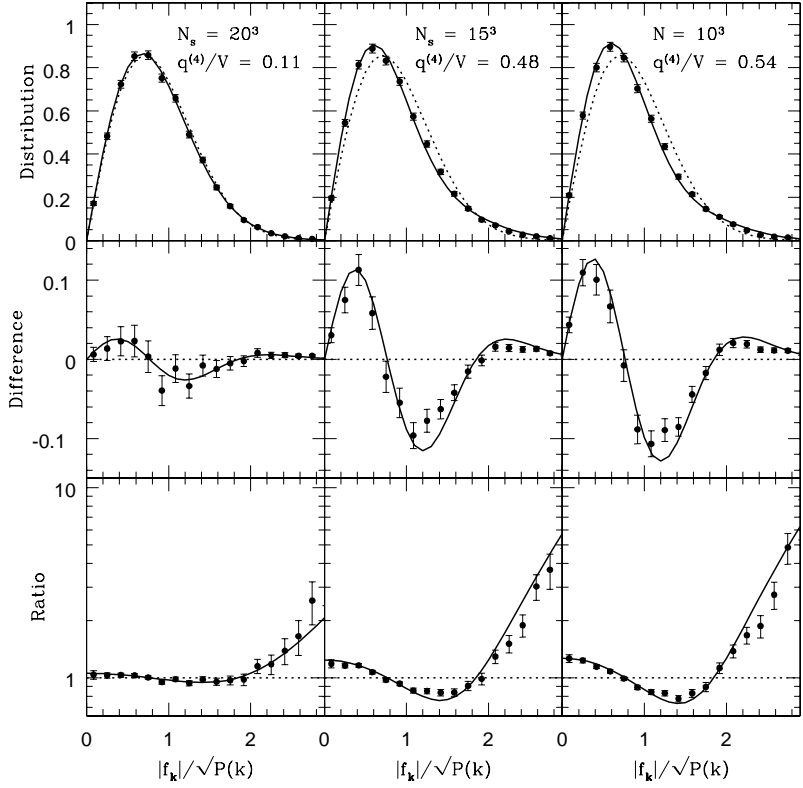}
\caption{\label{fig:modpdf_vt} The one-point distribution function of
  normalized Fourier modulus with wavevectors $\bm{k}$ which absolute
  magnitudes are in a range of $|\bm{k}| = [15, 25]$ in units of the
  fundamental wavenumber $2\pi/L$. The number of initial random points
  of Voronoi tessellations are $N_{\rm r} = 20^3$, $15^3$, and $10^3$
  from left to right panels, respectively. In top panels, the
  distribution functions are shown. Symbols represent results of
  numerical realizations. Error bars correspond to Poisson errors of
  counts. Dotted lines represent the Gaussian distribution. Solid
  lines include the lowest-order correction of non-Gaussianity. The
  degree of the non-Gaussianity $q^{(4)}/V$ is indicated in each
  panel. The middle panels show differences from Gaussian
  distributions. The bottom panels show ratios to the Gaussian
  distributions.}
\end{figure}

First, the one-point probability distribution function of a Fourier
modulus is calculated. Instead of considering many realizations, we
consider the distribution of Fourier modulus in a single realization.
We take all the modes in which absolute lengths of wavevectors are in
a certain range, and calculate the distribution of Fourier modulus of
these modes. Since the field does not have any statistically preferred
direction, the function $q^{(2n)}(\bm{k})$ of Eq.~(\ref{eq:2-41}) does
not depend on the direction of $\bm{k}$. Therefore, averaging over
directions of wavevector does not change the one-point distribution
function of Fourier modulus. If the function $q^{(2n)}(\bm{k})$ does
not change much in a certain range of absolute length of wavevectors,
the one-point distribution function of Fourier modes in a single
realization is still given by Eq.~(\ref{eq:2-44}). In
Fig.~\ref{fig:modpdf_vt}, the one-point distribution functions of
Fourier modulus are plotted.
Three cases with different numbers
$N_{\rm r}$ of initial random points of Voronoi tessellations are
considered in this figure. The numerical results are compared with the
analytic formula of Eq.~(\ref{eq:2-37}). The trispectrum $P^{(4)}$,
which we need in plotting the analytic curve, is numerically
calculated from each realization.

The agreement between the numerical results and analytic formula is
quite well. The deviation from the Gaussianity is larger for larger
$N_{\rm r}$. When the number $N_{\rm r}$ is small, the scale of
characteristic clustering pattern is large, and the non-Gaussian
correction factor $q^{(4)}/V$ of the corresponding scale becomes
large. In this example, the lowest-order correction works quite well
in each case. Even in the case of $N_{\rm r} = 10^3$, in which the
clustering pattern is quite irregular as seen in the upper right panel
of Fig.~\ref{fig:slices_vt}, the lowest-order correction is sufficient
to describe the non-Gaussianity. When the number $N_{\rm r}$ becomes
large, deviations from the Gaussian distribution becomes small. This
is manifestation of the Theorem~\ref{th:2-2}, since increasing the
number $N_{\rm r}$ corresponds to increasing the box size relative to
the characteristic scales of the structure.

Second, we consider the three-point distributions of phases. When
there are three wavevectors satisfying $\bm{k}_1 + \bm{k}_2 =
\bm{k}_3$ and there is not any other relations among the three, the
three-point phase distribution is given by Eq.~(\ref{eq:3-10}). When
the field is statistically isotropic, the bispectrum should be
rotationally invariant. Therefore, averaging over directions of
wavevector triplets does not change the three-point distribution
function of phases. If the normalized bispectrum $p^{(3)}$ does not
change much in a certain range of wavevector triplets, the three-point
distribution function of phases in a single realization is still given
by Eq.~(\ref{eq:3-10}). We first consider a wavevector $\bm{k}_1$, the
length of which is in a certain range $[k_{1}^{\rm min},k_{1}^{\rm
  max}]$. Next we consider a wavevector $\bm{k}_2$, the length of
which is in a certain range $[k_{2}^{\rm min},k_{2}^{\rm max}]$. The
angle $\theta_{12}$ between $\bm{k}_1$ and $\bm{k}_2$ should also be
in a certain range $[\theta_{12}^{\rm min},\theta_{12}^{\rm max}]$.
The third wavevector $\bm{k}_3$ is automatically determined by
$\bm{k}_3 = \bm{k}_1 + \bm{k}_2$. We find all the triplets in a
realization according this procedure, and calculate the distribution
of the phase closure $\theta_{\bm{k}_1} + \theta_{\bm{k}_2} -
\theta_{\bm{k}_1 + \bm{k}_2}$.

\begin{figure}
\includegraphics[width=25pc]{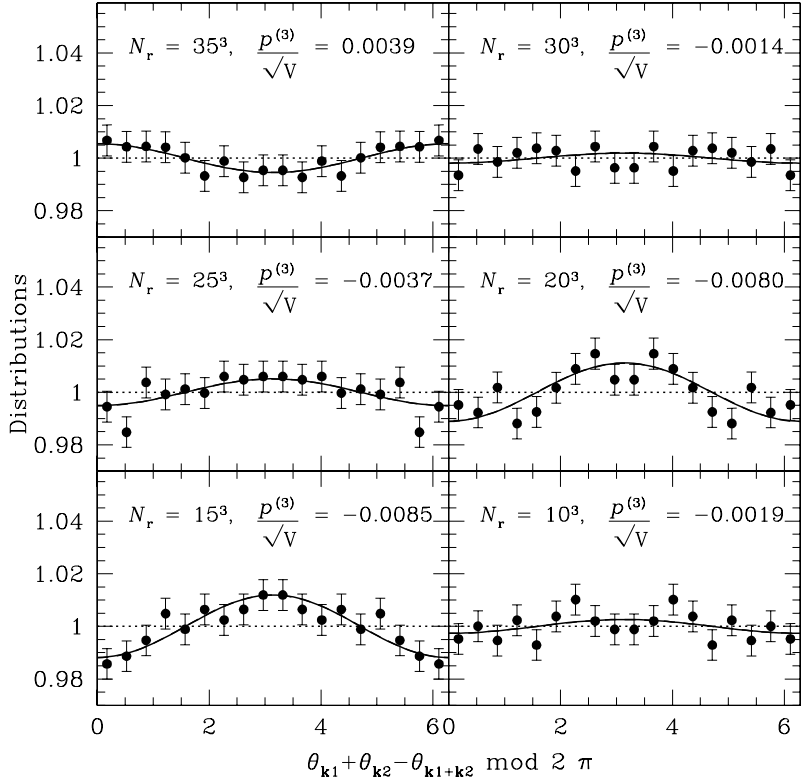}
\caption{\label{fig:phase3_vt} Relative distributions of the phase
  closure $\theta_{\bm{k}_1} + \theta_{\bm{k}_2} - \theta_{\bm{k}_1 +
    \bm{k}_2}$. The distributions are calculated for wavevectors
  satisfying $|\bm{k}_1| = [19,21]$, $|\bm{k}_2| = [19,21]$ in
  units of the fundamental wavenumber, and 
  $\theta_{12} = [118^\circ, 122^\circ]$, where $\theta_{12}$ is an
  angle between the two wavevectors. The number of
  initial random points of Voronoi tessellations $N_{\rm r}$, and
  parameters $p^{(3)}/\sqrt{V}$ are indicated in each panel.
  Symbols represent the distributions in numerical realizations.
  Dotted lines represent the Gaussian distribution (i.e., constant).
  Dashed and solid lines represent the first-order and second-order
  corrections, respectively, where dashed lines are indistinguishable
  from solid lines.
}
\end{figure}

In Fig.~\ref{fig:phase3_vt}, the distributions of phase closure are
plotted.
In these examples, the three wavevectors are not aligned, and the
Eq.~(\ref{eq:3-10}) is applicable. Configurations of the three
wavevectors, $\bm{k}_1$, $\bm{k}_2$, $- \bm{k}_3 = - \bm{k}_1 -
\bm{k}_2$, are approximately equilateral triangles. The lowest-order
and next-order corrections are indicated by dashed and solid lines,
respectively. The non-Gaussian correction terms are quite small in
each case, and the distribution is almost homogeneous (note the scales
of vertical axes).

\begin{figure}
\includegraphics[width=25pc]{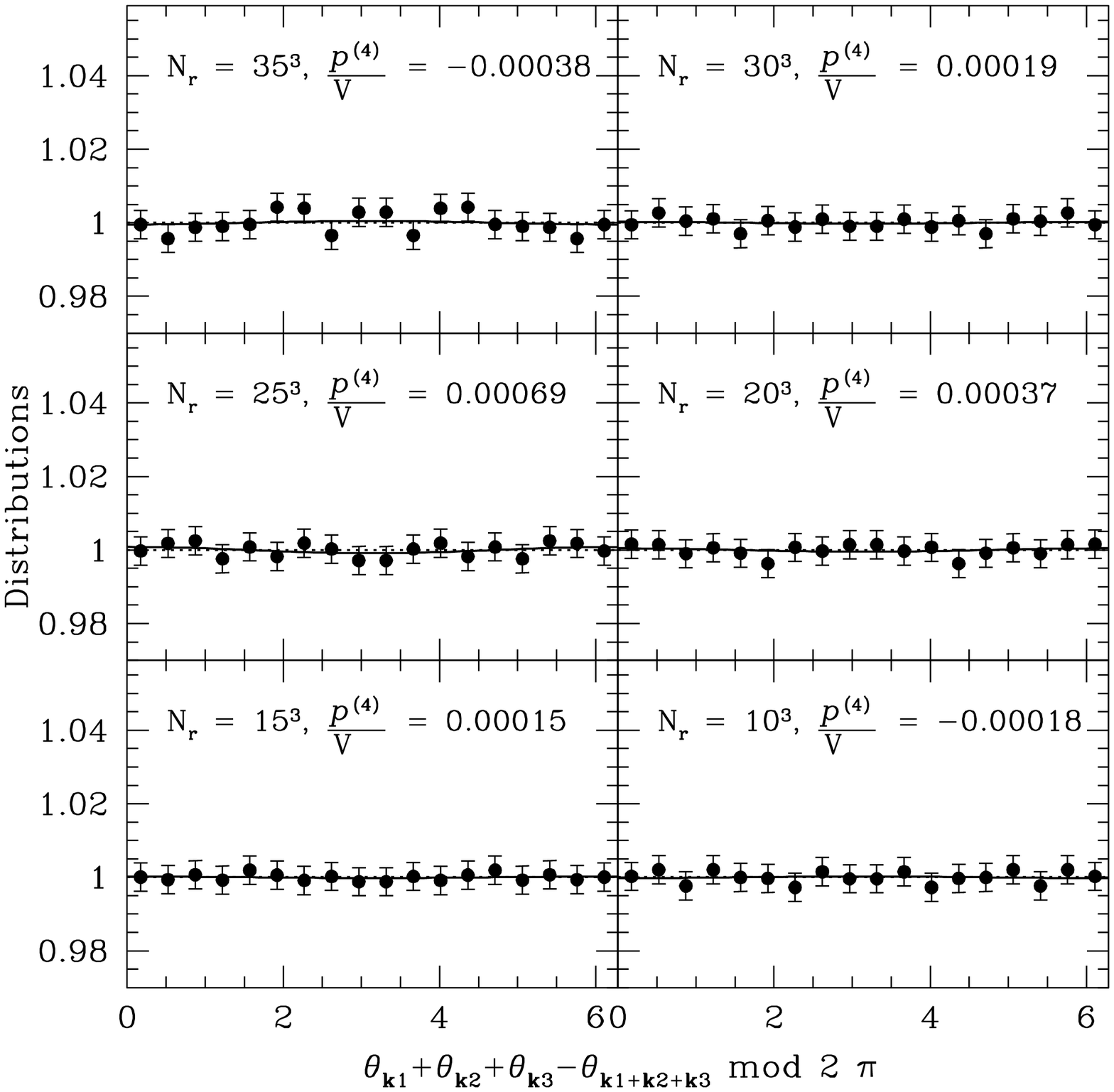}
\caption{\label{fig:phase4_vt} Relative distributions of the phase
  closure $\theta_{\bm{k}_1} + \theta_{\bm{k}_2} + \theta_{\bm{k}_3} -
  \theta_{\bm{k}_1 + \bm{k}_2 + \bm{k}_3}$. The distributions are
  calculated for wavevectors satisfying $|\bm{k}_1| = [19,21]$,
  $|\bm{k}_2| = [19,21]$, $|\bm{k}_3| = [19,21]$ in units of the
  fundamental wavenumber, and $\theta_{12} = [98^\circ, 102^\circ]$,
  $\theta_{23} = [98^\circ, 102^\circ]$, $\theta_{31} = [98^\circ,
  102^\circ]$, where $\theta_{ij}$ is an angle between the two
  wavevectors $\bm{k}_i$ and $\bm{k}_j$.
  The number of initial random
  points of Voronoi tessellations $N_{\rm r}$, and parameters
  $p^{(4)}/V$ are indicated in each panel. Symbols and lines are the
  same as Fig.~\ref{fig:phase3_vt}.
  }
\end{figure}

Similarly, we consider a phase closure of $\theta_{\bm{k}_1} +
\theta_{\bm{k}_2} + \theta_{\bm{k}_3} - \theta_{\bm{k}_1 + \bm{k}_2 +
  \bm{k}_3}$. Certain ranges for lengths of wavevectors $\bm{k}_1$,
$\bm{k}_2$, $\bm{k}_3$, and their mutual angles are considered, and
we calculate distributions of the phase closure of quadruplets in
these ranges. In Fig.~\ref{fig:phase4_vt}, the distributions are
plotted. 
The four wavevectors are not aligned, and the Eq.~(\ref{eq:3-18}) is
applicable in these examples. The distributions are very accurately
homogeneous.

In the Voronoi Tessellation fields, distribution functions of Fourier
modulus can be significantly different from Gaussian, while the phase
closures are almost homogeneous even though the distributions are
strongly non-Gaussian.

\subsection{Lognormal fields}

\begin{figure}
\includegraphics[width=18pc]{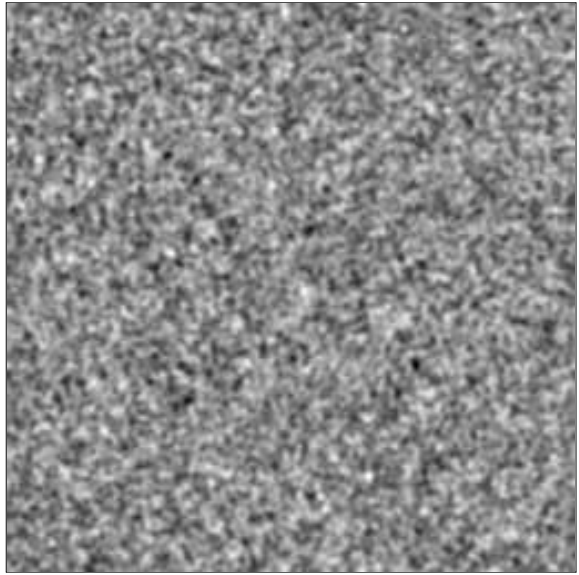}
\caption{\label{fig:slices_ga} A gray-scale image of a 2-dimensional slice of a
  3-dimensional random Gaussian field.
}
\end{figure}

\begin{figure}
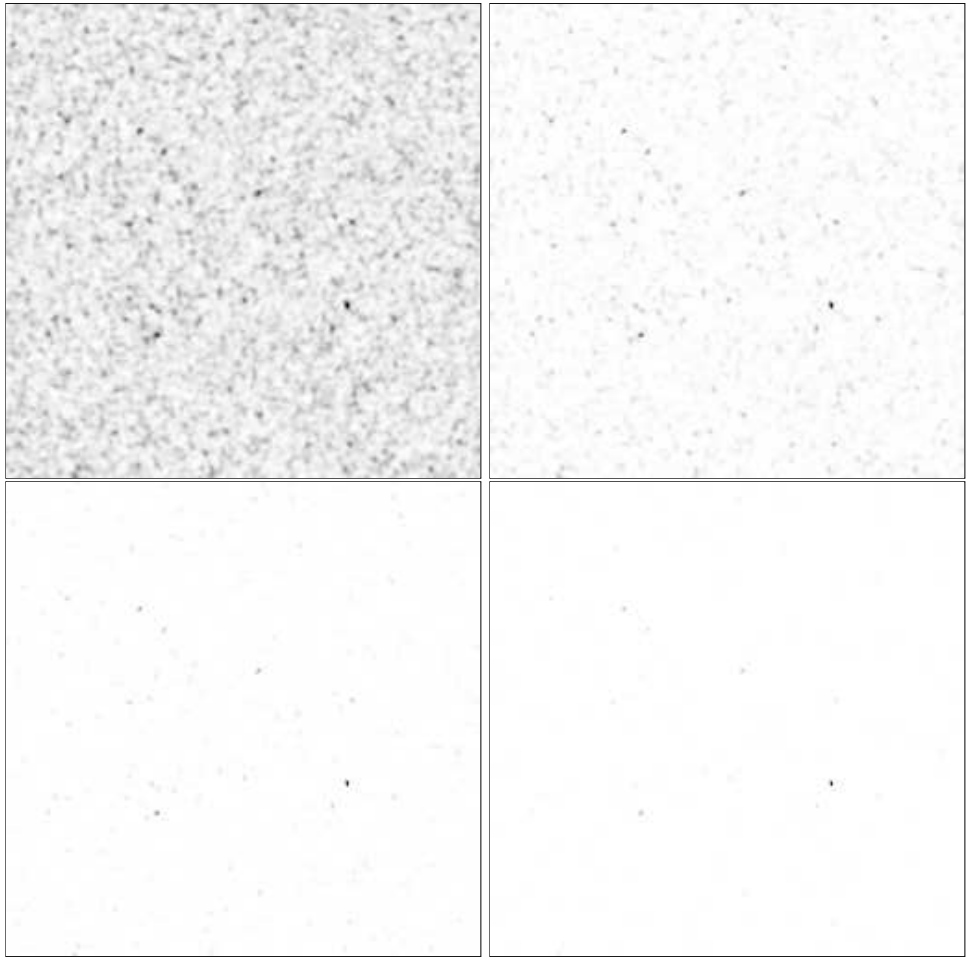

\includegraphics[width=15pc]{fig_7a.ps}
\includegraphics[width=15pc]{fig_7b.ps}\\
\includegraphics[width=15pc]{fig_7c.ps}
\includegraphics[width=15pc]{fig_7d.ps}
\caption{\label{fig:slices_ln} Gray-scale images of 2-dimensional
  slices of the 3-dimensional lognormal fields. Each panel has
  different non-Gaussianity parameter $g$. Upper left: $g=0.5$, upper
  right: $g=1.0$, lower left: $g=1.5$, lower right: $g=2.0$.}
\end{figure}

Our second example of non-Gaussian fields are lognormal fields. A
lognormal field is generated from a random Gaussian field by an
exponential mapping \cite{cj91}. First we consider  a random Gaussian field
$\phi(\bm{x})$ which has zero mean $\langle\phi(\bm{x})\rangle = 0$ and unit
variance $\langle[\phi(\bm{x})]^2\rangle = 1$, where $\langle\cdots\rangle$
represents a spatial average. This field can have a spatial correlation
$\xi_{\phi}(\bm{x}-\bm{x}') = \langle\phi(\bm{x})\phi(\bm{x}')\rangle$.
From this field, we consider a lognormal field
\begin{equation}
  \rho(\bm{x}) = e^{g \phi(\bm{x})},
\label{eq:4-2}
\end{equation}
where $g$ is an arbitrary parameter. Since the distribution of the
field $\phi(\bm{x})$ is Gaussian, the spatial mean of the generated
field can be analytically calculated to give $\langle \rho(\bm{x})
\rangle = e^{g^2/2} \equiv \bar{\rho}$. We define a
shifted lognormal field
\begin{equation}
  f(\bm{x}) = \frac{\rho(\bm{x}) - \bar{\rho}}{g\bar{\rho}} = 
  \frac{\exp\left[ g \phi(\bm{x}) - g^2/2\right] - 1}{g}.
\label{eq:4-3}
\end{equation}
In a limit $g \rightarrow 0$, the above field reduces to a random
Gaussian field. In this sense, the parameter $g$ controls the
non-Gaussianity of the distribution. The non-Gaussianity is large when
$g$ is large. We numerically generate a realization of random Gaussian
field from a given initial power spectrum $P_{\phi}(k)$, and a
realization of a lognormal field is obtained by the above equation. In
the examples below, we consider a power-law power spectrum with a
high-pass filter $P_{\phi}(k) \propto k^n e^{-k^2 \lambda^2/2}$, where
we set a spectral index $n=0$ and a smoothing scale $\lambda=0.03L$
($L$ is a box size). In Fig.~\ref{fig:slices_ga}, a gray-scale image
of 2-dimensional slice of the generated 3-dimensional random Gaussian
field $\phi(\bm{x})$ is shown.
In Fig.~\ref{fig:slices_ln}, gray-scale images of 2-dimensional slices
of the generated 3-dimensional lognormal fields are plotted.
The lognormal mapping of Eq.~(\ref{eq:4-2}) enhances rare peaks of
the initial Gaussian field. For cases the parameter $g$ is of order
unity or larger, a few peaks are prominent, and other structures are
diminished.

\begin{figure}
\includegraphics[width=25pc]{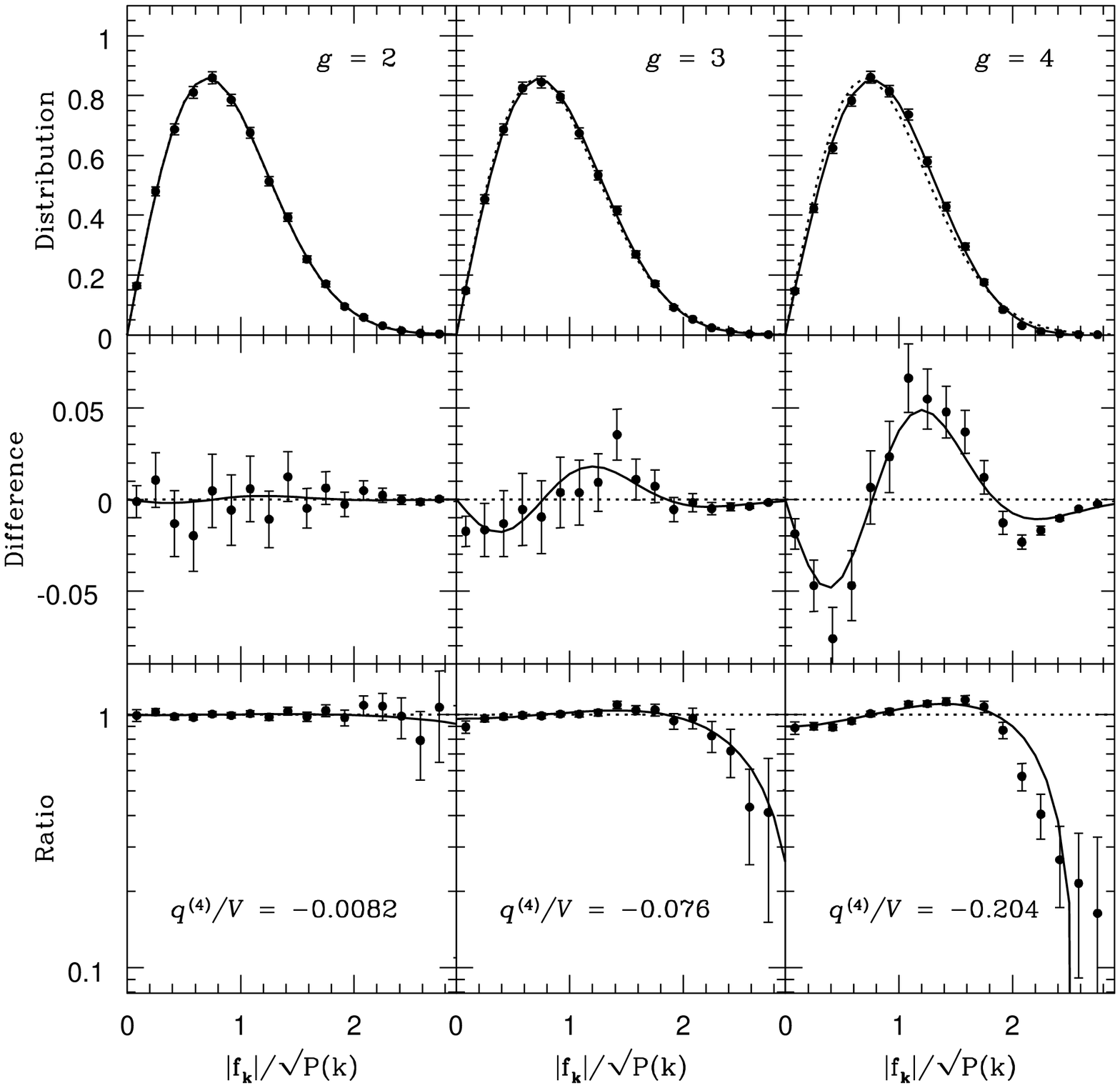}
\caption{\label{fig:modpdf_ln} Same as Fig.~\ref{fig:modpdf_vt}, but
  for the lognormal fields.
}
\end{figure}

The one-point distribution functions of a Fourier modulus are
similarly calculated as in the previous case. The results are plotted
in Fig.~\ref{fig:modpdf_ln}. 
The distribution for the case $g=2$ is almost Gaussian. We calculate
the cases $g \leq 2$, and they are also indistinguishable from
Gaussian. Although the field distribution for the case $g=2$ is very
different from Gaussian as seen in Fig.~\ref{fig:slices_ln}, the
distribution of Fourier modulus is almost Gaussian. Therefore, the
Theorem~\ref{th:2-2} is efficiently achieved in this case.

\begin{figure}
\includegraphics[width=25pc]{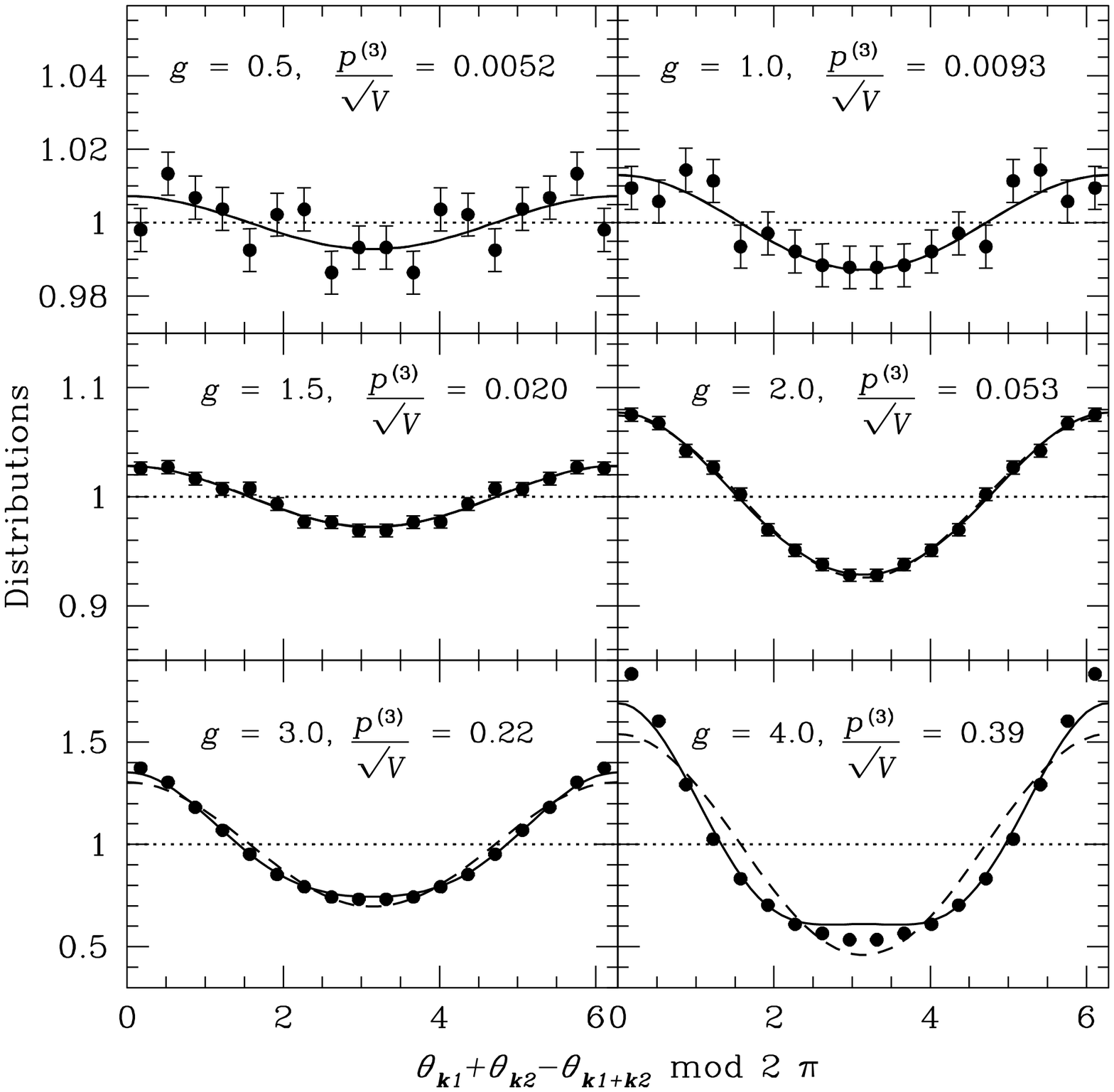}
\caption{\label{fig:phase3_ln} Same as Fig.~\ref{fig:phase3_vt}, but
  for lognormal fields. It should be noted that scales of the
  horizontal axes are different. }
\end{figure}

In Fig.~\ref{fig:phase3_ln}, the distributions of phase closure of
three modes, $\theta_{\bm{k}_1} + \theta_{\bm{k}_2} - \theta_{\bm{k}_1
  + \bm{k}_2}$, are plotted.
Choices of three wavevectors are the same as in
Fig.~\ref{fig:phase3_vt}. The phase closures exhibit significant
deviations from homogeneous distribution for $g \geq 1.5$. The
first-order corrections are sufficient for most of the cases, except
when the non-Gaussianity is very strong such as $g=3, 4$ cases.

\begin{figure}
\includegraphics[width=25pc]{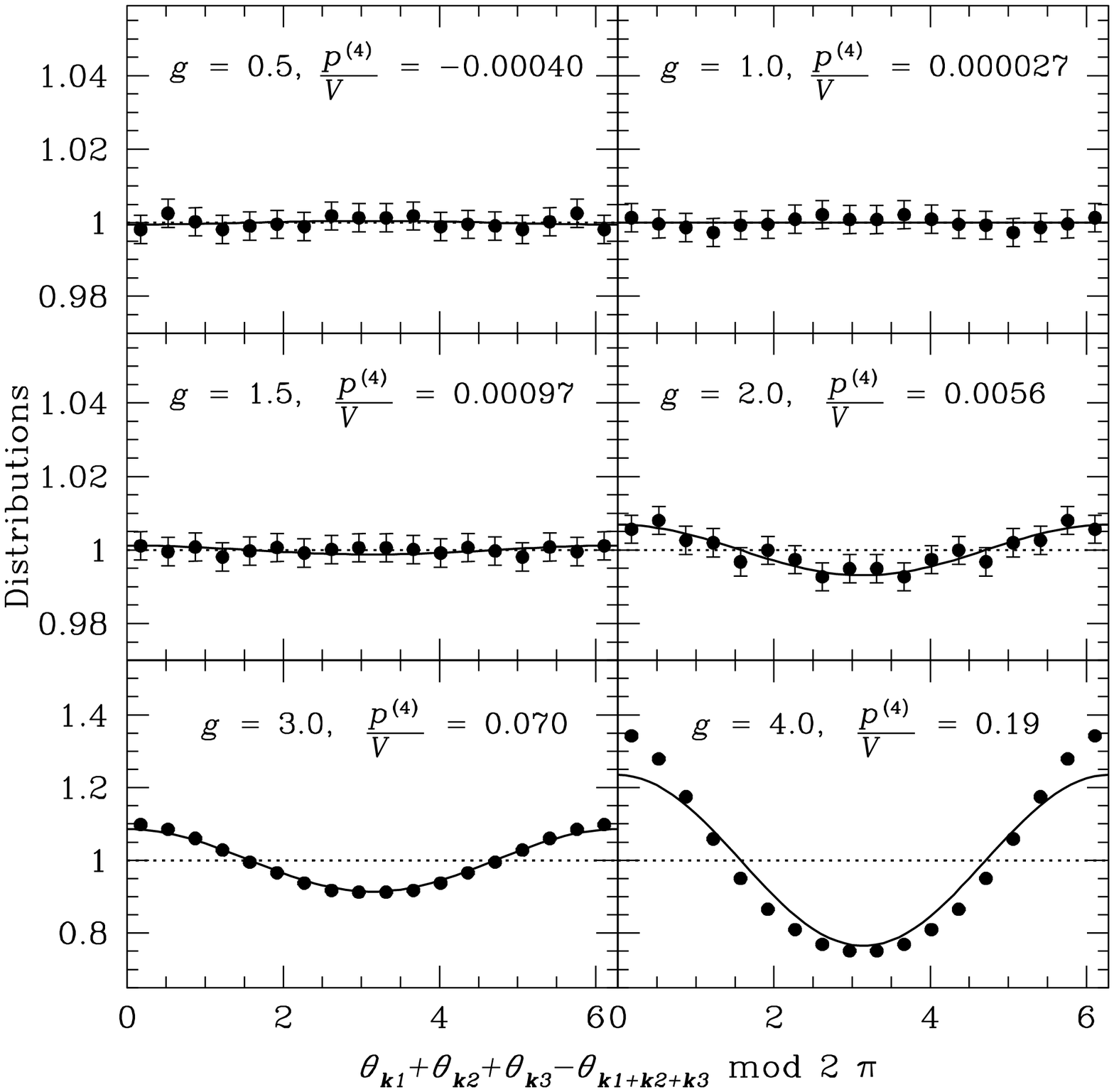}
\caption{\label{fig:phase4_ln} Same as Fig.~\ref{fig:phase4_vt},
but for lognormal fields. It should be noted that scales of the
horizontal axes are different. 
}
\end{figure}

In Fig.~\ref{fig:phase4_ln}, the distributions of phase closure of
four modes, $\theta_{\bm{k}_1} + \theta_{\bm{k}_2} + \theta_{\bm{k}_3}
- \theta_{\bm{k}_1 + \bm{k}_2 + \bm{k}_3}$, are plotted.
There are again deviations from homogeneous distributions. The degree
of the deviations is much less than the phase closures of three modes,
because the order of non-Gaussian correction terms is second.

In the lognormal fields, distribution functions of Fourier modulus can
be different from Gaussian when the non-Gaussianity is strong.
Moreover, the phase closures can be inhomogeneous when the
non-Gaussianity is strong.

\subsection{Quadratic fields}

Our last example of non-Gaussian fields are quadratic fields, which
are defined below. A lognormal field in the previous subsection is
generated from a random Gaussian field by an exponential mapping.
We similarly define a quadratic field, but by a quadratic mapping:
\begin{equation}
  f(\bm{x}) = \phi(\bm{x}) + h \left[\phi^2(\bm{x}) - 1\right],
\label{eq:4-4}
\end{equation}
where $\phi(\bm{x})$ is an random Gaussian field which satisfy
$\langle\phi(\bm{x})\rangle = 0$, $\langle\phi^2(\bm{x})\rangle =
1$, and $h$ is an arbitrary parameter. We numerically generate a
realization of random Gaussian field just in the same way as in the
lognormal case, using the same input power spectrum $P_{\phi}(k)
\propto k^n e^{-k^2 \lambda^2/2}$, where $n=0$ and $\lambda=0.03L$.
The parameter $h$ controls the non-Gaussianity. For sufficiently small
$h$, the field is essentially random Gaussian. For sufficiently large
$h$, the field approaches to a pure quadratic one, $f \propto \phi^2
- 1$.

\begin{figure}
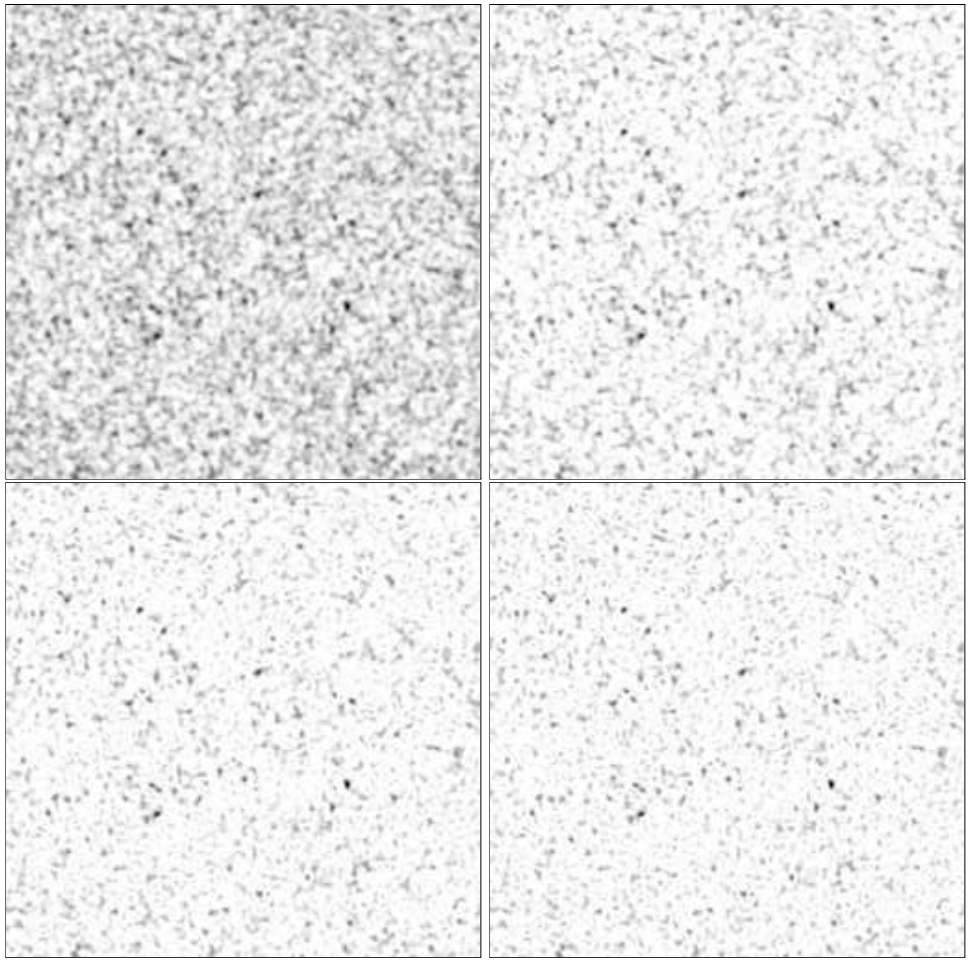

\includegraphics[width=15pc]{fig_11a.ps}
\includegraphics[width=15pc]{fig_11b.ps}\\
\includegraphics[width=15pc]{fig_11c.ps}
\includegraphics[width=15pc]{fig_11d.ps}
\caption{\label{fig:slices_qd} 2-dimensional slices of the
  3-dimensional quadratic fields. Each panel has different
  non-Gaussianity parameter $h$. Upper left: $h=0.2$, upper
  right: $h=0.5$, lower left: $h=1.0$, lower right: $h=2.0$.}
\end{figure}

In Fig.~\ref{fig:slices_qd}, gray-scale images of 2-dimensional slices
of the generated 3-dimensional lognormal fields are plotted.
The quadratic mapping of Eq.~(\ref{eq:4-4}) enhances peaks of the
initial Gaussian field. However, the enhancement is not so strong as
that in lognormal mappings.

\begin{figure}
\includegraphics[width=25pc]{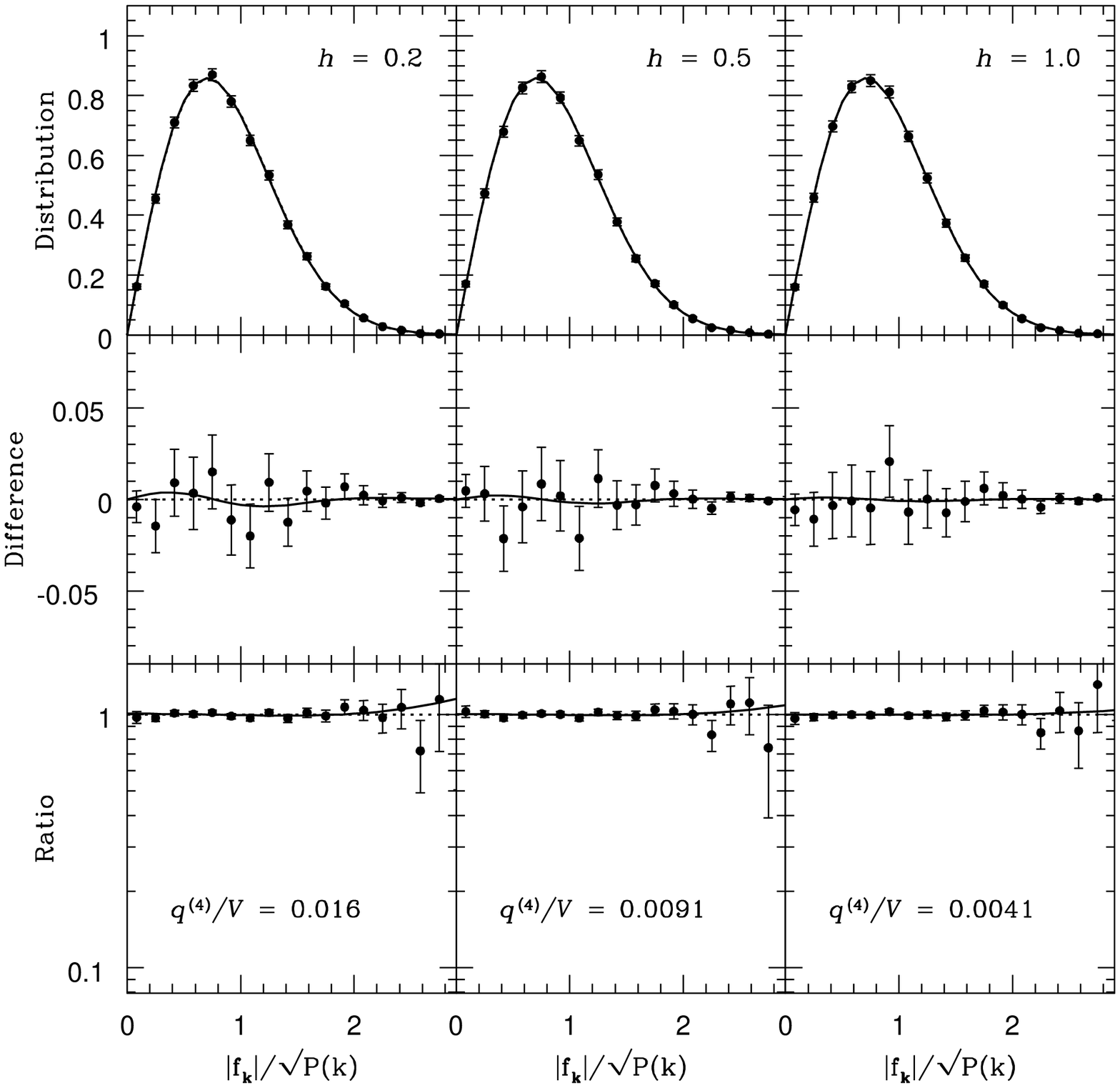}
\caption{\label{fig:modpdf_qd} Same as Fig.~\ref{fig:modpdf_vt}, but
  for the lognormal fields. It should be noted that scales of the
horizontal axes are different. 
}
\end{figure}

The one-point distribution functions of a Fourier modulus are
calculated, and the results are plotted in Fig.~\ref{fig:modpdf_qd}.
The distributions for all the cases are almost Gaussian, in spite of
the non-Gaussianity in the fields. Theorem~\ref{th:2-2} is
efficiently achieved also in this case.

\begin{figure}
\includegraphics[width=25pc]{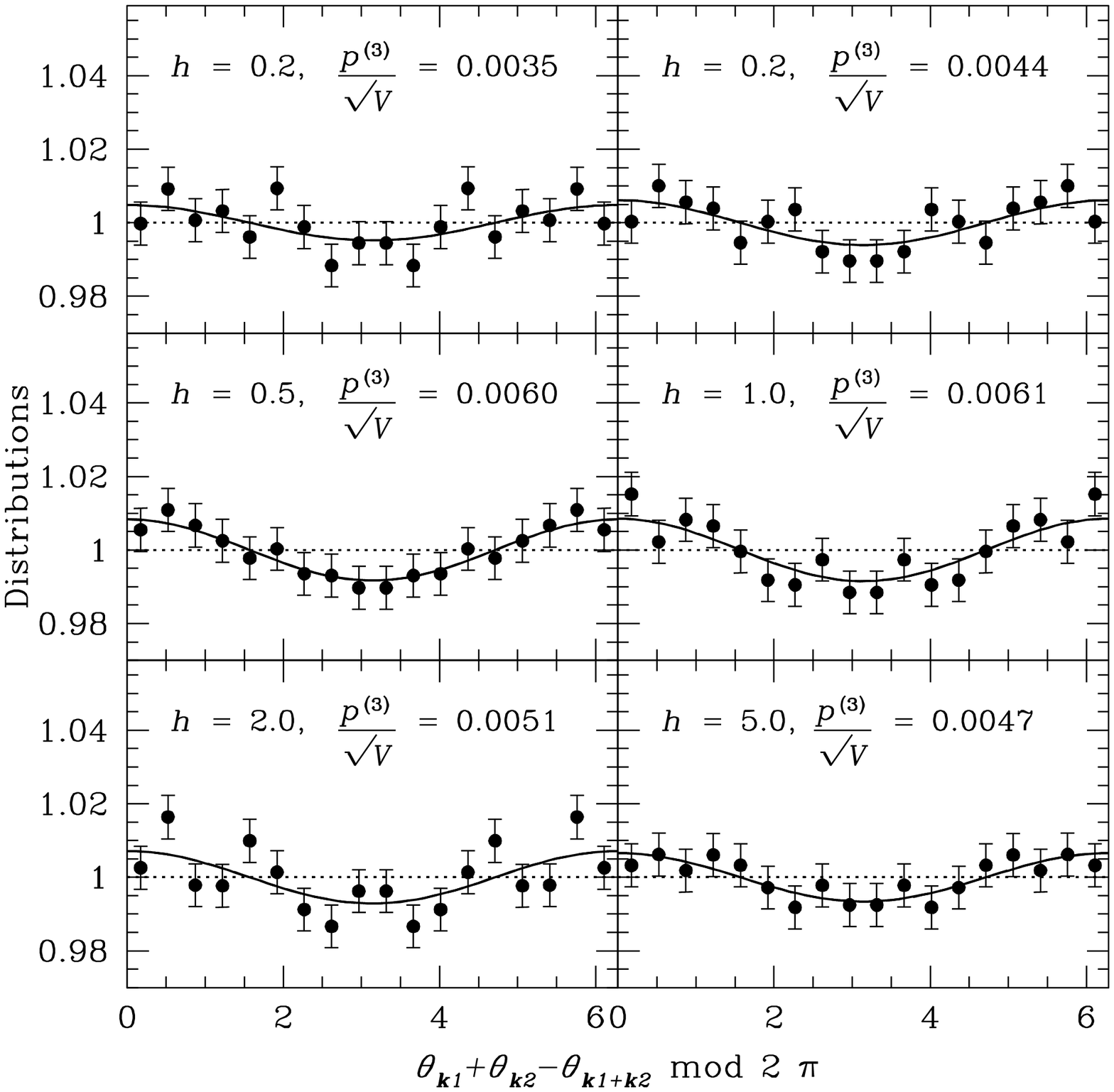}
\caption{\label{fig:phase3_qd} Same as Fig.~\ref{fig:phase3_vt}, but
  for lognormal fields. It should be noted that scales of the
horizontal axes are different. 
}
\end{figure}

In Fig.~\ref{fig:phase3_qd}, the distributions of phase closure of
three modes, $\theta_{\bm{k}_1} + \theta_{\bm{k}_2} - \theta_{\bm{k}_1
  + \bm{k}_2}$, are plotted.
Choices of three wavevectors are the same as in
Fig.~\ref{fig:phase3_vt}. The phase closures exhibit small
deviations from homogeneous distribution.

\begin{figure}
\includegraphics[width=25pc]{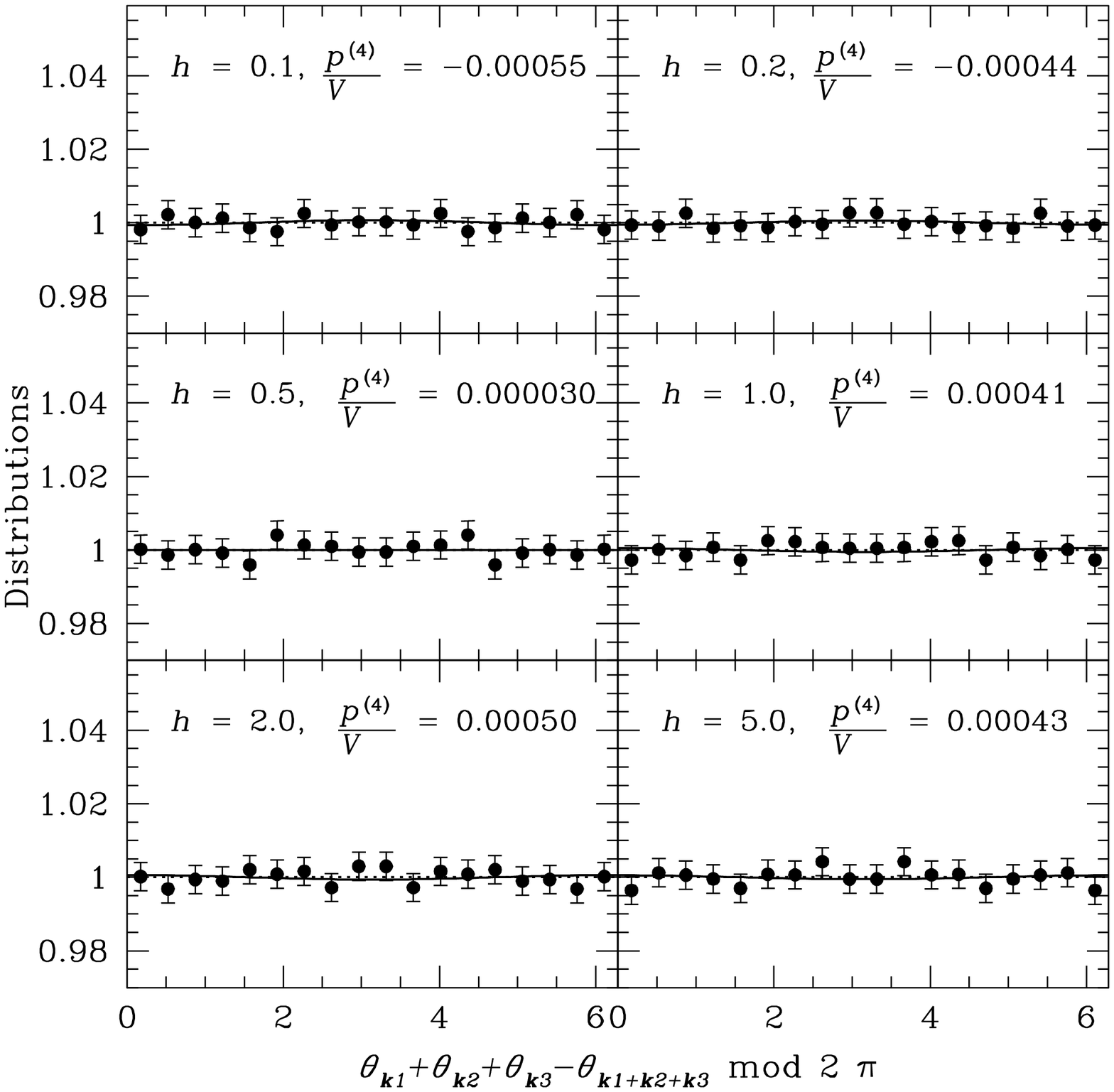}
\caption{\label{fig:phase4_qd} Same as Fig.~\ref{fig:phase4_vt},
but for lognormal fields.
}
\end{figure}

In Fig.~\ref{fig:phase4_qd}, the distributions of phase closure of
four modes, $\theta_{\bm{k}_1} + \theta_{\bm{k}_2} + \theta_{\bm{k}_3}
- \theta_{\bm{k}_1 + \bm{k}_2 + \bm{k}_3}$, are plotted.
The distributions are almost homogeneous.

In the quadratic fields, distribution functions of both Fourier
modulus and phase closures are similar to random Gaussian fields.
This result is partly because the volume $V$ is much larger than the
characteristic scales of clustering in our example. When the volume is
effectively small, small deviations seen in Fig~\ref{fig:phase3_qd},
for example, will be enhanced.


\section{Conclusions}

In this paper, statistical behavior of Fourier modes in general
non-Gaussian fields is studied by explicitly deriving the joint
distribution function of the Fourier coefficients for the first time.
A distribution function for a random Gaussian field is very simple:
each Fourier coefficients are independently distributed and Gaussian.
In a non-Gaussian field, there are complex couplings among all modes,
and we provide a general framework of entangling this complexity. The
distribution function is generally expanded by a hierarchy of
polyspectra in Eq.~(\ref{eq:2-25}), which has full information of
statistical properties of the mode couplings.

The distribution function is formally considered as a series expansion
of $V^{-1/2}$, where $V$ is a total volume of the field. If we take a
sufficiently large volume, the joint distribution for a particular set
of modes approaches to the Gaussian distribution. However, this does
not mean that the field itself is Gaussian. Non-Gaussianity enters the
distribution function in a volume-dependent way. This is a reason why
there appears a total volume in relations to define polyspectra by
Fourier coefficients.

The general distribution function is explicitly calculated up to
second order of $V^{-1/2}$, which is given in Appendix~\ref{app:a}.
The distribution function up to this order depends only on bispectra
and trispectra. Information of higher-order polyspectra is contained
in higher-order terms of $V^{-1/2}$.

We derive $N$-point distribution functions from the general equations.
A closed form of expression for the one-point distribution function
[Eq.~(\ref{eq:2-44})] is derived using all the hierarchy of collapsed
polyspectra, $q^{(2n)}(\bm{k})$. As a consequence, one-point
distribution of Fourier phase is shown to always be homogeneous for a
random field in a spatially homogeneous space (Theorem~\ref{th:2-1}).
As another consequence, the probability distribution function of a
particular mode is Gaussian in a large-volume limit
(Theorem~\ref{th:2-2}). 

For higher-point distribution functions, contributions from
lower-point functions are separated, and reduced $N$-point
distribution functions are introduced. Explicit equations for the
reduced functions up to second order are given for $N=2, 3, 4$.
Structures of mode couplings in non-Gaussian fields are generally
given for the first time in an analytically tractable way. These
equations of the general joint distribution function and $N$-point
distribution functions in terms of polyspectra are fundamental
equations that can be used to investigate the statistics of
Fourier modes in non-Gaussian fields, in general.

The statistics of phase correlations are focused on as an application
of the general results. The structure of the phase correlations in
non-Gaussian fields has been a long-standing issue. We believe our
analysis in this work also provide a breakthrough in this respect. The
distribution function of Fourier phases are straightforwardly obtained
from our general results. We derive analytic expressions for
$N$-point distribution functions of phases in terms of polyspectra.
Explicit expressions up to second order are given.

Regarding the phase correlations, we obtain several theorems which are
proven by using the full expression of the joint distribution
function, i.e., the results without assuming truncated expression of
the series. The essence of the theorems is that phase correlations
among $N$ modes $\bm{k}_1, \ldots \bm{k}_N$ are present only when
there is a relation
\begin{equation}
  j_1 \bm{k}_1 + j_2 \bm{k}_2 + \cdots + j_N \bm{k}_N
  = \bm{0},
\label{eq:5-1}
\end{equation}
where $j_1,\ldots,j_N$ are integers and at least two of them is not
zero. Since the zero mode $\bm{k} = \bm{0}$ is excluded, there should
be at least one linear relation among the wavevectors with integral
coefficients. As a result, there should not be any inhomogeneous
distribution of a particular phase, because Eq.~(\ref{eq:5-1}) with
$N=1$ is not possible (This is another aspect of
Theorem~\ref{th:2-1}). For the two-point phase distributions, there
are phase correlations only between modes whose wavevectors are
parallel to each other and corresponding proportional factor should be
a rational number (Theorem~\ref{th:3-2}).

Some results beyond the second order are obtained for phase
correlations. A lowest-order contribution to the two-point phase
distributions is analytically given by Eq.~(\ref{eq:3-8}), even when
the proportional factor of the two wavevectors are not simple and
correlations are of arbitrarily higher order. Similar expressions for
$N$-point distribution functions in a limited case that there is
only one linear relation of Eq.~(\ref{eq:5-1}) among wavevectors
[Eq.~(\ref{eq:3-26})]. When there are more than one linear relations
among the wavevectors, the expression could be more tedious as in
Eq.~(\ref{eq:3-21}), for example.

As numerical checks of the derived equations, we compare some of the
analytic equations with numerical realizations of three types of
non-Gaussian random fields. We consider the Voronoi tessellation
fields, the lognormal fields, and the quadratic fields. The statistics
of Fourier modes differently deviates from Gaussian distributions,
depending on which type of non-Gaussian field is analyzed. The
distributions of Fourier modulus are easily distorted in Voronoi
tessellation fields, while that in quadratic fields are almost
Gaussian. In the three- and four-point phase distributions, deviations
from Gaussianity in Voronoi tessellation fields and in quadratic
fields are small, while that of lognormal fields can be large.
Although these tendency is not general and depends on our choice of
scales and configurations of wavevectors, the deviations from
Gaussianity appear quite differently from fields to fields. 

The derived analytic results describe the numerical results very
well. As for the simple distribution functions considered in this
work, lowest-order results are sufficient to describe the numerical
results in most of the cases. This is partly because the volume $V$ is
sufficiently larger than the scales of modes we analyze.

This work provides a basic framework toward understanding the
statistical nature of the Fourier analysis. In general, the joint
distribution function of random variables has the full information on
statistical properties of the variables. Therefore, all the
statistical information on the Fourier modes are contained in the
derived joint distribution function. We have analyzed limited number
of consequences from this function. Various aspects of the mode
couplings in non-Gaussian fields, which arises as dynamical effects in
physical situations, might be investigated in future work.




\begin{acknowledgments}
    I wish to thank Alex Szalay and Chiaki Hikage for discussion. I
    acknowledge support from the Ministry of Education, Culture,
    Sports, Science, and Technology, Grant-in-Aid for Encouragement of
    Young Scientists, 15740151, 2003, and Grant-in-Aid for Scientific
    Research, 18540260, 2006.
\end{acknowledgments}

\appendix

\section{
\label{app:a}
General Joint Distribution Function of Fourier Coefficients Up to
Second Order
}


In this appendix, the joint probability distribution function of
Fourier modes in terms of amplitude $A_{\bm{k}}$ and phase
$\theta_{\bm{k}}$ in non-Gaussian fields up to second order is
expressed in terms of summations of independent modes. The form of
Eq.~(\ref{eq:2-30}) is not useful because the modes in each summation
with different labels can be the same as described in the main text.
We expand each summation in Eq.~(\ref{eq:2-30}) so that the different
labels refer different modes. The calculation requires careful
classification of overlapping wavevectors in each summation. After
tedious manipulations, the result has a form,
\begin{equation}
  \frac{P}{P_{\rm G}} =
    1 + 
    \frac{1}{\sqrt{V}} \sum_{i=1}^2 \mathcal{Q}^{(1)}_i +
    \frac{1}{V} \sum_{i=1}^{22} \mathcal{Q}^{(2)}_i
    + \mathcal{O}(V^{-3/2}),
\label{eq:a-1}
\end{equation}
where $\mathcal{Q}^{(1)}_i$ and  $\mathcal{Q}^{(2)}_i$ are given by
\begin{flalign}
  \mathcal{Q}^{(1)}_1 =
&
  \Sumprime_{\bm{k}_1, \bm{k}_2}^\uhs
  {A_{\bm{k}_1}}^2 A_{\bm{k}_2}
  \cos\left(2\theta_{\bm{k}_1} - \theta_{\bm{k}_2}\right)
  p^{(3)}\left(\bm{k}_1,\bm{k}_1, -\bm{k}_2\right),
&
\label{eq:11a}
\end{flalign}
\begin{flalign}
  \mathcal{Q}^{(1)}_2 =
&  \Sumprime_{\bm{k}_1, \bm{k}_2, \bm{k}_3}^\uhs
  A_{\bm{k}_1} A_{\bm{k}_2} A_{\bm{k}_3}
  \cos\left(
    \theta_{\bm{k}_1} + \theta_{\bm{k}_2} - \theta_{\bm{k}_3}\right)
  p^{(3)}\left(\bm{k}_1,\bm{k}_2, -\bm{k}_3\right),
&
\label{eq:11b}
\end{flalign}
\begin{flalign}
  \mathcal{Q}^{(2)}_{1} =
&
  \sum_{\bm{k}_1}^\uhs
  \left(
    \frac14 {A_{\bm{k}_1}}^4 - {A_{\bm{k}_1}}^2 + \frac12
  \right)
  p^{(4)}\left(\bm{k}_1,\bm{k}_1,-\bm{k}_1,-\bm{k}_1\right),
&
\label{eq:12u}
\end{flalign}
\begin{flalign}
  \mathcal{Q}^{(2)}_{2} =
&
  \Sumprime_{\bm{k}_1, \bm{k}_2}^\uhs
  \frac12
  \left(
    {A_{\bm{k}_1}}^2 {A_{\bm{k}_2}}^2 - {A_{\bm{k}_1}}^2 - {A_{\bm{k}_2}}^2 + 1
  \right)
  p^{(4)}\left(\bm{k}_1,\bm{k}_2,-\bm{k}_1,-\bm{k}_2\right),
&
\label{eq:12v}
\end{flalign}
\begin{flalign}
  \mathcal{Q}^{(2)}_{3} =
&
  \Sumprime_{\scriptstyle \bm{k}_1, \bm{k}_2}^\uhs
  \frac13
  {A_{\bm{k}_1}}^3 {A_{\bm{k}_2}}
  \cos\left(3\theta_{\bm{k}_1}-\theta_{\bm{k}_2}\right)
  p^{(4)}\left(\bm{k}_1,\bm{k}_1,\bm{k}_1, -\bm{k}_2\right),
&
\label{eq:12r}
\end{flalign}
\begin{flalign}
  \mathcal{Q}^{(2)}_{4} =
&
  \Sumprime_{\bm{k}_1, \bm{k}_2, \bm{k}_3}^\uhs
  {A_{\bm{k}_1}}^2 {A_{\bm{k}_2}} {A_{\bm{k}_3}}
  \cos\left(2\theta_{\bm{k}_1} + \theta_{\bm{k}_2} - \theta_{\bm{k}_3}\right)
  p^{(4)}\left(\bm{k}_1,\bm{k}_1,\bm{k}_2,-\bm{k}_3\right),
&
\label{eq:12s}
\end{flalign}
\begin{flalign}
  \mathcal{Q}^{(2)}_{5} =
&
  \Sumprime_{\scriptstyle \bm{k}_1, \bm{k}_2, \bm{k}_3}^\uhs
  \frac12 {A_{\bm{k}_1}}^2 {A_{\bm{k}_2}} {A_{\bm{k}_3}}
  \cos\left(2\theta_{\bm{k}_1} - \theta_{\bm{k}_2} -
      \theta_{\bm{k}_3}\right)
  p^{(4)}\left(\bm{k}_1,\bm{k}_1,-\bm{k}_2,-\bm{k}_3\right),
&
\label{eq:12p}
\end{flalign}
\begin{flalign}
  \mathcal{Q}^{(2)}_{6} =
&
  \Sumprime_{\bm{k}_1, \bm{k}_2, \bm{k}_3, \bm{k}_4}^\uhs
  \frac13
  {A_{\bm{k}_1}} {A_{\bm{k}_2}} {A_{\bm{k}_3}} {A_{\bm{k}_4}}
  \cos\left(\theta_{\bm{k}_1} + \theta_{\bm{k}_2} + \theta_{\bm{k}_3}
    - \theta_{\bm{k}_4}\right)
  p^{(4)}\left(\bm{k}_1,\bm{k}_2,\bm{k}_3,-\bm{k}_4\right),
&
\label{eq:12t}
\end{flalign}
\begin{flalign}
&
  \mathcal{Q}^{(2)}_{7} =
  \Sumprime_{\bm{k}_1, \bm{k}_2, \bm{k}_3, \bm{k}_4}^\uhs
  \frac14
  {A_{\bm{k}_1}} {A_{\bm{k}_2}} {A_{\bm{k}_3}} {A_{\bm{k}_4}}
  \cos\left(\theta_{\bm{k}_1} + \theta_{\bm{k}_2} - \theta_{\bm{k}_3}
    - \theta_{\bm{k}_4}\right)
  p^{(4)}\left(\bm{k}_1,\bm{k}_2,-\bm{k}_3,-\bm{k}_4\right),
&
\label{eq:12q}
\end{flalign}
\begin{flalign}
  \mathcal{Q}^{(2)}_{8} =
&
  \Sumprime_{\bm{k}_1, \bm{k}_2}^\uhs
  \left[
    \frac12 {A_{\bm{k}_1}}^4 {A_{\bm{k}_2}}^2
    \cos^2\left(2\theta_{\bm{k}_1} - \theta_{\bm{k}_2}\right) -
    {A_{\bm{k}_1}}^2 {A_{\bm{k}_2}}^2 -
    \frac14 {A_{\bm{k}_1}}^4 + {A_{\bm{k}_1}}^2 +
    \frac12 {A_{\bm{k}_2}}^2  - \frac12
  \right]
\nonumber\\
& \hspace{22pc} \times
  \left[p^{(3)}\left(\bm{k}_1,\bm{k}_1,-\bm{k}_2\right)\right]^2,
&
\label{eq:12a}
\end{flalign}
\begin{flalign}
  \mathcal{Q}^{(2)}_{9} =
&
  \Sumprime_{\bm{k}_1, \bm{k}_2, \bm{k}_3}^\uhs
  \biggl[
    {A_{\bm{k}_1}}^2 {A_{\bm{k}_2}}^2 {A_{\bm{k}_3}}^2
    \cos^2
    \left(\theta_{\bm{k}_1}+\theta_{\bm{k}_2}-\theta_{\bm{k}_3}\right)
    - \frac12
    \left(
      {A_{\bm{k}_1}}^2 {A_{\bm{k}_2}}^2
      + {A_{\bm{k}_1}}^2 {A_{\bm{k}_3}}^2
      + {A_{\bm{k}_2}}^2 {A_{\bm{k}_3}}^2
    \right.
\nonumber\\
& \hspace{11.5pc}
    \left.
      - {A_{\bm{k}_1}}^2 - {A_{\bm{k}_2}}^2
      - {A_{\bm{k}_3}}^2 + 1
    \right)
  \biggr]
  \left[p^{(3)}\left(\bm{k}_1,\bm{k}_2,-\bm{k}_3\right)\right]^2,
&
\label{eq:12f}
\end{flalign}
\begin{flalign}
  \mathcal{Q}^{(2)}_{10} =
&  \Sumprime_{\bm{k}_1, \bm{k}_2, \bm{k}_3}^\uhs
  \left[
    {A_{\bm{k}_1}}^2 {A_{\bm{k}_2}}^3 {A_{\bm{k}_3}}
    \cos\left(2\theta_{\bm{k}_1} - \theta_{\bm{k}_2}\right)
    \cos\left(2\theta_{\bm{k}_2} - \theta_{\bm{k}_3}\right)
  \right.
\nonumber\\
& \hspace{5pc}
  \left. -
    {A_{\bm{k}_1}}^2 {A_{\bm{k}_2}} {A_{\bm{k}_3}}
    \cos\left(2\theta_{\bm{k}_1}+\theta_{\bm{k}_2} - \theta_{\bm{k}_3}\right)
  \right]
  p^{(3)}\left(\bm{k}_1,\bm{k}_1,-\bm{k}_2\right)
  p^{(3)}\left(\bm{k}_2,\bm{k}_2,-\bm{k}_3\right),
&
\label{eq:12c}
\end{flalign}
\begin{flalign}
  \mathcal{Q}^{(2)}_{11} =
&
  \Sumprime_{\bm{k}_1, \bm{k}_2, \bm{k}_3}^\uhs
  \biggl[
    2 {A_{\bm{k}_1}}^3 {A_{\bm{k}_2}}^2 A_{\bm{k}_3}
    \cos\left(2\theta_{\bm{k}_1} - \theta_{\bm{k}_2}\right)
    \cos\left(\theta_{\bm{k}_1} + \theta_{\bm{k}_2} - \theta_{\bm{k}_3}\right) -
    {A_{\bm{k}_1}}^3 A_{\bm{k}_3}
    \cos\left(3\theta_{\bm{k}_1} - \theta_{\bm{k}_3}\right)
&
\nonumber \\
& \hspace{5pc}
   -\; 2 {A_{\bm{k}_1}} A_{\bm{k}_2}^2 {A_{\bm{k}_3}}
    \cos\left(\theta_{\bm{k}_1} - 2\theta_{\bm{k}_2} + \theta_{\bm{k}_3}\right)
  \biggr]
  p^{(3)}\left(\bm{k}_1,\bm{k}_1,-\bm{k}_2\right)
  p^{(3)}\left(\bm{k}_1,\bm{k}_2,-\bm{k}_3\right), 
&
\label{eq:12b}
\end{flalign}
\begin{flalign}
  \mathcal{Q}^{(2)}_{12} =
&
  \Sumprime_{\bm{k}_1, \bm{k}_2, \bm{k}_3, \bm{k}_4}^\uhs
  \frac12
  {A_{\bm{k}_1}}^2 {A_{\bm{k}_2}}^2 {A_{\bm{k}_3}} {A_{\bm{k}_4}}
  \cos\left(2\theta_{\bm{k}_1}-\theta_{\bm{k}_3}\right)
  \cos\left(2\theta_{\bm{k}_2}-\theta_{\bm{k}_4}\right)
\nonumber \\
& \hspace{16pc} \times
  p^{(3)}\left(\bm{k}_1,\bm{k}_1,-\bm{k}_3\right)
  p^{(3)}\left(\bm{k}_2,\bm{k}_2,-\bm{k}_4\right),
&
\label{eq:12e}
\end{flalign}
\begin{flalign}
  \mathcal{Q}^{(2)}_{13} =
&
  \Sumprime_{\bm{k}_1, \bm{k}_2, \bm{k}_3, \bm{k}_4}^\uhs
  \biggl[
    2{A_{\bm{k}_1}}^3 {A_{\bm{k}_2}} {A_{\bm{k}_3}} {A_{\bm{k}_4}}
    \cos\left(2\theta_{\bm{k}_1}-\theta_{\bm{k}_3}\right)
    \cos\left(\theta_{\bm{k}_1}-\theta_{\bm{k}_2}+\theta_{\bm{k}_4}\right)
\nonumber\\
&
\hspace{11pc}
  -\;
    2{A_{\bm{k}_1}} {A_{\bm{k}_2}} {A_{\bm{k}_3}} {A_{\bm{k}_4}}
    \cos\left(\theta_{\bm{k}_1} + \theta_{\bm{k}_2}
      - \theta_{\bm{k}_3} - \theta_{\bm{k}_4}\right)
  \biggr]
\nonumber\\
& \hspace{16pc} \times
  p^{(3)}\left(\bm{k}_1,\bm{k}_1,-\bm{k}_3\right)
  p^{(3)}\left(\bm{k}_1,-\bm{k}_2,\bm{k}_4\right),
&
\label{eq:12d}
\end{flalign}
\begin{flalign}
  \mathcal{Q}^{(2)}_{14} =
&
  \Sumprime_{\bm{k}_1, \bm{k}_2, \bm{k}_3, \bm{k}_4}^\uhs
  \Bigl[
    {A_{\bm{k}_1}}^3 {A_{\bm{k}_2}} {A_{\bm{k}_3}}
    {A_{\bm{k}_4}}
    \cos\left(2\theta_{\bm{k}_1}-\theta_{\bm{k}_4}\right)
    \cos\left(\theta_{\bm{k}_1}-\theta_{\bm{k}_2}-\theta_{\bm{k}_3}\right)
\nonumber \\
& \hspace{11pc}
    - {A_{\bm{k}_1}} {A_{\bm{k}_2}} {A_{\bm{k}_3}} {A_{\bm{k}_4}}
    \cos\left(
      \theta_{\bm{k}_1}+\theta_{\bm{k}_2}+\theta_{\bm{k}_3} -
      \theta_{\bm{k}_4}
    \right)
  \Bigr]
\nonumber\\
& \hspace{16pc} \times
  p^{(3)}\left(\bm{k}_1,\bm{k}_1,-\bm{k}_4\right)
  p^{(3)}\left(-\bm{k}_1,\bm{k}_2,\bm{k}_3\right),
&
\label{eq:12g}
\end{flalign}
\begin{flalign}
  \mathcal{Q}^{(2)}_{15} =
&
  \Sumprime_{\bm{k}_1, \bm{k}_2, \bm{k}_3, \bm{k}_4}^\uhs
  \Bigl[
    2{A_{\bm{k}_1}}^2 {A_{\bm{k}_2}} {A_{\bm{k}_3}} {A_{\bm{k}_4}}^2
    \cos\left(2\theta_{\bm{k}_1} - \theta_{\bm{k}_4}\right)
    \cos\left(\theta_{\bm{k}_2} - \theta_{\bm{k}_3}
      + \theta_{\bm{k}_4}\right)
\nonumber \\
& \hspace{11pc}
    - {A_{\bm{k}_1}}^2 {A_{\bm{k}_2}} {A_{\bm{k}_3}}
    \cos\left(2\theta_{\bm{k}_1} + \theta_{\bm{k}_2} -
    \theta_{\bm{k}_3}\right)
  \Bigr]
\nonumber \\
&
\hspace{16pc}
\times
  p^{(3)}\left(\bm{k}_1,\bm{k}_1,-\bm{k}_4\right)
  p^{(3)}\left(\bm{k}_2,-\bm{k}_3,\bm{k}_4\right),
&
\label{eq:12h}
\end{flalign}
\begin{flalign}
  \mathcal{Q}^{(2)}_{16} =
&
  \Sumprime_{\bm{k}_1, \bm{k}_2, \bm{k}_3, \bm{k}_4}^\uhs
  \biggl[
    {A_{\bm{k}_1}}^2 {A_{\bm{k}_2}} {A_{\bm{k}_3}} {A_{\bm{k}_4}}^2
    \cos\left(2\theta_{\bm{k}_1}-\theta_{\bm{k}_4}\right)
    \cos\left(\theta_{\bm{k}_2}+\theta_{\bm{k}_3}-\theta_{\bm{k}_4}\right)
\nonumber \\
&
\hspace{11pc}
    - \frac12 {A_{\bm{k}_1}}^2 {A_{\bm{k}_2}} {A_{\bm{k}_3}}
    \cos\left(2\theta_{\bm{k}_1}-\theta_{\bm{k}_2}-\theta_{\bm{k}_3}\right)
  \biggr]
\nonumber \\
&
\hspace{16pc}
\times
  p^{(3)}\left(\bm{k}_1,\bm{k}_1,-\bm{k}_4\right)
  p^{(3)}\left(\bm{k}_2,\bm{k}_3,-\bm{k}_4\right),
&
\label{eq:12i}
\end{flalign}
\begin{flalign}
  \mathcal{Q}^{(2)}_{17} =
&
  \Sumprime_{\bm{k}_1, \bm{k}_2, \bm{k}_3, \bm{k}_4}^\uhs
  \biggl[
    4{A_{\bm{k}_1}}^2 {A_{\bm{k}_2}} {A_{\bm{k}_3}} {A_{\bm{k}_4}}^2
    \cos\left(\theta_{\bm{k}_1}+\theta_{\bm{k}_2}-\theta_{\bm{k}_4}\right)
    \cos\left(\theta_{\bm{k}_1}-\theta_{\bm{k}_3}+\theta_{\bm{k}_4}\right)
\nonumber \\
&
\hspace{3pc}
  -\; 2{A_{\bm{k}_1}}^2 {A_{\bm{k}_2}} {A_{\bm{k}_3}}
    \cos\left(2\theta_{\bm{k}_1}+\theta_{\bm{k}_2}-\theta_{\bm{k}_3}\right)
  - 2{A_{\bm{k}_2}} {A_{\bm{k}_3}} {A_{\bm{k}_4}}^2
    \cos\left(\theta_{\bm{k}_2}+\theta_{\bm{k}_3}-2\theta_{\bm{k}_4}\right)
  \biggr]
\nonumber \\
&
\hspace{16pc} \times
  p^{(3)}\left(\bm{k}_1,\bm{k}_2,-\bm{k}_4\right)
  p^{(3)}\left(\bm{k}_1,-\bm{k}_3,\bm{k}_4\right),
&
\label{eq:12k}
\end{flalign}
\begin{flalign}
  \mathcal{Q}^{(2)}_{18} =
&
  \Sumprime_{\bm{k}_1, \bm{k}_2, \bm{k}_3, \bm{k}_4, \bm{k}_5}^\uhs
  {A_{\bm{k}_1}}^2 {A_{\bm{k}_2}} {A_{\bm{k}_3}} {A_{\bm{k}_4}} {A_{\bm{k}_5}}
  \cos\left(2\theta_{\bm{k}_1}-\theta_{\bm{k}_4}\right)
  \cos\left(\theta_{\bm{k}_2}+\theta_{\bm{k}_3}-\theta_{\bm{k}_5}\right)
\nonumber \\
&
\hspace{16pc} \times
  p^{(3)}\left(\bm{k}_1,\bm{k}_1,-\bm{k}_4\right)
  p^{(3)}\left(\bm{k}_2,\bm{k}_3,-\bm{k}_5\right),
&
\label{eq:12j}
\end{flalign}
\begin{flalign}
  \mathcal{Q}^{(2)}_{19} =
&
  \Sumprime_{\bm{k}_1, \bm{k}_2, \bm{k}_3, \bm{k}_4, \bm{k}_5}^\uhs
  \biggl[
  2{A_{\bm{k}_1}} {A_{\bm{k}_2}} {A_{\bm{k}_3}} {A_{\bm{k}_4}} {A_{\bm{k}_5}}^2
  \cos\left(\theta_{\bm{k}_1}-\theta_{\bm{k}_3}-\theta_{\bm{k}_5}\right)
  \cos\left(\theta_{\bm{k}_2}-\theta_{\bm{k}_4}+\theta_{\bm{k}_5}\right)
\nonumber \\
&
\hspace{11pc}
  - \;
  {A_{\bm{k}_1}} {A_{\bm{k}_2}} {A_{\bm{k}_3}} {A_{\bm{k}_4}}
  \cos\left(\theta_{\bm{k}_1}+\theta_{\bm{k}_2}
    -\theta_{\bm{k}_3}-\theta_{\bm{k}_4}\right)
  \biggr]
\nonumber \\
&
\hspace{15pc} \times
  p^{(3)}\left(\bm{k}_1,-\bm{k}_3,-\bm{k}_5\right)
  p^{(3)}\left(\bm{k}_2,-\bm{k}_4,\bm{k}_5\right),
&
\label{eq:12l}
\end{flalign}
\begin{flalign}
  \mathcal{Q}^{(2)}_{20} =
&
  \Sumprime_{\bm{k}_1, \bm{k}_2, \bm{k}_3, \bm{k}_4, \bm{k}_5}^\uhs
  \Bigl[
    2{A_{\bm{k}_1}} {A_{\bm{k}_2}} {A_{\bm{k}_3}} {A_{\bm{k}_4}} {A_{\bm{k}_5}}^2
    \cos\left(\theta_{\bm{k}_1}+\theta_{\bm{k}_2}-\theta_{\bm{k}_5}\right)
    \cos\left(\theta_{\bm{k}_3}-\theta_{\bm{k}_4}+\theta_{\bm{k}_5}\right)
\nonumber \\
&
\hspace{11pc}
   - {A_{\bm{k}_1}} {A_{\bm{k}_2}} {A_{\bm{k}_3}} {A_{\bm{k}_4}}
    \cos\left(\theta_{\bm{k}_1} + \theta_{\bm{k}_2} + \theta_{\bm{k}_3}
      - \theta_{\bm{k}_4}\right)
  \Bigr]
\nonumber \\
&
\hspace{16pc} \times
  p^{(3)}\left(\bm{k}_1,\bm{k}_2,-\bm{k}_5\right)
  p^{(3)}\left(\bm{k}_3,-\bm{k}_4,\bm{k}_5\right),
&
\label{eq:12m}
\end{flalign}
\begin{flalign}
  \mathcal{Q}^{(2)}_{21} =
&
  \Sumprime_{\bm{k}_1, \bm{k}_2, \bm{k}_3, \bm{k}_4, \bm{k}_5}^\uhs
  \biggl[
    \frac12 {A_{\bm{k}_1}} {A_{\bm{k}_2}} {A_{\bm{k}_3}}
    {A_{\bm{k}_4}} {A_{\bm{k}_5}}^2
    \cos\left(\theta_{\bm{k}_1}+\theta_{\bm{k}_2}-\theta_{\bm{k}_5}\right)
    \cos\left(\theta_{\bm{k}_3} + \theta_{\bm{k}_4} - \theta_{\bm{k}_5}\right)
\nonumber \\
&
\hspace{11pc}
  - \frac14 {A_{\bm{k}_1}} {A_{\bm{k}_2}} {A_{\bm{k}_3}} {A_{\bm{k}_4}}
    \cos\left(\theta_{\bm{k}_1} + \theta_{\bm{k}_2} - \theta_{\bm{k}_3}
      - \theta_{\bm{k}_4}\right)
  \biggr]
\nonumber \\
&
\hspace{16pc} \times
  p^{(3)}\left(\bm{k}_1,\bm{k}_2,-\bm{k}_5\right)
  p^{(3)}\left(\bm{k}_3,\bm{k}_4,-\bm{k}_5\right),
&
\label{eq:12n}
\end{flalign}
\begin{flalign}
  \mathcal{Q}^{(2)}_{22} =
&
  \Sumprime_{\bm{k}_1,\bm{k}_2,\bm{k}_3,\bm{k}_4,\bm{k}_5,\bm{k}_6}^\uhs
  \frac12
  {A_{\bm{k}_1}} {A_{\bm{k}_2}} {A_{\bm{k}_3}} {A_{\bm{k}_4}}
  {A_{\bm{k}_5}} {A_{\bm{k}_6}}
  \cos\left(\theta_{\bm{k}_1}+\theta_{\bm{k}_2}-\theta_{\bm{k}_3}\right)
  \cos\left(\theta_{\bm{k}_4}+\theta_{\bm{k}_5}-\theta_{\bm{k}_6}\right)
&
\nonumber \\
&
\hspace{16pc} \times
  p^{(3)}\left(\bm{k}_1,\bm{k}_2,-\bm{k}_3\right)
  p^{(3)}\left(\bm{k}_4,\bm{k}_5,-\bm{k}_6\right),
&
\label{eq:12o}
\end{flalign}
In the above equations, the abbreviated symbol
${\sum}'^\uhs_{\bm{k}_1,\bm{k}_2,\ldots}$ indicates that any two of
wavevectors $\bm{k}_1,\bm{k}_2,\ldots$ appeared in the summation
should be mutually different. It is important to note that each
quantity ${\cal Q}^{(i)}_j$ vanishes when any one of the modes
appearing in a summation is integrated over, because of
Eqs.~(\ref{eq:2-34a})--(\ref{eq:2-34c}).

\newcommand{\apjl}{Astrophys. J. Letters}
\newcommand{\apjs}{Astrophys. J. Suppl.}
\newcommand{\mnras}{Mon. Not. Roy. Astron. Soc.}
\newcommand{\pasj}{Publ. Astron. Soc. Japan}


\end{document}